\documentclass[11pt]{article}

\pdfoutput=1 
\usepackage{./aux/jheppub}
\usepackage[utf8]{inputenc}
\usepackage{graphicx,tikz}
\usepackage{amsfonts,amsmath,amssymb}
\usepackage{array,mathrsfs,yfonts,dsfont,bbm,colonequals,amscd,mathtools}
\usepackage{relsize,suffix,cancel,bbm,capt-of,upgreek,verbatim,caption,subcaption}
\usetikzlibrary{patterns}
\usetikzlibrary{math}

\newcommand{\mS}{\mathrm{s}}
\newcommand{\mT}{\mathrm{t}}
\newcommand{\mU}{\mathrm{u}}

\makeatletter
\newcommand{\thickhline}{%
    \noalign {\ifnum 0=`}\fi \hrule height 1pt
    \futurelet \reserved@a \@xhline
}
\makeatother

\newcommand{\be}{\begin{equation}}
\newcommand{\ee}{\end{equation}}
\newcommand{\ba}{\begin{equation}\begin{aligned}}
\newcommand{\ea}{\end{aligned}\end{equation}}

\def\tildeDf{{\widetilde{\Delta}_\phi}}
\def\etaAdS{{\eta_{\rm AdS}}}
\newcommand\SO{\mathrm{SO}}
\newcommand\tl[1]{\widetilde{#1}}
\newcommand\gegenbauerconst{q}
\newcommand\gegenbauermeasure{\rho}

\newcommand{\zb}{\bar{z}}
\newcommand{\wb}{\bar{w}}

\newcommand{\Df}{\Delta_{\phi}}

\newcommand{\cO}{\mathcal{O}}
\newcommand{\cP}{\mathcal{P}}
\newcommand{\cG}{\mathcal{G}}

\newcommand{\dDisc}{\mathrm{dDisc}}
\newcommand{\Dgap}{\Delta_{\text{gap}}}
\newcommand\imp{\mathrm{imp}}
\newcommand\AdS{\mathrm{AdS}}
\newcommand\cJ{\mathcal{J}}
\def\MM{\mathcal{M}}
\def\GG{\mathcal{G}}

\def\legP{\mathcal{P}}
\newcommand\taushift{\alpha}

\usepackage{mathtools}          

\DeclareMathOperator*{\Res}{Res}
\DeclareMathOperator*{\vol}{vol}

\colorlet{darkblue}{blue!70!black}
\colorlet{darkgreen}{green!70!black}
\colorlet{darkred}{red!70!black}

\newcommand\simon[1]{\textcolor{blue}{simon: {\it #1}}}

\title{AdS Bulk Locality from Sharp CFT Bounds}

\preprint{
YITP-SB-21-9\\
\vspace{-0.34in}
\begin{flushright}CALT-TH 2021-024\end{flushright}
}

\author[a]{Simon Caron-Huot,\!}
\author[b]{Dalimil Maz\'{a}\v{c},\!}
\author[c]{Leonardo Rastelli,\!}
\author[d]{and David Simmons-Duffin}

\affiliation[a]{Department of Physics, McGill University, 3600 Rue University, Montr\'eal, QC Canada}
\affiliation[b]{Institute for Advanced Study, Princeton, NJ 08540, USA}
\affiliation[c]{C. N. Yang Institute for Theoretical Physics, Stony Brook University,
Stony Brook, NY 11794, USA}
\affiliation[d]{Walter Burke Institute for Theoretical Physics, Caltech, Pasadena, CA 91125, USA }

\abstract{
It is a long-standing conjecture that any CFT with a large central charge and a large gap $\Delta_\mathrm{gap}$ in the spectrum of higher-spin single-trace operators must be dual to a local effective field theory in AdS. We prove a sharp form of this conjecture by deriving numerical bounds on bulk Wilson coefficients in terms of $\Delta_\mathrm{gap}$ using the conformal bootstrap. Our bounds exhibit the scaling in $\Delta_\mathrm{gap}$ expected from dimensional analysis in the bulk. Our main tools are dispersive sum rules that provide a dictionary between CFT dispersion relations and S-matrix dispersion relations in appropriate limits. This dictionary allows us to apply recently-developed flat-space methods to construct positive CFT functionals. 
We show how AdS$_4$ naturally resolves the infrared divergences present in 4D flat-space bounds.
Our results imply the validity of twice-subtracted dispersion relations for any S-matrix arising from
the flat-space limit of AdS/CFT.
}

\makeatletter
\gdef\@fpheader{}
\makeatother

\usepackage[T1]{fontenc} 
\usepackage{dsdshorthand}

\begin{document} 

\maketitle
\flushbottom



\section{Introduction}\label{sec:Introduction}

It has long been appreciated that causality and unitarity impose restrictions on  the space of allowed effective field theories (EFTs).
For an EFT to arise as the low-energy approximation to a consistent quantum theory, its Wilson coefficients must obey certain inequalities~\cite{Adams:2006sv}.
These constraints,
which have a long history originating in pion physics (see for example \cite{Martin:1969ina,Pham:1985cr, PhysRevD.51.1093}),
can be derived from canonical assumptions about the S-matrix: analyticity, crossing, boundedness and a positive partial wave decomposition. While these properties have not been rigorously established in all cases, they
are believed to encode the fundamental axioms of unitarity and causality 
(the notion that one cannot send information faster than light).
For quantum field theories in flat space, inequalities that follow from 2 $\to$ 2 scattering have been systematically analyzed in a series of recent papers \cite{Bellazzini:2020cot, Tolley:2020gtv, Arkani-Hamed:2020blm, Caron-Huot:2020cmc, Sinha:2020win, Chiang:2021ziz}. 
The basic strategy is to write a (suitably subtracted) dispersion relation for the amplitude, which expresses low-energy parameters in terms of an unknown but {\it positive} UV spectral density.

An important conclusion is that generic low-energy parameters obey two-sided bounds
compatible with dimensional analysis \cite{Tolley:2020gtv, Caron-Huot:2020cmc}: higher dimensional operators are suppressed by inverse powers of the cut-off.
In effect, in a causal EFT, power-counting rules are implied by causality.
The incorporation of a massless spin two particle (i.e.~dynamical gravity) 
presents an apparent difficulty due to the singular nature of forward scattering,
which was recently overcome in \cite{Caron-Huot:2021rmr}.
The idea was to measure couplings instead at small impact parameter  $b \lesssim 1/M$,
where $M$ is the UV cut-off.
In this way one can derive sharp bounds compatible with dimensional analysis, even in the presence of gravity.

In this paper, we extend this program to weakly coupled gravitational EFTs in asymptotically anti-de Sitter space.  In the purest 
model, we would assume that the massless graviton is the only light particle, with the first massive state appearing at a high scale $M \gg 1/R_{\rm AdS}$. The effective action that captures the physics at energies below the cut-off $M$, written in a schematic notation that ignores the various possible index contractions,
is
\be \label{gravityEFT}
 S_{\rm gravity} =  \frac{1}{16 \pi G} \int d^Dx   \sqrt{-g} \left(- 2 \Lambda + {\cal R} + \alpha_2  {\cal R}^2 + \alpha_3  {\cal R}^3 + \dots \right)\, ,
 \ee
where $ \Lambda = - (D-1) (D-2)/(2R_{\rm AdS}^{2})$ and where the dots indicate an infinite tower of higher dimensional operators. 
For  the EFT to be weakly coupled, the effective gravitational coupling $G E^{D-2}$ must be small at energies $E \lesssim M$. We are then assuming the following hierarchy of scales:
\be \label{hierarchy}
  \frac{1}{R_{\rm AdS}} \ll M \ll M_{\rm Planck}  \equiv G^{\frac{1}{2-D}} \,.
\ee

Effective field theory intuition suggests that the Wilson coefficients  $\alpha_n$ should be suppressed by inverse powers of the cut-off according to their naive dimension, $\alpha_n \sim 1/M^{2n-2}$, so that the theory is local and well approximated by Einstein gravity at energies $E \ll M$.
It is a longstanding open problem to confirm this expectation with precise numerical bounds.
A  beautiful approach to this problem was discussed in \cite{Camanho:2014apa}, where it was shown,
at the level of three-point couplings,
that in order to maintain causality (produce time delays rather than time advances in graviton scattering)
higher derivative corrections to Einstein gravity must be parametrically suppressed by the expected inverse powers of $M$.

An important task left open has been to turn these parametric estimates into sharp bounds with precise numerical coefficients:
\def\lsim{\mathrel{\hbox{\rlap{\lower.55ex \hbox{$\sim$}} \kern-.3em \raise.4ex \hbox{$<$}}}}
\be
 |\alpha_2| \lsim \frac{1}{M^2} \quad\mbox{versus}\quad 
|\alpha_2| \leq \frac{\#}{M^2}\,. 
\ee
Besides ruling out the possibility of numerically large ``$O(1)$'' factors,
sharp bounds are important for several reasons.
For example, from an experimental 
 perspective, relating the size of (hypothetically nonzero) EFT coefficients and the mass of new physical states is clearly valuable.
From a formal perspective, the bootstrap program taught us that many interesting physical theories
live at the boundary of the space of consistent theories; exploring this space requires sharp bounds.

Since dealing with the graviton polarization adds an extra layer of complication orthogonal to the central question, we will find it to convenient to introduce an additional light scalar. For concreteness, let us consider a model where the massless graviton and a scalar $\varphi$ (of mass $m_\varphi \sim 1/R_{\rm AdS}$) are the only single-particle states below the cutoff $M$. Schematically again,
\be \label{scalarmodel}
S =  S_{\rm gravity} +      \int d^Dx   \sqrt{-g} \left( \frac{1}{2}  \varphi (\Box - m_{\varphi}^2) \varphi   +\sum_{n}
  g_{n}  \, D^{2n}  \varphi^4   + \dots \right) \, .
\ee  
We assume that 
 {\it all} EFT interaction are weak, with the overall strength of the scalar quartic couplings controlled by the Newton constant, $g_n \sim  G$. 
The goal is to find precise bounds for $g_{n}$ that reflect the expected EFT scaling, $g_{n}/G \sim M^{2-2n}$.  

We focus in this paper on AdS space because the axioms of gravity are clearest: we take
the perspective that an AdS quantum gravity theory is non-perturbatively \emph{defined} by its
dual CFT, and regard the program of carving out the space of consistent AdS$_{d+1}$ EFTs as
a special corner of the  $d$-dimensional conformal bootstrap.
We follow a boundary CFT approach, in the same spirit as several previous works~\cite{Heemskerk:2009pn, Hofman:2009ug, Afkhami-Jeddi:2016ntf, Caron-Huot:2017vep, Kulaxizi:2017ixa, Costa:2017twz, Meltzer:2017rtf, Kologlu:2019bco, Belin:2019mnx}.
The lofty goal is a derivation of bulk locality from a rigorous bootstrap perspective. 

This problem was cleanly formulated in CFT language in~\cite{Heemskerk:2009pn}. 
The authors envisioned a family of CFTs, parametrized by a small coupling $\epsilon$ that controls the approach to mean field theory and abstracts the rules of large-$N$ expansions.\footnote{In the quintessential example of SU($N_c$) $\mathcal{N}=4$ super Yang-Mills,
$\epsilon= G / R^{D-2}_{\rm AdS}\sim \frac{1}{N_c^2}$.
}
The limiting mean field theory at $\epsilon = 0$
contains
``single-trace'' operators with spin $J_i \leq 2$ and twists $\tau_i  < \Delta_{\rm gap}$, where 
$\Delta_{\rm gap} = M R_{\rm AdS} \gg 1$ is a large scale.
The central conjecture of~\cite{Heemskerk:2009pn} is that these are sufficient conditions for locality of the bulk AdS theory.

There is a large body of evidence for this conjecture, from a variety of CFT arguments, e.g.~\cite{Afkhami-Jeddi:2016ntf, Caron-Huot:2017vep, Kulaxizi:2017ixa, Costa:2017twz, Meltzer:2017rtf,  Kologlu:2019bco, Belin:2019mnx}.
A common thread is to impose causality constraints on CFT  four-point functions,  in kinematic regimes that probe bulk locality. (A precise characterization of one such regime, the ``bulk-point limit'', was given in \cite{Maldacena:2015iua}.)  It was argued in \cite{Afkhami-Jeddi:2016ntf}
that the stress tensor exchange by itself violates causality unless the three-point functions are tuned to be those of Einstein gravity; under certain assumptions about the other contributions, this shows  parametric  suppression of the higher derivative terms.
An approach based on computing the bulk phase shift from CFT and imposing that it obeys causality was described in \cite{Kulaxizi:2017ixa, Costa:2017twz}, leading again to parametric suppression. 
  
The Lorentzian inversion formula \cite{Caron-Huot:2017vep} would seem like a promising start to obtain more quantitative, sharp bounds.
Indeed, it leads to a rigorous non-perturbative upper bound for bulk contact interactions, of the form
\be \label{J bound}
c(\Delta, J) <  \frac{f(\Delta, d) }{ (\Delta_{\rm gap}^2)^{J - 2} } \, ,
\ee
where $J$ is the spin of the interaction (defined as the highest spin in the partial wave expansion of the contact diagram) and  $f(\Delta, d)$ a computable function.
These bounds confirm a key insight of \cite{Heemskerk:2009pn}:
that the higher-derivative corrections that can arise at a given
order in $1/\Delta_{\rm gap}^2$ are supported on finitely many spins.
They are however incomplete, since the spin of an interaction is usually less than its scaling dimension.
This can be illustrated by inspecting the scattering amplitudes corresponding to
the first few scalar self-interactions (the notation follows Section \ref{Smat sum rules}):
\be
\mathcal{M} \supset g_2 (s^2+t^2+u^2) + g_3 stu + g^4(s^2+t^2+u^2)^2+ \ldots\,.
\ee
The first two interactions are spin-2: in the Regge limit $s,t\to\pm\infty$ with $u$ fixed,
they grow like $\sim s^J$ with $J=2$.  Therefore the bound \eqref{J bound} gives the same suppression for $g_2$ and $g_3$,
whereas a trained dimensional analyst would expect $g_3/g_2 \lsim 1/\Delta_{\rm gap}^2$.
The $g_4$ interaction, being spin-4, enjoys a stronger suppression and generally only a finite number of interactions exist below a certain spin.
In weakly coupled AdS EFTs, where tree-level interactions are suppressed by $\epsilon$
(which we assume to be much smaller than any power of $1/\De_\mathrm{gap}$),
\eqref{J bound} is also too weak to be useful.
It was suggested in \cite{Caron-Huot:2017vep} to use stress-tensor sum rules to bound other 
couplings, but this strategy has not led to sharp bounds (and cannot, as we will see).

Finally,  commutativity of lightray operators \cite{Kravchuk:2018htv, Kologlu:2019mfz} (local operators integrated over a null direction) leads to interesting sum rules, known as ``superconvergence sum rules'', that to order $O(\epsilon)$ relate EFT couplings to the heavy {\it single-trace} data; under plausible physical assumptions, they have been  used to argue for parametric suppression of higher derivative couplings \cite{Kologlu:2019bco} (see also \cite{Belin:2019mnx}).
Turning those estimates into sharp bounds has however remained a stumbling block.
This is what we accomplish in the present paper.
 
\subsection{Our approach}

\subsubsection{Setup and assumptions}

Let us state our assumptions more  precisely, focusing on the  CFT version of
the  model~(\ref{scalarmodel}). In this example, 
the only single-trace operators with  twist $\tau  < \Delta_{\rm gap}$ are the stress tensor $T_{\mu \nu}$ and a scalar primary $\phi$ of  dimension $\Delta_\phi \ll \Delta_{\rm gap}$.\footnote{Note that $\Delta_\mathrm{gap}$ is conceptually a twist gap (not a dimension gap) for us. Still, we follow the literature and use the name $\Delta_\mathrm{gap}$.} 
The conformal block decomposition of the  four-point function of $\phi$ (in any channel) takes the form
\be
\langle \phi \phi \phi \phi \rangle = 
\underbrace{  G_{\mathds{1} } + \sum  G_{ [\phi\phi]_{n,\ell}} + G_{T_{\mu \nu}} + G_{[\text{composites}]} }_{\tau<\Dgap}+ \underbrace{\sum G_{{\text{heavy}}}}_{\tau>\Dgap}\,.
\ee
The exchanged primaries with $\tau<\Dgap$ comprise the identity, the double-trace composites of~$\phi$, the stress tensor 
and additional multi-trace composites built out of $\phi$ and $T_{\mu \nu}$. In the strict $\epsilon \to 0$ limit, only the identity and the $\phi$ double traces survive, with exact mean field theory dimensions and OPE coefficients.
We are interested in the leading $O(\epsilon)$ deviation from mean field theory.  At this order, the  stress  tensor  block appears, while
 the double traces acquire $O(\epsilon)$ anomalous dimensions and corrections to their squared OPE coefficients. 
The other light composites  are subleading 
and can be ignored.
The setup could be modified to include other light single-traces (for example, towers of Kaluza-Klein modes),
as long as they all have spin $J\leq 2$.
On the other hand, we make no assumptions about the primaries with $\tau >\Dgap$, other than that they obey the usual unitarity constraints.
While the physics below $\Dgap$ is assumed to be weakly coupled, the physics above $\Dgap$ can be strongly coupled. 

Our objective is to derive constraints on the $O(\epsilon)$ double-trace data. These data are in one-to-one correspondence~\cite{Heemskerk:2009pn} with the quartic couplings
$g_n$ of the AdS effective field theory (\ref{scalarmodel}) (defined precisely in \eqref{eq:MellinEFT}).
We will
establish  precise two-sided bounds that scale with the expected inverse powers of $\Delta_{\rm gap}$.
In essence, we will succeed in  uplifting to AdS
  the  flat-space bounds  of our previous paper \cite{Caron-Huot:2021rmr}, with the  four-point correlator of $\phi$ playing the role of the $2 \to 2$ scalar scattering amplitude. 
While the physical picture is clear,  its implementation has  become technically possible only  thanks to new advances in the analytic conformal bootstrap.
Traditional bootstrap methods (such as the standard numerical oracle \cite{Rattazzi:2008pe}) are inadequate to constrain weakly coupled AdS EFTs, due to the dominant contributions of double-trace composites.
The right tools are the recently derived dispersive CFT sum rules \cite{Mazac:2019shk, Carmi:2019cub,  Penedones:2019tng, Carmi:2020ekr, Caron-Huot:2020adz, Gopakumar:2021dvg, Meltzer:2021bmb}, which 
are ideally suited to study perturbative expansions around mean field theory.\footnote{See \cite{Mazac:2016qev,Mazac:2018mdx} for some early examples of dispersive CFT sum rules and \cite{Mazac:2018ycv,Mazac:2018qmi,Kaviraj:2018tfd,Mazac:2018biw,Paulos:2019fkw,Hartman:2019pcd,Paulos:2019gtx,Paulos:2020zxx} for their further applications.}

\subsubsection{Reminder: dispersive sum rules}

Dispersive CFT sum rules are rooted in Lorentzian kinematics  and the notion of the double discontinuity (dDisc) or double commutator. Let us give an informal sketch of their origin, referring to \cite{Caron-Huot:2020adz} for a detailed treatment.   
The crossing equation for $\langle\phi \phi \phi \phi \rangle$ is precisely the causality constraint
\be
\label{eq:vanishingcommutatorintro}
\langle\Omega|\phi(x_4)[\phi(x_1),\phi(x_3)]\phi(x_2)|\Omega\rangle = 0 \quad \text{for} \quad x_1-x_3\quad\text{spacelike}\, ,
\ee
expanded using vacuum OPEs (meaning that a complete set of states is inserted to the left of the two rightmost operators).
One might intuit that the strongest causality constraints stem from approaching the lightcone,
and dispersive sum rules do just that.
They are derived by integrating $x_1$ and $x_3$
along space-like separated null rays 
against certain meromorphic kernels $f(x_1,x_3)$,
\begin{align}
0&=\quad\int_{-\oo}^\oo dx_1^+\!\!\int_{-\oo}^\oo dx_3^+ f(x_1,x_3)\langle\Omega|\phi(x_4)\phi(x_3)\phi(x_1)\phi(x_2)|\Omega\rangle 
\nn\\
&\quad-\int_{-\oo}^\oo dx_1^+\!\!\int_{-\oo}^\oo dx_3^+ f(x_1,x_3)\langle\Omega|\phi(x_4)\phi(x_1)\phi(x_3)\phi(x_2)|\Omega\rangle\,.
\label{eq:integratedCrossing}
\end{align}
In the absence of the kernel $f(x_1,x_3)$, each term in (\ref{eq:integratedCrossing}) would become a double-commutator, since null-integrated operators kill the vacuum, for example:
\begin{align}
&\int_{-\oo}^\oo dx_1^+\!\!\int_{-\oo}^\oo dx_3^+ \langle\Omega|\phi(x_4)\phi(x_3)\phi(x_1)\phi(x_2)|\Omega\> \nn\\
&= \int_{-\oo}^\oo dx_1^+\!\!\int_{-\oo}^\oo dx_3^+ \langle\Omega|[\phi(x_4),\phi(x_3)][\phi(x_1),\phi(x_2)]|\Omega\>
\qquad (\textrm{without $f(x_1,x_3)$}).
\label{eq:doublecommutator} 
\end{align}
Importantly, the double-commutator (\ref{eq:doublecommutator}) does not get contributions from states with double-trace dimensions $\De=2\De_\f+2n+J$ in the $12\to 34$ OPE.

However, in general $f(x_1,x_3)$ is needed to suppress the endpoints of the null integral (\ref{eq:integratedCrossing}) and ensure convergence.\footnote{No kernel is needed for certain correlators of spinning operators, giving superconvergence relations \cite{Kologlu:2019bco}.} The poles of $f(x_1,x_3)$ then introduce additional contributions not proportional to a double-commutator. Overall, expanding (\ref{eq:integratedCrossing}) using vacuum OPEs, we obtain a sum rule
\begin{align}
\sum_{\De,J} p_{\De,J} \w[G_{\De,J}^s] &= 0,
\end{align}
where $\omega$ is a functional and $\omega[G_{\De,J}^s]$ exhibits double zeros on all double-trace dimensions above some twist $\tau_\mathrm{min}$, ensuring that the heavy and light contributions are separately $O(\epsilon)$. 
There are several other equivalent ways to derive dispersive sum rules \cite{Caron-Huot:2020adz} (notably from dispersion relations in Mellin space \cite{ Penedones:2019tng, Caron-Huot:2020adz}).
Here we have chosen to emphasize their conceptual kinship with classic
physical arguments underlying S-matrix crossing symmetry and dispersion relations \cite{Itzykson:1980rh},
and with superconvergence sum rules, further elaborated upon in Appendix \ref{app:superconvergence}.

\subsubsection{Application to holographic CFTs}

Consider now a dispersive sum rule $\omega$ with $\tau_{\text{min}}<\Dgap$, and let us apply it to a holographic family of CFTs parametrized by $\epsilon$ as described above. We can split the contributions to the sum rule into the light ones with $\tau\leq\Dgap$ and heavy ones with $\tau>\Dgap$. Defining
\be
\omega|_{\text{light}} = \sum\limits_{\tau\leq\Dgap} p_{\De,J}\,\omega [G^s_{\Delta,J}]\,,
\qquad \omega|_{\text{heavy}} = \sum\limits_{\tau>\Dgap} p_{\De,J}\,\omega [G^s_{\Delta,J}]\,,
\ee
the sum rule reads
\be
-\omega|_{\text{light}} = \w|_\mathrm{heavy}\,.
\label{eq:dispSumRule}
\ee
It is easy to see that $\omega|_{\text{light}}= O(\epsilon)$ as $\epsilon\rightarrow 0$. Indeed, in mean field theory ($O(\epsilon^ 0)$), the heavy contribution to $\omega$ vanishes. At $O(\epsilon)$, $\omega|_{\text{light}}$ comes entirely from the anomalous dimensions and OPEs of the double traces with $\tau<\Dgap$ and can be computed from the bulk effective field theory. If we manage to construct 
a functional $\omega$ that is non-negative for $\tau>\Dgap$, we conclude from \eqref{eq:dispSumRule} that
\be
-\omega|_{\text{light}}\geq 0\,.
\label{eq:EFTBoundCFT}
\ee
 In this way, UV consistency gives rise to inequalities satisfied by the low-energy observables. This is in analogy with how dispersive sum rules constrain flat-space EFTs. 
  
  A distinct advantage of the CFT approach to AdS theories is that it is fully rigorous.  The requisite analyticity and causality 
  properties of the four-point correlator are direct consequences of the bootstrap axioms (see \cite{Kravchuk:2021kwe} for a recent discussion of how  Lorentzian CFT axioms follow from the standard Euclidean axioms), while boundedness in the Regge limit
  (with Regge intercept $\leq 1$) is a consequence of unitarity and the OPE at the non-perturbative level \cite{Caron-Huot:2017vep}.

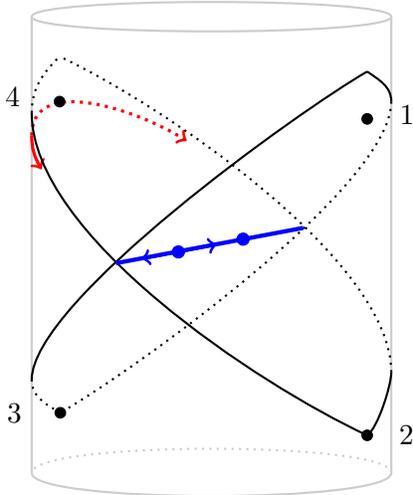
\begin{figure}
\centering
\begin{tikzpicture}
    \draw[looseness=0.14,thick,black!20!white] (0.94,7.03) to[out=-90,in=-90] (5.72,7.03) to[out=90,in=90] (0.94,7.03);
    \draw[looseness=0.22,thick,black!20!white] (0.94,0.96) to[out=-90,in=-90] (5.72,0.96);
    \draw[looseness=0.23,dotted,thick,black!20!white] (5.72,0.96) to[out=90,in=90] (0.94,0.96); 
    \draw[thick,black!20!white] (0.94,0.96) -- (0.94,7.03);
    \draw[thick,black!20!white] (5.72,0.96) -- (5.72,7.03);
    \draw[looseness=1,->,very thick,red,dotted] (0.94,5.45) to[out=90,in=148] (3,5.38);
    \draw[looseness=1,->,very thick,red] (0.94,5.45) to[out=-90,in=120] (1.07,5);
    \draw[fill] (1.32, 1.76) circle (2pt);
    \draw[fill] (5.4,1.46) circle (2pt);
    \draw[fill] (1.31,5.9) circle (2pt);
    \draw[fill] (5.4,5.67) circle (2pt);
    \draw[looseness=0.5,thick] (5.4,1.46) to[out=35,in=-90] (5.72,2.3);
    \draw[looseness=0.63,thick,dotted] (5.72,2.3) to[out=90,in=-25]  (1.31,6.5);
    \draw[looseness=0.67,thick] (5.4,1.46) to[out=155,in=-90] (0.94,5.75);
    \draw[looseness=1,thick,dotted] (0.94,5.75) to[out=90,in=-135] (1.31,6.5);
    \draw[looseness=1,thick,dotted] (1.32, 1.76) to[out=150,in=-90] (0.94,2.2);
    \draw[looseness=0.44,thick] (0.94,2.2) to[out=90,in=-150] (5.4,6.3);
    \draw[looseness=1,thick] (5.4,6.3) to[out=-35,in=90] (5.72,5.9);
    \draw[looseness=0.48,thick,dotted] (5.72,5.9) to[out=-90,in=30] (1.32, 1.76);
    \draw[blue,ultra thick] (2.06,3.75) -- (4.55,4.22);
    \draw[fill,blue] (2.89,3.9) circle (2.3pt);
    \draw[fill,blue] (3.75,4.07) circle (2.3pt);
    \draw[<->,blue,very thick] (2.40,3.81) -- (3.4,4.0);
    \node [left] at (0.94,1.75) {$3$};
    \node [right] at (5.7,5.72) {$1$};
    \node [right] at (5.69,1.47) {$2$};
    \node [left] at (0.92,5.97) {$4$};
\end{tikzpicture}\caption{
A four-point function near the Regge limit. Bulk excitations localize along two null sheets, which intersect in a transverse  hyperbolic space $H_{d-1}$ (thick blue line). The red and blue arrows show the boundary and bulk actions of $\SO(d-1,1)$ isometries, which we exploit to focus in impact parameter. 
\label{fig:etaAdSintro}}
\end{figure}

To construct dispersive functionals that will give sharp bounds for the bulk couplings, it is helpful to develop the following physical picture. We can think of dispersive sum rules as expressing the causality of 2 $\to$ 2 scattering in AdS$_{d+1}$. A choice of $\omega$ corresponds to choosing wavefunctions for the incoming and outgoing particles. We will choose these to mimic the derivation of the corresponding bounds in flat space \cite{Caron-Huot:2021rmr}. There, the key idea was to use wavefunctions which ensure that highly energetic intermediate states scatter at small impact parameter $b \sim 1/M$. The corresponding physical picture of a high energy collision in AdS is sketched in Figure \ref{fig:etaAdSintro}. The high energy (Regge) limit of the AdS/CFT correlator localizes along two null sheets, intersecting in the transverse hyperbolic space $H_{d-1}$ (impact parameter space).\footnote{Previous analyses of CFT correlators in the Regge limit include \cite{Cornalba:2006xk,Cornalba:2006xm,Cornalba:2007fs,Cornalba:2007zb,Costa:2012cb,Li:2017lmh,Caron-Huot:2020nem}.} In order to probe local bulk physics, we must construct dispersive functionals  whose action on heavy blocks $G^s_{\Delta, J}$ is sharply localized at small  impact parameter $\beta \cong 2J/\Delta \ll 1$.
   
We will achieve this starting from the $B_{k, v}$ functionals of \cite{Caron-Huot:2020adz}. Here $k\geq 2$ denotes the number of subtractions used in the dispersion relation, and $v$ is an additional continuous parameter. The action of $B_{k,v}$ on the heavy states is spread over impact parameters of order one, i.e.~comparable to the AdS radius. We will explain how to construct linear combinations of these sum rules which are localized at small impact parameter. Specifically, we will use harmonic analysis on the transverse $H_{d-1}$ 
to find a basis of sum rules with a fixed transverse momentum $\nu$.  
This defines a new family of dispersive sum rules, which we call $C_{k,\nu}$.
They provide a CFT version of the flat-space sum rules ${\cal C}_{k, u}$ introduced in our previous paper~\cite{Caron-Huot:2021rmr}, to which they will reduce in the bulk point limit $J / \Delta \ll 1$.

The $C_{k,\nu}$ sum rules are absolutely convergent
integrals over a ``spacelike kinematics'' region where the commutator in \eqref{eq:vanishingcommutatorintro} vanishes.
For $\nu$ large and in a certain range, strong oscillations make the integral dominated
by a complex saddle point where $x_1{-}x_3$ is effectively timelike.
On the saddle point, the kinematics are effectively timelike, and thus able to causally focus onto a bulk point!
Moments of the S-matrix can thus be measured from spacelike kinematics. We refer to this phenomenon as {\it spacelike scattering}.

Several additional technical hurdles must be cleared. We must control
the action of the functionals also in the Regge limit $J/\Delta = O(1)$, check heavy positivity, and finally construct ``improved'' versions of the sum rules to deal with the graviton pole. When the dust settles, we are able to uplift to AdS the two-sided flat space bounds of \cite{Caron-Huot:2021rmr}. A nice feature of AdS is that it provides an infrared regulator, so the story goes through also in the $D=4$ case. This is in contrast with the situation in flat space, where our argument is precluded in $D=4$ by soft graviton divergences.

For concreteness, we carry out detailed calculations in the model  (\ref{scalarmodel}) of a single light scalar coupled to gravity.  Additional intermediate light states of  spin $J \leq 2$ would have minor effects:
adding  $J =0$ states is of no consequence, because they drop out of any twice-subtracted sum rules, while additional light $J=2$ states would change the precise numerical bounds.
The analysis will turn out to be controlled by saddle points with transparent physical interpretation, suggesting that the method would straightforwardly generalize to spinning external operators.

The remainder of the paper is organized as follows. In Section \ref{sec:CFTSumRules}, we overview the ingredients of our argument. In particular, we review the derivation flat-space bounds of \cite{Caron-Huot:2021rmr} based on the dispersive sum rules ${\cal C}_{k, u}$. We also explain how to construct the analogous dispersive CFT sum rules $C_{k,\nu}$. In Section \ref{sec:heavyAction}, we develop techniques to evaluate actions of dispersive functionals on heavy operators, and apply these techniques to $C_{k,\nu}$.  Section \ref{sec:lightAction} is in turn devoted to the computation of the light contributions to dispersive sum rules. The pieces are put together in Section \ref{sec:holo}, where we explain how to uplift the flat-space bounds to AdS bounds. We conclude in Section \ref{sec:Conclusions}. Several technical appendices complement the main text. 

A dictionary summarizing the translation between relevant quantities in flat space and AdS appears in
Table \ref{tab:dictionary}.

\begin{table}[t]
\centering
  \begin{tabular}{l l l  c}
\thickhline
 & flat space & AdS & eq.\\\hline
amplitude & $\mathcal{M}$ & $\cG$ &\\
discontinuity & $\mathrm{Im}[\mathcal{M}]$ & $\dDisc[\cG]$&\\
energy & $s=m^2$ & $m^2 = (\Delta-J-d+1)(\Delta+J-1)$ & \eqref{eq:massdefinition}\\
angular momentum & $J$ & $J$ &\\
impact parameter & $b=\frac{2J}{m}$ & $\beta=\cosh^{-1}(\eta_{\text{AdS}})  = \log \frac{\Delta+J-1}{\Delta-J-d+1}$ & \eqref{eq:impactparameter}\\
transverse momentum & $u=-p^2$ & $u = -\nu^2$ &\\
Regge limit & $m\gg 1$ at fixed $b$ & $m\gg1$ at fixed $\beta$ &\\
bulk-point limit & $m\gg 1$ at fixed $J$ & $m\gg 1$ at fixed $J$ &\\
number of subtractions & $k$ & $k$ &\\
dispersive sum rule & $\mathcal{C}_{k,u}$ & $C_{k,\nu}$ & \eqref{eq:magicmoment}\\
improved sum rule & $\mathcal{C}^\mathrm{imp}_{k,u}$ & $C^\mathrm{imp}_{k,\nu}$ & \eqref{eq:definitionofk}\\
\thickhline
  \end{tabular}
  \caption{A dictionary between relevant notions in flat space and AdS.}
  \label{tab:dictionary}
\end{table}

\noindent {\bf Note:} While this work was being completed, \cite{Kundu:2021qpi} appeared which has some partial overlap, for example with the non-gravitational bounds from the forward limit discussed in section~\ref{sec:positiveforwardlimit}.


\section{CFT sum rules and their flat space limit}\label{sec:CFTSumRules}

Our key object of study will be a correlation function of four identical scalar operators in a $d$-dimensional CFT:
\be
  \big\langle \phi(x_1)\cdots \phi(x_4)\big\rangle = \frac{\mathcal{G}(u,v)}{(x_{13}^2x_{24}^2)^{\Df}}\,,
\ee
which is a function of two cross-ratios
\be
 u=z\zb=\frac{x_{12}^2x_{34}^2}{x_{13}^2x_{24}^2},\qquad
  v=(1{-}z)(1{-}\zb)=\frac{x_{23}^2x_{14}^2}{x_{13}^2x_{24}^2}\,.
\ee
The correlator is equal to a sum of conformal blocks in any of the three channels $s,t,u$
\be \cG(u,v) =  \sum\limits_{\De,J}
p_{\De,J}\, G^{s}_{\Delta,J}(z,\bar z)
= 
\sum\limits_{\De,J}p_{\De,J}\, G^{t}_{\Delta,J}(z,\bar z)
=
\sum\limits_{\De,J}p_{\De,J}\, G^{u}_{\Delta,J}(z,\bar z)\,,
\end{equation}
where importantly all coefficients $p_{\De,J}$ are positive.
We use the following normalization conventions for $s$-channel conformal blocks,
\be \label{normalization}
G^{s}_{\Delta,J}(z,\zb) \sim (z\zb)^{-\Df} \times z^{\frac{\Delta-J}{2}}\zb^{\frac{\Delta+J}{2}}\quad\textrm{for}\quad 0<z  \ll   \zb  \ll 1\,. 
\ee
The $t$- and $u$-channel blocks are defined by
\ba
G^t_{\De,J}(z,\bar z) &= G^s_{\De,J}(1-z,1-\bar z)\\
G^u_{\De,J}(z,\bar z) &= [(1-z)(1-\bar z)]^{-\De_\f} G^s_{\De,J}\left(\tfrac{1}{1-z},\tfrac{1}{1-\bar z}\right)\,.
\ea

The spectrum of a holographic CFT (we take this to define a holographic CFT) contains
single-trace operators of low spin $J\leq 2$, double-trace operators of any spin and twist $\Delta-J\approx 2\Df+2m$,
together with ``heavy'' higher-spin operators with twist above some large gap $\tau=\Delta-J>\Delta_{\rm gap}$.
In the holographic picture, the low-spin single-traces represent light fields whose interactions are expected to be controlled by a local effective theory
up to a length scale $R_{\rm AdS}/\Delta_{\rm gap}$.  When attempting to relate the light and heavy sectors, however, one runs
into the difficulty that double-trace exchanges contribute strongly to any correlator. 
A solution to this problem was presented in~\cite{Penedones:2019tng,Caron-Huot:2020adz}, which constructed ``dispersive'' sum rules:
\be
 0 = \sum\limits_{\De,J}p_{\De,J} \w[G^s_{\Delta,J}] \label{eq:examplesumrule}
\ee
with the property that the action $\w[G^s_{\Delta,J}]$ has double-zeros in $\De$ on all but a finite number of double-trace families.

The idea of holographic sum rules is to rewrite \eqref{eq:examplesumrule}
by separating out the contribution of heavy states with twist $\tau=\De-J>\De_\mathrm{gap}$:
\be\begin{aligned}
\label{eq:heavylightsplit}
\left.\w\right|_{\mathrm{heavy}} &\equiv \sum_{\tau>\De_\mathrm{gap}} p_{\De,J}\w[G_{\De,J}^s] \\ &= - \left.\w\right|_{\mathrm{light}}.
\end{aligned}\ee
The second line collects together terms with $\tau\leq \De_\mathrm{gap}$.

The light contribution to (\ref{eq:heavylightsplit}) can be computed using low energy effective field theory (EFT) in the bulk, as will be detailed in section \ref{sec:lightAction}.  The heavy contribution
is in general unknown and the idea is to look for functionals such that
$\left.\w\right|_{\mathrm{heavy}}$ is manifestly a sum of positive terms: any such $\omega$ gives
an inequality that the bulk EFT must satisfy. In this section, we review the flat-space sum rules
$\mathcal{C}_{k,u}$ as used in~\cite{Caron-Huot:2020cmc,Caron-Huot:2021rmr}
and $B_{k,v}$ CFT sum rules from \cite{Caron-Huot:2020adz}, and we outline the various steps
which will relate them in the next section.

The key concept will be that the $s$-channel blocks for heavy operators decay exponentially away from the $u$-channel Regge limit.
Thus, the action of $\w$ on heavy blocks is determined by the expansion of its kernel around the Regge limit.
We characterize this expansion in terms of {\it Regge moments}. We will show that, order by order in $1/\Delta_{\rm gap}$, 
the $B_{k,v}$ sum rules provide a complete basis of moments, and furthermore, that they exist in sufficient number that each
can be localized at small AdS \emph{impact parameters}.  These localized sum rules will be direct analogs of familiar flat space sum rules, which we begin by reviewing.

\subsection{Review of S-matrix sum rules}
\label{Smat sum rules}

We consider $2\to 2$ scattering of identical real scalars in $D=d+1$ spacetime dimensions.
For an EFT to be relevant, the scalar must be light compared to the cutoff scale $M$
and we will thus treat it as massless.
The amplitude $\MM(s,t)$ is a function of the usual Mandelstam invariants satisfying $s+t+u=0$.
It is strongly constrained by causality and unitarity,
which together imply analyticity, positivity, and boundedness properties.

Specifically, for fixed momentum transfer $u<0$,
the amplitude is analytic for sufficiently large complex $|s|$ and 
enjoys spin-2 convergence in the Regge limit $|s|\to\infty$,\footnote{Exchange of a particle of spin $J$ grows like $\MM\sim s^J$ in the Regge limit.} meaning that  the following limit vanishes
along any line of constant argument:
\be
 \lim_{|s|\to\infty} \frac{\MM(s,u)}{s^2} = 0\qquad (u<0)\,.
 \label{eq:ReggeBound}
\ee
These conditions amount to convergence of $k$-subtracted dispersion relations with $k\geq 2$, 
and can be summarized by the ${\cal C}_{k,u}$ sum rules defined as:
\be \label{Ckudirect}
{\cal C}_{k, u} \equiv  -\oint_{\infty} \frac{ds'}{2\pi i}  \frac{1}{s'} \frac{\MM(s',u)}{[s'(s' + u)]^{k/2}} = 0 \, ,\qquad u<0,\ k =2, 4, \dots\,.
\ee
These naturally split into low- and high-energy contributions as in \eqref{eq:heavylightsplit}:
using a standard contour deformation argument, one can
replace arcs at infinity by real segments at $s>M^2$ and $u>M^2$,
proportional to the imaginary part of the amplitude, plus a light contribution from arcs with $|s|\sim M$.
One usually doesn't know the amplitude at $s>M^2$ except that it should admit a partial wave expansion;
this gives (see for example \cite{Caron-Huot:2020cmc} for details):
\be\begin{aligned}\label{eq:heavylightsplitC}
\left.-{\cal C}_{k,u}\right|_{\mathrm{light}} \equiv \oint_{|s|\sim M^2} \frac{ds'}{2\pi i}
\frac{1}{s'} \frac{\MM(s',u)}{[s'(s' + u)]^{k/2}}
&= \left\< \frac{2m^2+u}{m^2+u} \frac{\cP_{J}\big(1+\frac{2u}{m ^2}\big)}{(m ^2 (m ^2+u))^{k/2}}\right\> \\ &
\equiv \left.{\cal C}_{k,u}\right|_{\mathrm{heavy}}\,,
\end{aligned}\ee
where the averaging symbol denotes a sum over all heavy states of energy $m$ and spin $J$:
\be
 \left\<(\cdots)\right\> \equiv \frac{1}{\pi} \sum_{J\, {\rm even}}\int_{M^2}^\infty
 \frac{dm^2}{m^2} m^{4-D}n_J \rho_J(m^2) \;(\cdots)\,,
\ee
in which, crucially, the spectral density is positive by unitarity (conservation of probability):
\be
 0\leq \rho_J(m^2)\leq 2\,.
\ee
The remaining ingredients are kinematical: $1+2u/m^2=\cos\theta$ represents 
the scattering angle in the rest frame of the pair,
 $n_J$ is a normalization, and $\legP_J(x)$ are proportional to Gegenbauer polynomials which we will sometimes refer to as Legendre polynomials (they are Legendre polynomials in $D=4$):
 \begin{align}
  \legP_J(x)= {}_2F_1\p{-J,J+D-3,\tfrac{D-2}{2},\tfrac{1-x}{2}}\,,\label{def legP} 
\\
n_J= \frac{2^{D} \pi^{\frac{D-2}{2}}}{\Gamma(\tfrac{D-2}{2})} (J+1)_{D-4} (2J+D-3)\,.
\label{def nJ}
\end{align}
Thus, in short, the heavy contribution \eqref{eq:heavylightsplitC} is a sum
of Legendre polynomials with positive but unknown coefficients.
The polynomials themselves are not positive in the region $u<0$ in which the spin sum converges.

\subsection{Review of EFT constraints from S-matrix sum rules}\label{ssec:SMatrixReview}

Let us illustrate in a concrete example how these sum rules constrain EFTs.
Suppose our scalar is weakly interacting at energies below $M$.
To tree-level accuracy, including gravity and possible self-interactions,
the amplitude can be parametrized as
\begin{align} 
\MM_{\rm low}(s,t) &=\ 8\pi G \left[ \frac{st}{u} + \frac{su}{t} + \frac{tu}{s}\right]
-\lambda_3^2\left[ \frac{1}{s}+\frac{1}{t}+\frac{1}{u}\right]-\lambda_4 \nn\\& \quad + g_2 (s^2+t^2+u^2) + g_3 (stu) +\ldots\,,\label{Mlow param}
\end{align}
where $g_2, g_3$ and $\ldots$ represent higher-derivative corrections to $\phi^4$ self-interactions.
The light contribution to \eqref{eq:heavylightsplitC} can be readily computed by residues:
\be
\left.{-}{\cal C}_{2,u}\right|_{\mathrm{light}} = \frac{8\pi G}{-u} +2g_2 -g_3 u+\ldots
\label{S light}
\ee
and the heavy action \eqref{eq:heavylightsplitC} can be similarly
expanded in small $u$.  The first few terms will be useful below:
\begin{align}
 \left.{\cal C}_{k,u}\right|_{\mathrm{heavy}} &=\left\<
\frac{2}{m^{2k}}\Bigg(1 - \frac{u}{m^{2}}\left[-\frac{2\mathcal{J}^2}{d-1}+\frac{k+1}{2}\right]+
\right. \nonumber\\
& \qquad\quad \left.
\,\,\,\,\,\,\,\,\,\,\,\,\,\,\,\,
+\frac{u^2}{m^{4}}\left[\frac{2\cJ^4}{d^2-1}-\frac{(d(k+3)+k-1)\mathcal{J}^2}{d^2-1} +\frac{(k+2)^2}{8}\right]
+\ldots
\Bigg)
\right\>, \label{S heavy}
\end{align}
where $\mathcal{J}^2 = J(J+d-2)$ is the rotation Casimir invariant, and $D=d+1$.

In principle, both sides are infinite series,
however we would like to constrain just the first few terms in $\cC_{2,u}|_\mathrm{light}$.
In the absence of gravity, one could simply expand around the forward limit
$u=0$ and match eqs.~\eqref{S light}-\eqref{S heavy} term-by-term. For example
this expresses $g_2=\< \frac{1}{m^4}\>$ as a positive sum rule \cite{Pham:1985cr,Adams:2006sv}
and suggests that all higher-derivative corrections must disappear as $g_2\to 0$.
Remarkably, by combining sum rules involving different subtraction degrees,
one finds two-sided bounds on ratios $g_k/g_2$ which justify the EFT power-counting
logic from the sole principle of causality \cite{Tolley:2020gtv,Caron-Huot:2020cmc,Arkani-Hamed:2020blm,Chiang:2021ziz}.

It is important to note that the method applies to interactions which grow like $s^2$ or faster in the Regge limit.
In the multi-field case (see \cite{Trott:2020ebl,Li:2021cjv} for recent applications),
this leaves a finite number of exceptions, for example a two-derivative self-interaction
$(\Phi^*\partial \Phi)^2$ for a complex scalar field. (For a single real scalar field, this interaction
did not appear because it is proportional to equations of motion.)
One might of course impose parametric upper bounds on these coefficients,
to ensure that EFT loops remain under control below the cutoff,
but in this paper we focus on sharp bounds that are homogeneous in couplings.

The detailed results are modified in the presence of gravity.
The graviton, a massless spin-2 particle, causes spin-two sum rules to diverge in the forward limit!

From the viewpoint of \eqref{eq:heavylightsplitC} the pole means that the partial wave expansion diverges
at large spins. This is physically meaningful: the fact that the divergence is a sum of positive terms
readily fixes the sign of $G$ to be positive: gravity is universally attractive.
But just subtracting the pole will not yield reliable conclusions about higher order terms since
the sign of divergent sums cannot be reliably predicted.

Various work-around strategies have been proposed.
In \cite{Caron-Huot:2021rmr}, we pointed out that there is in fact no need to expand around $u=0$, because all the ``$\ldots$''  couplings in \eqref{S light} can be eliminated using the forward limit of the ${\cal C}_{k\geq 4}$ sum rules or the first $u$-derivative around them (which are nonsingular since our EFT, by assumption, doesn't contain light higher-spin particles):
\be\label{C2 improved def flat}
{\cal C}_{2, u}^\mathrm{improved} \equiv {\cal C}_{2, u} - \sum_{n=2}^\oo
\p{n\, u^{2n-2} {\cal C}_{2n,0} + u^{2n-1} {\cal C}_{2n,0}'}\,.
\ee
Explicitly, the light and heavy contributions give
\begin{align}
\frac{8\pi G}{-u} +2g_2 -g_3 u &= 
\left\< \frac{2m^2+u}{m^2+u} \frac{\legP_J(1+\frac{2u}{m^2})}{m^2(m^2+u)}
- \frac{u^2}{m^6}\left[\frac{(4m^2+3u)\legP_J(1)}{(m^2+u)^2}+\frac{4u \legP'_J(1)}{m^4-u^2}\right]\right\>
\label{B2 improved flat} \nn\\
&\equiv \left\<{\cal C}_{2, u}^{\rm improved}[m^2,J] \right\>\,.
\end{align}
The key is that the left-hand-side is \emph{exact} (when acting on any rational amplitude $\MM$).

Exchange of other light spin-2 particles of mass $m_i$, if the EFT contains them, would add terms
proportional to $1/(m_i^2-u)$ on the left-hand-side.
The method would be unchanged but the O(1) coefficients in the bounds would be modified.

\begin{figure}
\centering
\includegraphics[width=0.6\textwidth]{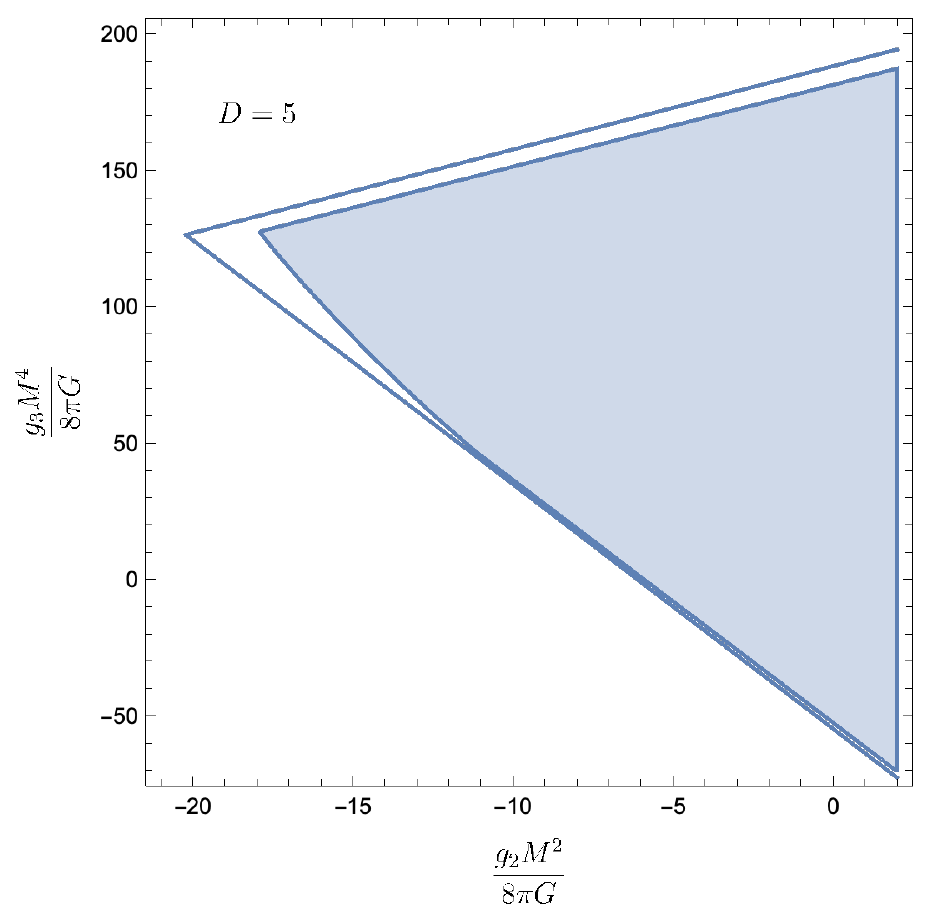}
\caption{Allowed region (shaded) for $g_2$ and $g_3$ from \cite{Caron-Huot:2021rmr}, in a theory of a scalar coupled to gravity in flat space in dimensions $D=5$ and heavy mass scale $M$. Asymptotically to the right, the region is a cone which matches the non-gravitational bounds; note the offset to the left.
The solid lines show the non-optimal bounds from the simple functionals \eqref{g2_g3_functionals}.
\label{fig:allowed g2g3}
}
\end{figure}

Liberated from small-$u$ series, we can create a new class of functionals
by integrating ${\cal C}^{\mathrm{improved}}_{k,u}$ against various functions of $u$ and search
for functions which make the heavy contribution manifestly positive.
Given the singularities of the integrand, a natural range is $-M^2<u<0$.
It is not obvious how to analytically construct suitable functions on this range, but
a numerical search algorithm was presented in \cite{Caron-Huot:2021rmr}.
Polynomials of surprisingly low degree turn out to work, although their coefficients are hard to interpret.
For the purposes of this paper, we will simply record two simple polynomials
found with this method in $D=5$:\footnote{The numbers show
rational approximants to numerical results: no attention should be paid to their number-theoretic properties.}
\be\begin{aligned}
\label{g2_g3_functionals}
\mathcal{C}^{(1)} &\equiv \int_0^{1} dp\ {\cal C}^{\mathrm{improved}}_{2,-M^2p^2}\times
p^2(1-p)^2 \left[2280 - 665 p + 2964 p^5 - 8280 p^6\right],\\
\mathcal{C}^{(2)} &\equiv \int_0^{1} dp\ {\cal C}^{\mathrm{improved}}_{2,-M^2p^2}\times
p^2(1-p)^2 \left[1785 - 3468 p - 18785 p^5 + 26187 p^6\right].
\end{aligned}\ee
We claim that the heavy actions of both these functionals are rigorously positive in $D=5$:
\be
 \mathcal{C}^{(i)}[m^2,J] \geq 0 \quad\forall m\geq M, J\ {\rm even}\,. \label{positivity}
\ee
Positivity of the left-hand-side of \eqref{B2 improved flat} then gives the following inequalities:
\be
-8.96\frac{g_2}{M^2} -54.7 \frac{8\pi G}{M^4} \leq g_3 \leq 3.06\frac{g_2}{M^2} +188\frac{8\pi G}{M^4} \qquad (D=5)\,,
\label{g2_g3_simple_bounds}
\ee
which are shown in Figure~\ref{fig:allowed g2g3} alongside with the optimal bounds found in \cite{Caron-Huot:2021rmr}.

\begin{figure}
\centering
\includegraphics[width=0.55\textwidth]{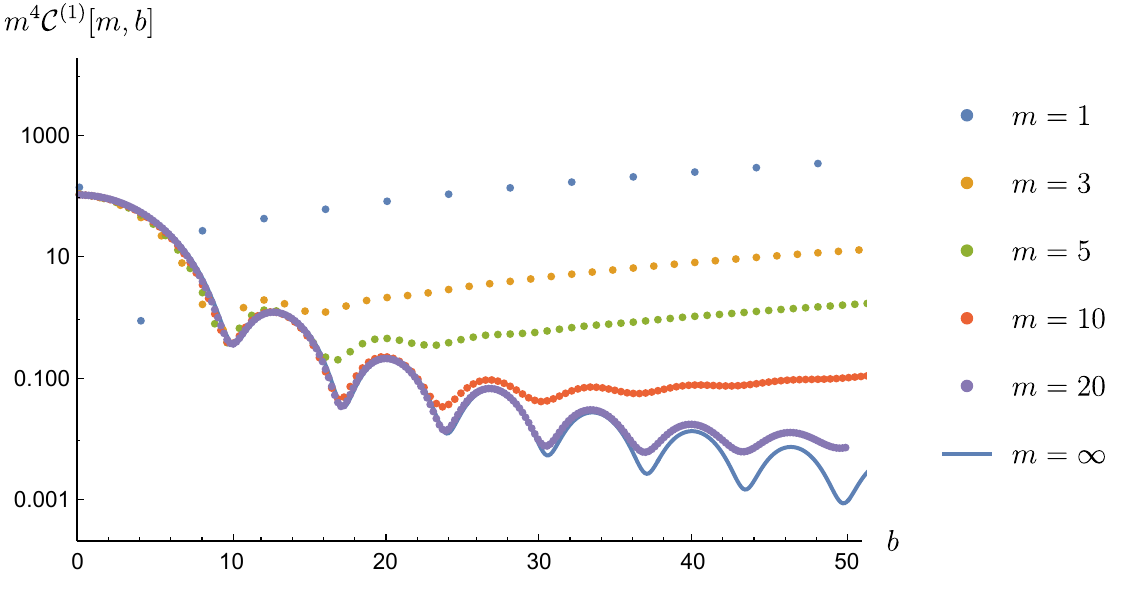}
\includegraphics[width=0.44\textwidth]{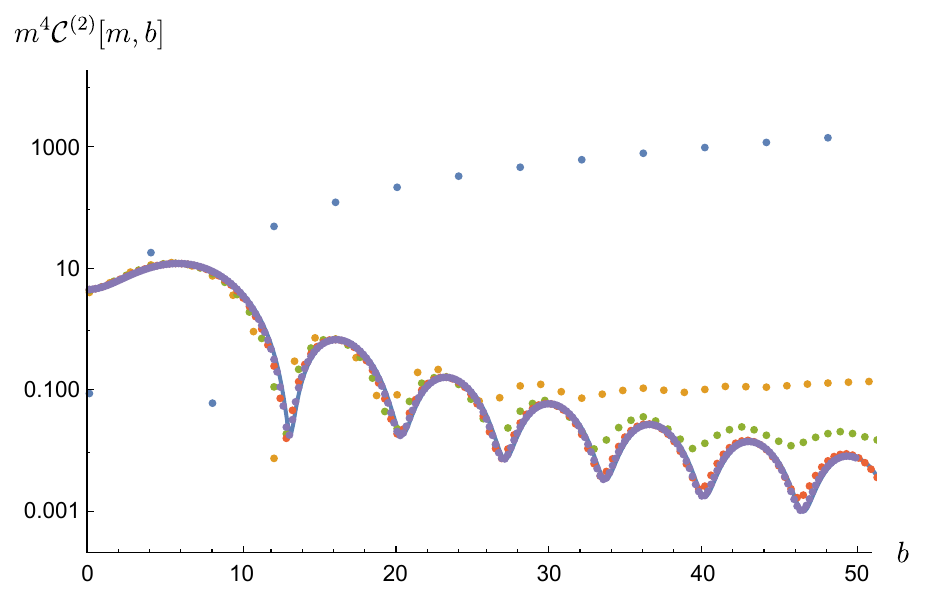}
\caption{Action of the ${\cal C}^{(i)}$ functionals in \eqref{g2_g3_functionals}
for $i=1,2$ respectively, on various heavy states with $m\geq M$ in units where $M=1$.
Individual points, representing different spins $J$, are lined up with constant impact parameter
$b=\frac{2J}{m}$ in order to highlight the regularity as $m\to\infty$ and $J\to\infty$.  This helps confirm
positivity on all heavy states.
The blue outliers (at $J=2$ and $J=0,4$ respectively)
are caused by single zeros slightly below threshold $m=M$. 
\label{fig:functionals}
}
\end{figure}

Positivity of the functionals is verifiable by exhaustion:
in Figure~\ref{fig:functionals} we plot the action of $\mathcal{C}^{(1)}$ on various
states of different masses $m\geq M$ and spin $J$ up to 500, grouped
in terms of their impact parameters $b=\frac{2J}{m}$.
The infinite mass limit at fixed impact parameter can be computed for arbitrary $J$ using
\be
 \lim_{m,J\to\infty} \legP_J\p{1-\frac{2p^2}{m^2}} =
 \Gamma(\tfrac{D-2}{2})\frac{J_{\frac{D-4}{2}}(pb)}{(pb/2)^{\frac{D-4}{2}}} \equiv \tl J(p b)\,.
 \label{eq:impactparamlimitofgeg}
\ee
The plots reveal a clear trend where the infinite-mass limit
is approached from above.

The bounds in \eqref{g2_g3_simple_bounds} show that, once the value of $g_2$ is fixed,
the value of $g_3$ lies in a range that conforms with dimensional analysis.
Furthermore, while $g_2$ isn't strictly positive due to gravity's attractive force, the magnitude of this effect is bounded and conforms with dimensional analysis at the scale $M$.
Our goal in the rest of the paper will be to uplift these bounds to EFTs in AdS; the methods will be general and will apply to any other dispersive sum rule on 2$\to 2$ scattering.

\subsection{What does it mean to probe local physics in AdS?} \label{ssec:what does it mean}

There are at least two ways to study the geometric dependence of a scattering process. Perhaps the most obvious is to scatter wavepackets that are designed to localize near a particular configuration --- for example near impact parameter $b$. 
An alternative approach, which will be crucial in the following, 
is to characterize the geometry of a process using the quantum numbers of intermediate states.

\begin{figure}
\centering{
\begin{subfigure}[t]{0.45\textwidth}\centering
\begin{tikzpicture}[scale=1]
	\draw[->,very thick] (-2,0) -- (-0.02,0);
	\draw[->,very thick] (2,1.5) -- (0.02,1.5);
	\draw[dotted,thick] (0,0) -- (0,1.5);
	\node [right] at (0,0.75) {$b$};
	\node [right] at (0,-0.1) {$m/2$};
	\node [left] at (0,1.6) {$m/2$};
	\draw[->,thick]  (0,2.8) -- (0.75,2.8);
	\draw[->,thick]  (0,2.8) -- (0,3.55);
	\node [above] at (0,3.55) {$x$};
	\node [right] at (0.75,2.8) {$z$};
\end{tikzpicture}
\caption{}
\end{subfigure}
\begin{subfigure}[t]{0.45\textwidth}\centering
\hspace{0.43in}
\begin{tikzpicture}[scale=0.65]
    \draw[looseness=0.14,thick,black!20!white] (0.94,7.03) to[out=-90,in=-90] (5.72,7.03) to[out=90,in=90] (0.94,7.03);
    \draw[looseness=0.22,thick,black!20!white] (0.94,0.96) to[out=-90,in=-90] (5.72,0.96);
    \draw[looseness=0.23,dotted,thick,black!20!white] (5.72,0.96) to[out=90,in=90] (0.94,0.96); 
    \draw[thick,black!20!white] (0.94,0.96) -- (0.94,7.03);
    \draw[thick,black!20!white] (5.72,0.96) -- (5.72,7.03);
    \draw[looseness=0.5,thick] (5.4,1.46) to[out=35,in=-90] (5.72,2.3);
    \draw[looseness=0.63,thick,dotted] (5.72,2.3) to[out=90,in=-25]  (1.31,6.5);
    \draw[looseness=0.67,thick] (5.4,1.46) to[out=155,in=-90] (0.94,5.75);
    \draw[looseness=1,thick,dotted] (0.94,5.75) to[out=90,in=-135] (1.31,6.5);
    \draw[looseness=1,thick,dotted] (1.32, 1.76) to[out=150,in=-90] (0.94,2.2);
    \draw[looseness=0.44,thick] (0.94,2.2) to[out=90,in=-150] (5.4,6.3);
    \draw[looseness=1,thick] (5.4,6.3) to[out=-35,in=90] (5.72,5.9);
    \draw[looseness=0.48,thick,dotted] (5.72,5.9) to[out=-90,in=30] (1.32, 1.76);
    \draw[fill] (1.32, 1.76) circle (2pt);
    \draw[fill] (5.4,1.46) circle (2pt);
    \draw[fill] (1.31,5.9) circle (2pt);
    \draw[fill] (5.4,5.67) circle (2pt);
    \draw[blue,ultra thick] (2.06,3.75) -- (4.55,4.22);
    \draw[fill,blue] (2.8,3.89) circle (2.3pt);
    \draw[fill,blue] (3.9,4.1) circle (2.3pt);
    \draw[blue,thick,->] (2.8,3.89) -- (3.22,4.21);
    \draw[blue,thick,->] (3.89,4.1) -- (3.51,4.42);
    \node [below,blue] at (3.35,4) {$\beta$};
    \node [right] at (6.3,4) {time};
    \draw [thick,->] (6.3,3.2) -- (6.33,4.8);
\end{tikzpicture}
\caption{}
\end{subfigure}}
\caption{Relation between impact parameter and angular momentum in the Regge limit.
(a) In flat space, a pair of massless particles with center-of-mass energy $m$
and transverse separation $b$ carries total angular momentum $J=m\frac{b}{2}$.
(b) In a holographic CFT${}_d$, the Regge limit of the four-point function localizes
along two null sheets.
They intersect in the transverse hyperbolic space $H_{d-1}$ depicted by the blue line.
A pair of massless particles with center-of-mass energy $\Delta/R_{\rm AdS}$ separated
in $H_{d-1}$ by geodesic distance $\beta R_{\rm AdS}$
carries total angular momentum $J=\Delta\tanh\frac{\beta}{2}$.
\label{fig:impact}}
\end{figure}
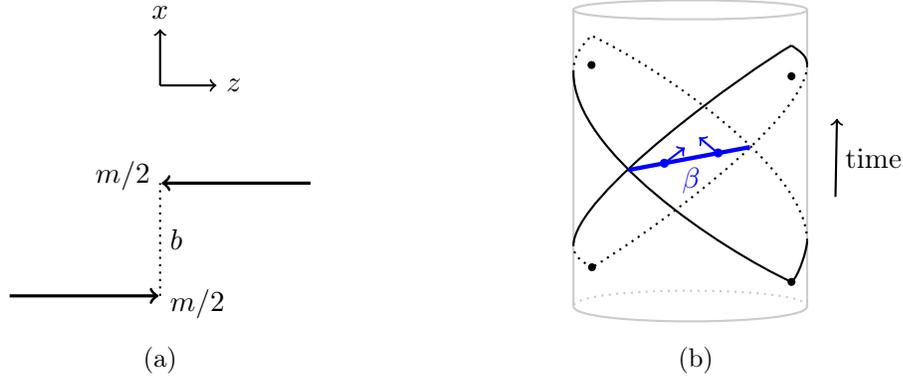

For example, consider a $2\to 2$ scattering process of massless scalars. If the particles create an intermediate massive state with mass $m$ and angular momentum $J$, then by conservation of energy and angular momentum, we deduce that the particles scattered with impact parameter $b=\frac{2J}{m}$, see Figure~\ref{fig:impact}: the asymptotic trajectories are translated perpendicularly
by the amount $b$.
In other words, we can use the ratio $b=\frac{2J}{m}$ to track contributions from different impact parameters. The relationship to wavepackets used in scattering is apparent in (\ref{eq:impactparamlimitofgeg}): at large $m$ and $J$, the transverse momentum $p$ becomes Fourier conjugate to $b$.

For our purposes, it will be useful to understand the analogous correspondence between conserved quantities and geometry in AdS. We write the AdS metric as
\be
 ds^2_{\rm AdS_D} =
 {-}dt^2\cosh^2 \beta + d\beta^2 +  d\Omega^2_{d-1}\sinh^2\beta\,.
\ee
Consider a massless particle in AdS, with action
\begin{align}
S = \int d\lambda\, L\,,\qquad L = \frac{1}{2e}(\cosh^2 \b\, \dot{t}^2 - \dot \b^2 - \sinh^2 \b\,\dot \theta^2)\,,
\end{align}
where $e(\lambda)$ is a 1-dimensional tetrad, $\l$ is a worldline parameter, and dots denote derivatives with respect to $\l$. For simplicity, we have restricted the particle to a single angular degree of freedom $\theta$ on the sphere $S^{d-1}$. The energy and angular momentum of the particle are
\begin{align}
p_t = \frac{\ptl L}{\ptl \dot t} = \frac{\cosh^2\b\, \dot t}{e}\,,\qquad p_\theta = \frac{\ptl L}{\ptl \dot \theta}=\frac{\sinh^2 \beta\,\dot \theta}{e}\,.
\end{align}
When the radial velocity $\dot\beta$ vanishes, the equations of motion imply $\cosh \beta\,\dot t=\sinh\beta\,\dot\theta$, so that $p_\theta/p_t = \tanh \beta$.

Consider now a pair of particles with large center-of-mass energy $\De/R_\mathrm{AdS}$, scattering with AdS impact parameter $\beta R_\mathrm{AdS}$, as shown in Figure~\ref{fig:impact}. At the point of closest approach (along their extrapolated asymptotic trajectories, not necessarily actual trajectories),
each particle has radial position $\beta/2$. Thus, their ratio of angular momentum and energy is 
\begin{align}
\frac{J}{\De}=\left.\frac{2p_\theta}{2p_t}\right|_{\beta/2}=\tanh\frac{\beta}{2}\,.
\end{align}
This is an key conceptual result, and it allows us to identify two important limits of the heavy operator spectrum:
\begin{itemize}
\item We define the {\it Regge limit} as large $\De$ with $J/\De$ fixed and order 1. This corresponds to high-energy bulk scattering with impact parameter comparable to $R_\mathrm{AdS}$.
\item We define the {\it bulk point limit} as large $\De$ with small $J/\De\ll 1$. This corresponds to scattering at impact parameters much smaller than $R_\mathrm{AdS}$.
\end{itemize}
 We would ultimately like to probe small impact parameters in AdS. Thus, we should construct observables that are dominated by heavy states in the bulk point limit.

To be more precise, we define the mass and AdS impact parameter of a heavy operator by
\begin{align}
\label{eq:massdefinition}
m^2 &= \taushift\bar{\taushift}\,, \\
\label{eq:impactparameter}
\etaAdS &\equiv \cosh\beta \equiv \frac{\taushift^2 + \bar\taushift^2}{2\taushift \bar\taushift}\,,
\end{align}
where 
\begin{align}
\label{eq:taushiftdefinition}
\taushift \equiv \De - J - d + 1\,, \qquad
\bar\taushift \equiv \De + J - 1\,.
\end{align}
For large $\De$, these reduce to
$m^2\approx \Delta^2-J^2$ and $\etaAdS\approx \frac{\Delta^2+J^2}{\Delta^2-J^2}$.\footnote{The constant terms in our definition of $m^2$ differ from others in the literature, e.g.\ \cite{Naqvi:1999va,Costa:2014kfa}.
Our choice of constants is motivated by the conformal group: they are the unique choice such that
the set $(\pm\taushift,\pm\bar\taushift)$ closes under all Weyl reflections. 
(The Weyl group of $\SO(d,2)$ is generated by shadow reflections $\De\mapsto d-\De$, spin shadow $J\mapsto 2-d-J$,
and ``light transform'' $\Delta \leftrightarrow 1-J$. These transformations preserve Casimir eigenvalues.) This will prove convenient for calculations and will
effectively remove odd powers of $1/m$ in large-mass expansions.}
The latter is equivalent to $\frac{J}{\De}\approx \tanh\frac \beta 2$. Although the AdS geodesic distance is $\beta$, we sometimes abuse terminology and refer to $\etaAdS$ as an ``impact parameter'' as well, since it encodes the same information as $\beta$.

Given a functional $\w$, it is useful to define the heavy density associated to $\w$ as its action on $s$-channel blocks $G_{\De,J}^s$, divided by some positive factors:
\begin{align}
\label{eq:definitionofheavydensity}
\omega[\De,J] &\equiv \frac{(-1)^J}{q_{\De,J}} \frac{\omega[G_{\De,J}^s]}{2\sin^2(\pi\frac{\tau-2\De_\f}{2})}\,.
\end{align}
The $\sin^2(\pi\frac{\tau-2\De_\f}{2})$ factor in the denominator cancels the sum of phases introduced by the dDisc. The positive factor $q_{\De,J}$ is defined by
\begin{align}
\label{eq:positivedensityfactors}
q_{\De,J} &\equiv \frac{1}{\pi p^\mathrm{MFT}_{\De,J}} \frac{2n_J}{m^{2d-4}}
\frac{\Gamma(\Delta-1)(2\Delta-d)}{\Gamma(\Delta-d+2)}\,,
\end{align}
where $m$ is given by (\ref{eq:massdefinition}), $n_J$ is given by (\ref{def nJ}), and we have divided by the OPE coefficients of Mean Field Theory \cite{Fitzpatrick:2011dm}:
\begin{align}
\label{eq:mftope}
p^\mathrm{MFT}_{\De,J} &= 
\frac{2 \Gamma (\Delta -1) \Gamma (J+\frac d 2)
  }{
  \Gamma (\Delta_\phi )^2 \Gamma (\Delta_\phi-\frac{d-2}{2})^2
   \Gamma (\Delta -\frac{d}{2})\Gamma (J+1)
}
\frac{
 \Gamma (\frac{J+\Delta }{2})^2 \Gamma (\frac{\Delta-J -d+2}{2})^2
}{
\Gamma (J+\Delta -1) \Gamma (\Delta-J-d +1)
}
\nn\\
&\quad \x
\frac{
 \Gamma (\frac{\Delta -J-2d+2\Delta_\phi+2}{2})
  \Gamma (\frac{\Delta+J-d +2 \Delta_\phi}{2})
}{
\Gamma (\frac{\Delta-J -2 \Delta_\phi +2}{2}) \Gamma (\frac{\Delta+J+d -2 \Delta_\phi}{2})
}\,.
\end{align}
The sign $(-1)^J$ will not affect our applications (we focus on identical scalars where only even-spin operators will enter the OPE) but we include it for completeness.
We can define a heavy expectation value by the positive measure
\begin{align}
\label{eq:cftpositivemeasure}
\<f(\De,J)\> &\equiv \sum_{\tau > \De_\mathrm{gap}}  2\sin^2\p{\pi\tfrac{\tau-2\De_\f}{2}} q_{\De,J} p_{\De,J}
 f(\De,J)\,.
\end{align}
The holographic sum rule (\ref{eq:heavylightsplit}) corresponding to $\omega$ then takes the form
\begin{align}
-\w|_\mathrm{light} &= \w|_\mathrm{heavy} = \<\w[\De,J]\>\,.
\end{align}

Our discussion of the relationship between conserved quantities and bulk geometry gives us a way to understand which bulk quantities a dispersive functional $\omega$ measures. We should study the heavy density $\w[\De,J]$ at large $m$ as a function of $\beta$. If the heavy density is localized in $\beta$, then $\w$ probes local physics in AdS.

\subsection{Physical CFT functionals and the $B_{k,v}$ sum rules}\label{ssec:Bkv}

The $B_{k,v}$ functionals introduced in ref.~\cite{Caron-Huot:2020adz} are double integrals:
\be
 B_{k,v}[\mathcal{F}] =
\frac{\Gamma\big(\tfrac{k}{2}\big)^2}{\Gamma(k-1)}\!\iint\limits_{C_-\times C_+}\!\!\frac{dwd\wb}{(2\pi i)^2}\,
\frac{(\wb-w)(v'-u')}{(u'v')^{\frac{k}{2}}[v^2-2(u'+v')v+(u'-v')^2]^{\frac{3-k}{2}}}
\mathcal{F}(w,\wb)\,.
\label{eq:BKVDefinition}
\ee
Here, $u'=w \bar w$ and $v'=(1-w)(1-\bar w)$ are conformal cross ratios that are being integrated over.
The integral has the following properties:
\smallskip
\begin{itemize}
\item It is antisymmetric under swapping the $s$ and $t$ channels, i.e.\ $(w,\bar w)\to (1-w,1-\bar w)$.
\item The integrand decays with Regge spin $k$, so for $k>1$ the contour can be swapped with the conformal block expansion in a physical correlator.
\end{itemize}
\smallskip
\noindent We refer to functionals satisfying these conditions as {\it physical}.

The contour $C_-\times C_+$ wraps the left- and right cut $w<0$ and $\wb>1$.
The kernel however has the remarkable property that the contour can be deformed so that both variables wrap
the left-cut, in such a way that one finds a double-discontinuity of the correlator:
\be
B_{k,v} [\mathcal{F}] = 
\frac{\Gamma\big(\tfrac{k}{2}\big)^2}{\pi^2\Gamma(k-1)}\int\limits_{v}^{\infty}dv'\!\!\!\!\!\!\!\!\int\limits_{0}^{(\sqrt{v'}-\sqrt{v})^2}\!\!\!\!\!\!\!\!\!du'
\frac{(v'-u')\,\dDisc_{s}[\mathcal{F}(u',v')]}{(u'v')^{\frac{k}{2}}[v^2-2(u'+v')v+(u'-v')^2]^{\frac{3-k}{2}}}\,.
\label{Bkv dDisc}
\ee
This is what makes these sum rules ``dispersive'' and is crucial for our applications.

A technical subtlety is that the collinear limit is singular, which is dealt with easily by
applying the sum rule to $\mathcal{F}=\cG-1$. The result is then \cite{Caron-Huot:2020adz}\footnote{This result relies on the technical assumption that $(\cG-1)$ vanishes in the $u$-channel collinear limit even
on the second sheet, which has never been rigorously proven \cite{Hartman:2015lfa,Kravchuk:2021kwe}.
However, we do not expect this assumption to affect the results of the present paper,
as long as the right-hand-side can be computed from the EFT correlator.}
\be
\sum\limits_{\De,J} p_{\De,J}\,B_{k,v} [G^s_{\Delta,J}] = (-1)^{\frac{k}{2}-1}\,. \label{Bkv sum rule}
\ee
When splitting between light and heavy contributions (see \eqref{eq:heavylightsplit}),
we assign the right-hand-side to the light sector and view it as part of the computable low-energy EFT.
The sum rule \eqref{Bkv sum rule} exists for all even $k=2,4,\ldots$ and real $v>0$.

Intuitively, within AdS$_{d+1}$, the $B_{k,v}$ sum rules are localized in light-cone times because they originate from the Regge limit; that is they localize to a codimension two impact parameter space space $H^{d-1}$.
However, they are still smeared over AdS-size impact parameters, and to fully localize
will require some amount of harmonic analysis on $H^{d-1}$.

\subsection{Regge moments}
\label{sec:reggemomentsheavy}

Given dispersive functionals like $B_{k,v}$, we would like to understand what aspects of bulk physics they measure. As discussed in section~\ref{ssec:what does it mean}, we can get a geometric picture for different functionals by studying their heavy density at large $m$ as a function of the AdS impact parameter $\beta$.

Note that $s$-channel blocks $G_{\De,J}^s(w,\bar w)$ with large $m$ are exponentially suppressed away from the $u$-channel Regge limit $w,\bar w\to \oo$. Thus, we expect the action of a dispersive functional on heavy blocks to be controlled by its expansion around this limit.\footnote{We will see shortly that dispersive functionals with large $\nu$, defined in (\ref{eq:inversetransforms}), can probe a finite distance away from the $u$-channel Regge limit, though still within the radius of convergence of the Regge moment expansion.} We will organize this expansion in terms of {\it Regge moments\/}: weighted integrals of the double-discontinuity along rays of constant angle of approach to the $u$-channel Regge limit:
\begin{align}
\label{eq:reggemomentdefinition}
\Pi_{k,\eta}[\cG] &\equiv  \int_0^{r_{\rm max}(\eta)} dr\,r^{k-2} \dDisc_s\,\cG(r,\eta)\,, \\
r&\equiv \sqrt{\rho \bar{\rho}}\,,\quad \eta \equiv \frac{\rho+\bar\rho}{2\sqrt{\rho \bar{\rho}}}\,,
\label{eq:definitionofreta}
\end{align}
where $\rho,\bar\rho$ are the radial coordinates of \cite{Hogervorst:2013sma} adapted to the $u$-channel:
\be \rho=\frac{1}{(\sqrt{1-w}+\sqrt{-w})^{2}}\,,\quad \bar\rho=\frac{1}{(\sqrt{1-\wb}+\sqrt{-\wb})^{2}}\,.
\ee
In terms of $r,\eta$, the $u$-channel Regge limit is $r\to 0$ with fixed $\eta$.
In (\ref{eq:reggemomentdefinition}), $r_{\rm max}(\eta) = \eta-\sqrt{\eta^2-1}$ is determined from the condition that $\rho\leq 1$. We refer to $k$ as the {\it spin\/} of the Regge moment.\footnote{Note that $\Pi_{k,\eta}$ is not a physical functional --- specifically $\Pi_{k,\eta}[\cG]$ does not vanish in general when $\cG$ is a physical four-point function.} Regge moments will be a useful organizational tool for the following reasons:
\begin{itemize}
\item Any dispersive functional $\w$ can be expanded in Regge moments.
\item The action of $\w$ on heavy blocks is determined order-by-order in $1/m^2$ (up to nonperturbative corrections) by its expansion in Regge moments.
\item Surprisingly, as we explain in section~\ref{sec:lightAction}, the contribution of light states $\left.\w\right|_{\mathrm{light}}$ is {\it also\/} determined by the expansion of $\w$ in Regge moments.
\end{itemize}

Let us explain how to expand a dispersive functional $\w$ in Regge moments.
We first write $\w$ as a weighted integral of the double-discontinuity
\begin{align}
\w[\cG] &=  \int_1^\oo d\eta\,\int_0^{r_{\rm max}(\eta)} dr\, K(r,\eta)\, \dDisc_s[\cG] + \dots\,,
\end{align}
where $K(r,\eta)$ is a distribution. Here, ``$\dots$'' represents non-$\dDisc$ contributions that do not contribute in the $u$-channel Regge limit. The expansion of $\w$ in Regge moments is defined by expanding $K(r,\eta)$ in powers of $r$. For example, if 
\begin{align}
\label{eq:equationwithbigo}
K(r,\eta) &= a(\eta) + b(\eta)r + O(r^2)\,,
\end{align}
then we write
\begin{align}
\w &\sim \int_1^\oo d\eta \, a(\eta) \Pi_{2,\eta} + \int_1^\oo d\eta \,b(\eta) \Pi_{3,\eta} + O(\Pi_{4})\,.
\end{align}
The notation $O(\Pi_n)$  indicates Regge moments with spins greater than or equal to $n$. In general, functionals with spin-$k$ Regge decay as defined in \cite{Caron-Huot:2020adz} have moment expansions starting with $\Pi_{k,\eta}$.

It turns out that physical functionals span the space of Regge moments with even $k\geq 2$ and $\eta\in[1,\oo)$.\footnote{The fact that odd $k$ does not appear is a consequence of antisymmetry under exchanging the $s$ and $t$ channels.}
To see this, consider the $B_{k,v}$ functional, which in $r,\eta$ coordinates takes the form
\be \label{Bkv r eta}
 B_{k,v}[\cG] = \frac{2^{3k-2}\Gamma\big(\tfrac{k}{2}\big)^2}{\pi^2\Gamma(k-1)}
\int\limits_{\sqrt{v}}^\infty \frac{\eta d\eta}{(\eta^2-v)^{\frac{3-k}{2}}}
 \int\limits_0^{r_{\rm max}(\eta)}
 dr\, r^{k-2}(1{-}r^4)
  \frac{\p{(1+r^2)^2-4v r^2}^{\frac{k-3}{2}}}{
 \p{(1+r^2)^2-4 \eta^2 r^2}^{k-1}
 }{\rm dDisc}_s[\cG]\,.
\ee
Expanding the kernel at small $r$, we have (see \eqref{eq:reggemomentdefinition})
\begin{align}
\label{eq:integraloverangularvariable}
B_{k,v} &\sim \frac{2^{3k-2}\Gamma\big(\tfrac{k}{2}\big)^2}{\pi^2\Gamma(k-1)}
\int_{\sqrt{v}}^\infty \frac{\eta d\eta}{(\eta^2-v)^{\frac{3-k}{2}}} \Pi_{k,\eta} + O(\Pi_{k+2})\,.
\end{align}
Recall that the continuous variable $v>0$ labels physical sum rules.
The claim is that, remarkably, the integral over $\eta$ in~(\ref{eq:integraloverangularvariable}) is invertible.

To see this, we note that it is a convolution in $\eta^2$, which in a conjugate Fourier space
is proportional to multiplication by $|p|^{\frac{1-k}{2}}$.
Its inverse is then division by the same thing. This can be written as the following convolution: 
\begin{align}
\label{eq:inversetoangularintegral}
 \frac{(-1)^{\frac{k}{2}}\pi \Gamma(k)}{2^{3k-2}\Gamma\big(\tfrac{k}{2}\big)^2}
\int_{\eta^2}^\infty \frac{dv} {\big[(v-\eta^2)^{\frac{k+1}{2}}\big]_+}B_{k,v}
&\sim
 \Pi_{k,\eta} + O(\Pi_{k+2})\,.
\end{align}
The $+$ distribution on the left-hand side is defined by subtracting singular terms in the Taylor series
around $v=\eta^2$.\footnote{For example, for $k=2$:
\be
 \frac{-\pi}{16}\int_{\eta^2}^\infty \frac{dv\ (B_{2,v}-B_{2,\eta^2})}{(v-\eta^2)^{\frac{3}{2}}} \sim \Pi_{2,\eta} + O(\Pi_4)\ .
\ee}
Swappability \cite{Rychkov:2017tpc} of the resulting functional will be confirmed directly in examples.

\subsection{Road map: harmonic transforms and the flat-space limit}

Using the inverse transform (\ref{eq:inversetoangularintegral}), we can construct physical
functionals that are pure $\Pi_{k,\eta}$ moments to leading order in the Regge moment expansion. However, such functionals are not localized in the bulk. Specifically, as we will see shortly, the $\Pi_{k,\eta}$ moment of a heavy block $G_{\De,J}^s$ is delocalized as a function of $\etaAdS=\frac{\De^2+J^2}{\De^2-J^2}$. To localize in $\etaAdS$, we need functionals that carry large momentum $\nu$ along the transverse $H_{d-1}$. More precisely, we seek functionals whose action on heavy blocks takes the form
\begin{align}
\label{eq:thingwewant}
\w[\De,J] &\sim \cP_{\frac{2-d}{2}+i\nu}(\etaAdS)\,,
\end{align}
where $\cP_{\frac{2-d}{2}+i\nu}(\etaAdS)$ is a Gegenbauer function (\ref{def legP}), which in this case plays the role of a harmonic function on $H_{d-1}$.

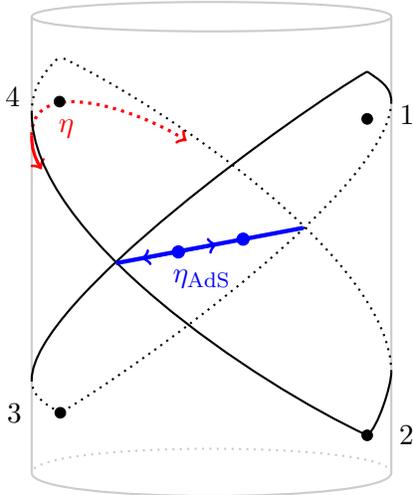
\begin{figure}
\centering
\begin{tikzpicture}
    \draw[looseness=0.14,thick,black!20!white] (0.94,7.03) to[out=-90,in=-90] (5.72,7.03) to[out=90,in=90] (0.94,7.03);
    \draw[looseness=0.22,thick,black!20!white] (0.94,0.96) to[out=-90,in=-90] (5.72,0.96);
    \draw[looseness=0.23,dotted,thick,black!20!white] (5.72,0.96) to[out=90,in=90] (0.94,0.96); 
    \draw[thick,black!20!white] (0.94,0.96) -- (0.94,7.03);
    \draw[thick,black!20!white] (5.72,0.96) -- (5.72,7.03);
    \draw[looseness=1,->,very thick,red,dotted] (0.94,5.45) to[out=90,in=148] (3,5.38);
    \draw[looseness=1,->,very thick,red] (0.94,5.45) to[out=-90,in=120] (1.07,5);
    \draw[fill] (1.32, 1.76) circle (2pt);
    \draw[fill] (5.4,1.46) circle (2pt);
    \draw[fill] (1.31,5.9) circle (2pt);
    \draw[fill] (5.4,5.67) circle (2pt);
    \draw[looseness=0.5,thick] (5.4,1.46) to[out=35,in=-90] (5.72,2.3);
    \draw[looseness=0.63,thick,dotted] (5.72,2.3) to[out=90,in=-25]  (1.31,6.5);
    \draw[looseness=0.67,thick] (5.4,1.46) to[out=155,in=-90] (0.94,5.75);
    \draw[looseness=1,thick,dotted] (0.94,5.75) to[out=90,in=-135] (1.31,6.5);
    \draw[looseness=1,thick,dotted] (1.32, 1.76) to[out=150,in=-90] (0.94,2.2);
    \draw[looseness=0.44,thick] (0.94,2.2) to[out=90,in=-150] (5.4,6.3);
    \draw[looseness=1,thick] (5.4,6.3) to[out=-35,in=90] (5.72,5.9);
    \draw[looseness=0.48,thick,dotted] (5.72,5.9) to[out=-90,in=30] (1.32, 1.76);
    \draw[blue,ultra thick] (2.06,3.75) -- (4.55,4.22);
    \draw[fill,blue] (2.89,3.9) circle (2.3pt);
    \draw[fill,blue] (3.75,4.07) circle (2.3pt);
    \draw[<->,blue,very thick] (2.40,3.81) -- (3.4,4.0);
    \node [below,blue] at (3.2,3.8) {$\eta_\mathrm{AdS}$};
    \node [below,red] at (1.4,5.8) {$\eta$};
    \node [left] at (0.94,1.75) {$3$};
    \node [right] at (5.7,5.72) {$1$};
    \node [right] at (5.69,1.47) {$2$};
    \node [left] at (0.92,5.97) {$4$};
\end{tikzpicture}\caption{
In the CFT, the angle of approach $\eta$ to the Regge limit (in which operators 3 and 4 become null from 1 and 2, respectively) acts as a smeared version of 
the bulk impact parameter $\etaAdS=\cosh\beta$.
Since both variables transform like hyperbolic cosines under $\SO(d-1,1)$
transverse isometries, shown as blue and red arrows,
harmonic analysis can be used to ``undo'' the smearing and achieve bulk focusing.
The operators always stay within the ``Rindler wedge'' in which $1$ is spacelike from $3$ (and $4$ is spacelike from $2$).
\label{fig:etaAdS}}
\end{figure}

Fortunately, symmetries provide a connection between bulk and boundary variables. Note that the stabilizer subgroup of the Regge limit is $\SO(1,1)\x \SO(d-1,1)\subset \widetilde{\SO}(d,2)$. The radial variable $r$ is essentially conjugate to the $\SO(1,1)$ generator (which moves us toward or away from the Regge limit). Meanwhile, $\eta$ is an hyperbolic cosine conjugate to the generators of the Lorentz group $\SO(d-1,1)$. This same Lorentz group acts as the isometry group of the transverse $H_{d-1}$ in the bulk, that is, on $\etaAdS$. Thus, by performing harmonic analysis with respect to $\SO(d-1,1)$, we can project onto particular momenta along the transverse $H_{d-1}$, as illustrated in Figure~\ref{fig:etaAdS}.

These considerations suggest that we study the harmonic transforms of $\Pi_{k,\eta}$ with respect to $\eta$, defined by
\begin{align}
 \widehat{\Pi}_{k,\nu} &= \int_1^\oo [d\eta] \cP_{\frac{2-d}{2}+i\nu}(\eta) \Pi_{k,\eta}\,,\qquad \textrm{where}\quad [d\eta]\equiv 2^{d-2}\,(\eta^2-1)^{\frac{d-3}{2}} d\eta\,,
 \label{eq:inversetransforms}
\end{align}
with inverse\footnote{To determine the relative normalization of the Gegenbauer transform and its inverse, one can use:
\begin{align}
\cP_{\frac{2-d}{2}-i\nu}(\eta) \cP_{\frac{2-d}{2}+i\nu'}(\eta) \sim
\frac{(2\eta)^{i(\nu-\nu')}}{\gegenbauerconst_{\frac{2-d}{2}-i\nu}\gegenbauerconst_{\frac{2-d}{2}+i\nu'}}+(\nu\leftrightarrow \nu'), \quad \mbox{for }\eta\to\infty\,.
\end{align}
}
\begin{align}
 \Pi_{k,\eta} &= \int\limits_{0}^{\infty} \frac{d\nu}{2\pi} \gegenbauermeasure(\nu)  \cP_{\frac{2-d}{2}-i\nu}(\eta)\, \widehat{\Pi}_{k,\nu}\,.
\label{harmonic Pi}
\end{align}
The measure $\gegenbauermeasure(\nu)$ is given by
\begin{align}
\rho(\nu) = \gegenbauerconst_{\frac{2-d}{2}-i\nu} \gegenbauerconst_{\frac{2-d}{2}+i\nu}\,,\quad\textrm{where}\quad
\gegenbauerconst_J\equiv \frac{\Gamma\big(\tfrac{d-2}{2}\big)\Gamma(J+d-2)}{
 \Gamma(d-2)\Gamma\big(J+\tfrac{d-2}{2}\big)}\,.
 \label{eq:cjdefinition}
\end{align}
A calculation will show that the action of $\widehat{\Pi}_{k,\nu}$ on a heavy block reads, up to $\nu$-independent factors:
\be
\widehat{\Pi}_{k,\nu}[\Delta,J] \propto \g_{2\De_\f+k-1}(\nu)^2\cP_{\frac{2-d}{2}+i\nu}(\etaAdS)\,,
\label{Pi heavy}
\ee
where 
\begin{align}
 \gamma_a(\nu) &\equiv \Gamma\Big(\tfrac{1+a-\frac d 2 - i\nu}{2}\Big) \Gamma\Big(\tfrac{1+a-\frac d 2 + i \nu}{2}\Big)
 \,.
 \label{eq:gammafactors}
\end{align}
Comparing with \eqref{eq:thingwewant},
the functional we seek is thus $\widehat{\Pi}$ divided by the factors in \eqref{Pi heavy}.
We can then use \eqref{eq:inversetoangularintegral} to turn it into a physical functional.
We are led to define the $C_{k,\nu}$ functional as a double integral transform:
\begin{align}
\label{eq:magicmoment}
C_{k,\nu}&\equiv
\frac{
a_{\De_\f}}{\g_{2\De_\f+k-1}(\nu)^2}
\frac{(-1)^{\frac{k}{2}}\pi \Gamma(k)}{2^{3k-2}\Gamma\big(\tfrac{k}{2}\big)^2}
\int_1^\oo [d\eta] \cP_{\frac{2-d}{2}+i\nu}(\eta) \int_{\eta^2}^\oo
\frac{dv} {\big[(v-\eta^2)^{\frac{k+1}{2}}\big]_+}B_{k,v}\,,
\end{align}
where we included a convenient normalization
\begin{align}
a_{\De_\f} &\equiv 
\frac{\pi^{\frac{d-3}{2}} \Gamma (\Delta_\phi )^2 \Gamma (\Delta_\phi -\frac{d-2}{2})^2}{ 2^{d-5} \Gamma (\frac{d-1}{2})}\,.
\end{align}
The $C_{k,\nu}$ functional will be key to our studies.
While the double transform may seem daunting, it has the virtue of being conceptually transparent:
it will lead to a simple dictionary between flat-space dispersion relations and holographic sum rules.
The pair $(\nu,\etaAdS)$ will play the same role for bulk AdS physics as the pair $(p,b)$ plays in flat space.

By construction, $C_{k,\nu}$ has the Regge moment expansion
\begin{align}
\label{eq:leadingreggemomentexpansionofc}
 C_{k,\nu} &\sim 
 \frac{a_{\De_\f}}{\g_{2\De_\f+k-1}(\nu)^2}\widehat{\Pi}_{k,\nu} + O(\Pi_{k+2})\,.
\end{align}
This can be interpreted as follows. Multiplication by $\g_{2\De_\f+k-1}(\nu)^2$ in \eqref{Pi heavy} represents convolution by a smoothing kernel: the CFT angle-of-approach $\eta$ is a smeared version of the bulk impact parameter $\etaAdS$, as depicted in Figure~\ref{fig:etaAdS}.
This smearing arises because in Rindler kinematics
(the causality sum rules $B_{k,v}$ exploit that commutators vanish when operators 1 and 3 are spacelike-separated), correlators do not exhibit any sharp feature attributable to bulk scattering; in holographic models, the boundary-to-bulk propagator of a light field
is spread all over the transverse $H_{d-1}$ in the Regge limit \cite{Cornalba:2006xk}.
This is in contrast with the ``bulk point'' singularity of ref.~\cite{Gary:2009ae,Maldacena:2015iua} where operators 1 and 4 are in the future of 3 and 2.
The role of the factors in \eqref{eq:leadingreggemomentexpansionofc} is to undo that smearing and enable bulk focusing. This is discussed formally below, along with limitations, in \eqref{eq:psidefinition}.

It will turn out that $C_{k,\nu}$ is conveniently computable.
In the next two sections, we show the following:
\begin{itemize}
\item {\bf Regge limit:} For $\nu\ll \De$, the action of $C_{k,\nu}$ on heavy blocks in the Regge limit (large $\De$ with fixed $\etaAdS$) is:
\begin{align}
\label{eq:bknuinreggelimit}
C_{k,\nu}[\De,J] &=
\frac{2\cP_{\frac{2-d}{2}+i\nu}(\etaAdS)}{m^{2k}}\x\p{1+ O\p{\frac {\nu^2}{m^2},\frac{1}{m^2}}}\,.
\end{align}
Recall that the heavy density $C_{k,\nu}[\De,J]$ is defined by (\ref{eq:definitionofheavydensity}), and $m$ is given by (\ref{eq:massdefinition}). The presence of the AdS harmonic function $\cP_{\frac{2-d}{2}+i\nu}(\etaAdS)$ confirms that $\nu$ plays the role of transverse momentum in AdS.

\item {\bf Bulk point limit:} The action of $C_{k,\nu}$ on heavy blocks in the the the bulk point limit (large $\De$ with small AdS impact parameter $\beta\sim \frac{2J}{\De}$) is:
\begin{align}
\label{eq:bulkpointbknu}
C_{k,\nu}[\De,J]
&= 
 \frac{2 m ^2-\nu ^2}{m ^2-\nu ^2}\frac{\cP_{J}\big(1-\frac{2 \nu ^2}{m ^2}\big)}{(m ^2 (m ^2-\nu ^2))^{k/2}} 
 \x\p{1  + O\p{\frac{J^2}{m^2}}}\,,
\end{align}
Unlike (\ref{eq:bknuinreggelimit}), equation~(\ref{eq:bulkpointbknu}) is valid for all $\nu\in[0,m)$.
The right-hand side of (\ref{eq:bulkpointbknu}) is precisely the contribution of a massive state to the $ \mathcal{C}_{k,u}$ dispersion relation in flat space (\ref{eq:heavylightsplitC}), via the dictionary 
\begin{align}
\label{eq:flatdictionaryunu}
u R_\mathrm{AdS}^2 &= -\nu^2\,.
\end{align}
Here, we have temporarily reintroduced $R_\mathrm{AdS}$, which we set to 1 in subsequent sections. 
Large $\nu$ will afford us high resolution at small impact parameters, corresponding to flat-space physics.

\item {\bf Light contributions:} Finally, 
consider a bulk EFT given by gravity plus a sum of contact interactions.  The contribution of light states to $C_{k,\nu}$ at large $\nu$ is
\begin{align}
\label{eq:lightbknu}
-\left.C_{k,\nu}\right|_\mathrm{light} &= \Res_{s=0} \left[\frac{2s-\nu^2}{s(s-\nu^2)}\frac{\mathcal{M}_\mathrm{flat}(s,-\nu^2)}{[s(s-\nu^2)]^{k/2}}\right]\x\p{1+O\p{\frac{1}{\nu^2}}},
\end{align}
where $\mathcal{M}_\mathrm{flat}(s,u)$ is the flat-space amplitude corresponding to the AdS interactions. (This amplitude is unique up to $1/R_\mathrm{AdS}$ corrections.) Up to corrections in $1/\nu^2$, (\ref{eq:lightbknu}) precisely matches the low-energy contribution to the flat-space sum rule $\mathcal{C}_{k,u}$  (\ref{eq:heavylightsplitC}) in the case of a tree-level low energy amplitude, again with the identification $u R_\mathrm{AdS}= -\nu^2$.
\end{itemize}

In terms of the heavy expectation value (\ref{eq:cftpositivemeasure}),
the holographic sum rule corresponding to $C_{k,\nu}$ has the form
\begin{align}
 C_{k,\nu}\ \ &:\ \ \Res_{s=0} \left[\frac{2s-\nu^2}{s(s-\nu^2)}\frac{\mathcal{M}_\mathrm{flat}(s,-\nu^2)}{[s(s-\nu^2)]^{k/2}}\right] + \dots
= \left\<
\frac{2 m ^2-\nu ^2}{m ^2-\nu ^2}\frac{\cP_{J}\big(1-\frac{2 \nu ^2}{m ^2}\big)}{(m ^2 (m ^2-\nu ^2))^{k/2}}
\right\> + \dots
\label{eq:holographicbknu}
\end{align}
On the left-hand side, ``$\dots$'' refers to $O(1/\nu^2)$ corrections; on the right-hand side, ``$\dots$'' refers to  $O(J^2/m^2)$ corrections as well as contributions from other regimes besides the bulk point limit. To obtain bounds on AdS interactions, we must treat these details with care; we do this in section~\ref{sec:holo}. 
However, the general outline of (\ref{eq:holographicbknu}) is clear: in an appropriate flat space limit, the CFT sum rule $C_{k,\nu}$~\eqref{eq:magicmoment} becomes the flat space sum rule $\mathcal{C}_{k,u}$ \eqref{Ckudirect}.

In the following sections, we derive these results in detail, starting with the Regge limit (sections~\ref{sec:impactparameters} and \ref{ssec:regge higher orders}), then the bulk point limit (section~\ref{sec:heavymomentbulkpoint}), and finally the light contributions (section~\ref{sec:lightAction}).


\section{Action of functionals on heavy blocks}\label{sec:heavyAction}

\subsection{Integrals against heavy blocks and AdS impact parameters}
\label{sec:impactparameters}

Our first step will be to find an effective way to compute integrals of kernels against $s$-channel blocks $G^s_{\Delta,J}$, and in particular to see the simplifications at large twist $\tau=\Delta-J$.
To this end, we consider an integral with a generic kernel:
\be
 I_{\De,J}^\Omega = \frac14\int_0^{-\infty}\int_0^{-\infty} dw\,d\wb\, |w-\wb|^{d-2}
 \Omega(u',v') G^s_{\Delta,J}(u',v')\,, \label{functional action}
\ee
where $u'=w\wb$ and $v'=(1-w)(1-\wb)$.
The factor $\frac 1 4$ was introduced for later convenience. 

We will see that large $\Delta$ pushes the integral toward the $u$-channel limit $u',v'\to \infty$.
Instead of trying to approximate the heavy $s$-channel blocks in this limit (a difficult task),
we will find a geometric realization of \eqref{functional action} as an integral over spacetime.
We first do so in a simple way that is sufficient for understanding the Regge limit of large $\De$ with fixed $\De/J$. We will subsequently change to a more sophisticated spacetime integral to analyze the bulk-point limit.

Denoting $e=(1,\vec{0})$ a unit vector in the \emph{time} direction, we set $(x_1,x_2,x_3)=(-e,-\oo e, 0)$, so that we may view the remaining
point $x_4$ as parametrizing the cross-ratios: $u'=-x_4^2$, $v'=-(x_4+e)^2$.
It is then straightforward to verify directly that the cross-ratio integral (\ref{functional action}) is equivalent to an $x_4$ integral:
\begin{align}
I_{\De,J}^\Omega &= 2^{d-2}\int_{0<x_4} \frac{d^d x_4}{\vol S^{d-2}} \frac{ \Omega(u',v')}{(-x_{13}^2)^{\tl \De_\f}(-x_{24}^2)^{\tl \De_\f}}
\langle \phi_1 \cdots\phi_4\rangle_{\Delta,J}^s \,,
\label{functional as x2 integral}
\end{align}
where the notation $x_j<x_i$ means that $x_i$ is in the future of $x_j$.\footnote{Note that we have chosen $x_2 < x_1 <x_3 <x_4$, which is a different set of causal relationships points from the ones shown in Figure~\ref{fig:etaAdS}. A change in causal relationships introduces phases into three-point functions, which we keep track of explicitly in our calculation.}
The bracket  denotes the contribution of an s-channel block to the correlator:
\begin{align}
\<\f_1\cdots \f_4\>_{\De,J}^s &= \frac{G_{\De,J}^s(u',v')}{(-x_{13}^2)^{\De_\f}(-x_{24}^2)^{\De_\f}}\,.
\end{align}

\begin{figure}
\centering{
\begin{subfigure}[t]{0.4\textwidth}\def\svgwidth{6cm}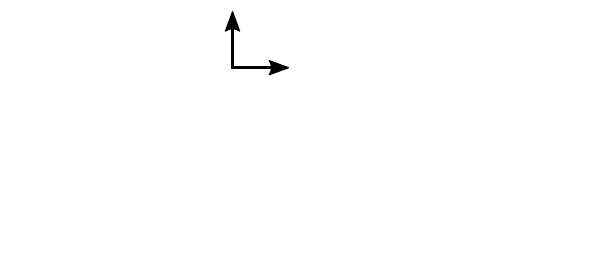\caption{}\end{subfigure}
\quad\begin{minipage}[t]{1cm} \raisebox{10mm}{$\Longrightarrow$}\end{minipage}\quad
\begin{subfigure}[t]{0.4\textwidth}\def\svgwidth{6cm}
\begingroup%
  \makeatletter%
  \providecommand\color[2][]{%
    \errmessage{(Inkscape) Color is used for the text in Inkscape, but the package 'color.sty' is not loaded}%
    \renewcommand\color[2][]{}%
  }%
  \providecommand\transparent[1]{%
    \errmessage{(Inkscape) Transparency is used (non-zero) for the text in Inkscape, but the package 'transparent.sty' is not loaded}%
    \renewcommand\transparent[1]{}%
  }%
  \providecommand\rotatebox[2]{#2}%
  \newcommand*\fsize{\dimexpr\f@size pt\relax}%
  \newcommand*\lineheight[1]{\fontsize{\fsize}{#1\fsize}\selectfont}%
  \ifx\svgwidth\undefined%
    \setlength{\unitlength}{160.72909981bp}%
    \ifx\svgscale\undefined%
      \relax%
    \else%
      \setlength{\unitlength}{\unitlength * \real{\svgscale}}%
    \fi%
  \else%
    \setlength{\unitlength}{\svgwidth}%
  \fi%
  \global\let\svgwidth\undefined%
  \global\let\svgscale\undefined%
  \makeatother%
  \begin{picture}(1,0.32975423)%
    \lineheight{1}%
    \setlength\tabcolsep{0pt}%
    \put(-0.00282526,0.08034922){\makebox(0,0)[lt]{\lineheight{1.25}\smash{\begin{tabular}[t]{l}$-e\infty$\end{tabular}}}}%
    \put(0.2991875,0.08284895){\makebox(0,0)[lt]{\lineheight{1.25}\smash{\begin{tabular}[t]{l}$-e$\end{tabular}}}}%
    \put(0.64664097,0.08187263){\makebox(0,0)[lt]{\lineheight{1.25}\smash{\begin{tabular}[t]{l}$0$\end{tabular}}}}%
    \put(0.86358213,0.26261155){\makebox(0,0)[lt]{\lineheight{1.25}\smash{\begin{tabular}[t]{l}$x_4$\end{tabular}}}}%
    \put(0,0){\includegraphics[width=\unitlength,page=1]{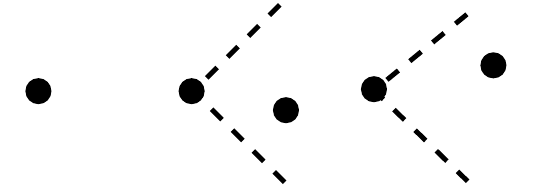}}%
    \put(0.03436081,0.22643044){\makebox(0,0)[lt]{\lineheight{1.25}\smash{\begin{tabular}[t]{l}$x_2$\end{tabular}}}}%
    \put(0.31480278,0.22424119){\makebox(0,0)[lt]{\lineheight{1.25}\smash{\begin{tabular}[t]{l}$x_1$\end{tabular}}}}%
    \put(0.63730782,0.22685933){\makebox(0,0)[lt]{\lineheight{1.25}\smash{\begin{tabular}[t]{l}$x_5$\end{tabular}}}}%
    \put(0.47431682,0.05957103){\makebox(0,0)[lt]{\lineheight{1.25}\smash{\begin{tabular}[t]{l}$x_3$\end{tabular}}}}%
    \put(0,0){\includegraphics[width=\unitlength,page=2]{fivepoints2.pdf}}%
  \end{picture}%
\endgroup%
\caption{}\end{subfigure}}
\caption{
The ``wavefunction trick'' to act with a functional on a conformal block.
(a) The conformal block $\<\f_1\cdots \f_4\>_{\De,J}^s$ is written as an integral over an auxiliary point $x_5$, 
restricted to a causal diamond, while $x_4$ is itself integrated over the future lightcone of $x_3$.
(b) The same integral after a conformal transformation that brings $x_5$ to the origin.
In the heavy block limit, $\Delta-J\sim \Delta_{\rm gap}\gg 1$
the integration points $x_3$ and $x_4$ repel each other as indicated by the arrows,
and the integral reduces to a Laplace transform of the functional (see eq~\eqref{I impact parameter}).
\label{fig:fivepoints}}
\end{figure}

We now evaluate the block using the Lorentzian shadow representation \cite{Kravchuk:2018htv}.
This involves integrating a fifth point over the causal diamond $x_3<x_5<x_4$ and also integrating over a null polarization vector $n$ in index-free notation:\footnote{The Lorentzian shadow representation (\ref{shadow}) converges for $J$ on the principal series $J\in \frac{2-d}{2}+i\mathbb{R}$, and can be analytically continued away from there. For integer spin $J$, there is an alternative shadow representation that involves contracting indices between two integer-spin three-point structures, which goes back to Polyakov \cite{Polyakov:1974gs}. We could use either representation, but (\ref{shadow}) will be more convenient.
}
\be
 \langle \phi_1\cdots \phi_4\rangle_{\Delta,J}^s = \frac{1}{b_{\Delta,J}}
 \int\limits_{x_3<x_5<x_4}  d^dx_5\, D^{d-2} n
\,|\langle \phi_1\phi_2 \mathcal{O}(x_5,n)\rangle|
 |\langle \mathcal{O}^\mathrm{S}(x_5,n)\phi_3\phi_4 \rangle|\,.
\label{shadow}
\ee
The measure $D^{d-2} n$ is the standard Lorentz-invariant measure on the projective null cone in $\mathbb{R}^{d-1,1}$ \cite{SimmonsDuffin:2012uy}, which is equivalent to an ordinary integral over the orientation of the spatial components of $n$:
\begin{align}
\int D^{d-2} n f(n) &= \int \frac{2d^d n \de(n^2)}{\mathop{\mathrm{vol}} \mathbb{R}^+} f(n) = \int d^{d-2} \Omega f(n)\big|_{n=(1,\vec \Omega)}\,.
\label{eq:projectivemeasure}
\end{align}
The operator $\cO$ has dimension $\De$ and spin $J$.
The quantum numbers of $\cO^\mathrm{S}$ are related to those of $\cO$ by the (Lorentzian) shadow $\mathrm{S}$: $(\Delta,J)\mapsto(d-\Delta,d-2-J)$.
The constant $b_{\Delta,J}$ is given by
\begin{align}
b_{\Delta,J} &= \frac{\pi ^{d-2} \Gamma (\Delta +2-d) \Gamma (J+\frac{d-2}{2}) \Gamma (\frac{J+\Delta }{2})^2 \Gamma (\frac{\Delta +2-d-J}{2})^2}{2 \Gamma (\Delta +\frac{2-d}{2}) \Gamma (J+d-2) \Gamma (J+\Delta ) \Gamma (\Delta +2-d-J)}\,.
\end{align}
The absolute values in (\ref{shadow}) indicate that all distances should be computed as $-x_{ij}^2$, so as to avoid phases since all distances are timelike.
Explicitly, the three-point structures are given by
\be
 |\langle \phi_1\phi_2 \mathcal{O}(x_5,n) \rangle| =
\frac{(-x_{12}^2)^{\frac{\Delta}2-\Df}}{(x_{15}^2x_{25}^2)^{\frac{\Delta}2}}
 \left( \frac{2n_\mu[ x_{15}^2 x_{25}^\mu-x_{25}^2 x_{15}^\mu]}{\sqrt{-x_{12}^2x_{15}^2x_{25}^2}}\right)^{J}\,,
\ee
and similarly for $|\langle \phi_3\phi_4 \mathcal{O}^{\mathrm S}(x_5,n)\rangle|$.

Substituting \eqref{shadow} into \eqref{functional as x2 integral} computes the functional
in terms of a triple integral $d^dx_4d^dx_5 D^{d-2}n$, where coordinates satisfy the causal ordering $x_2<x_1<x_3<x_5<x_4$.
The $n$ integral is in fact redundant after integrating over $x_4,x_5$, so we can fix $n=(1,\vec e_z)$ and cancel the factor $\frac{1}{\vol S^{d-2}}$.
The kinematics are shown in Figure~\ref{fig:fivepoints}.

More abstractly, we have now expressed
$I_{\Delta,J}^\Omega$ as a gauge-fixed version of a conformal integral over five causally-ordered points:
\be\begin{aligned}
I_{\De,J}^\Omega &= \frac{2^{2d-2}}{b_{\Delta,J}}\mathop{\mathrm{vol}} \mathrm{SO}(d-2)
\int\limits_{x_2<x_1<x_3<x_5<x_4} \frac{d^d x_1 d^d x_2 d^d x_3 d^d x_4d^d x_5 D^{d-2}n}{\mathop{\mathrm{vol}} \widetilde{\mathrm{SO}}(d,2)}
\\& \quad\times \frac{ \Omega(u',v')}{(-x_{13}^2)^{\tl \De_\f}(-x_{24}^2)^{\tl \De_\f}}\,|\langle \phi_1\phi_2 \mathcal{O}(x_5,n)\rangle|
 |\langle \mathcal{O}^\mathrm{S}(x_5,n)\phi_3\phi_4 \rangle|\,.
 \label{conformal measure}
\end{aligned}\ee
The meaning of $1/\vol\ \widetilde{\mathrm{SO}}(d,2)$ is that we must gauge-fix the action of the conformal group on all points and $n$, and introduce a Faddeev-Popov determinant. In the present gauge-fixing, the determinant is \cite{Kravchuk:2018htv} $1/2^d$ divided by the volume of the stabilizer group of $(0,1,\oo)$, namely $\SO(d-1)$, which again leads to (\ref{functional as x2 integral}) after using $\mathrm{vol}\ \mathrm{SO}(d-1)=\mathrm{vol}\ \mathrm{SO}(d-2)\times \mathrm{vol}\ S^{d-2}$.

The key idea is to now use conformal symmetry (only Lorentz transformations and rescalings about $-e$ are actually needed in this case)
to gauge-fix the position of the auxiliary point $x_5$ to the origin, so that the integration variables now become $x_3$ and $x_4$, instead of $x_4$ and $x_5$.
The range of $x_4$ is the future lightcone of the origin, and that of $x_3$ is the causal diamond $-e<x_3<0$.  
We may parametrize these in terms of positive timelike vectors $x,y$:
\be
\label{eq:firstconformalframe}
 (x_1,x_2,x_3,x_4;x_5)\mapsto \left(-e,-\oo e,\frac{-(e+y)}{-(e+y)^2},\frac{x}{-x^2};0\right),\qquad x,y>0\,.
\ee
Evaluating explicitly the three-point structures in \eqref{conformal measure} we find
\be \label{neat B integral}
  I_{\De,J}^\Omega = \frac{(-1)^J}{b_{\Delta,J}}\int_{x,y>0} \frac{d^dx}{(-x^2)^{\tildeDf}}\frac{d^dy}{(-y^2)^{\tildeDf}}
  \Omega(u',v')
  \frac{\big[{-}n{\cdot}(e+x+y)\big]^{2-d-J}}{\big[{-}(e+x+y)^2\big]^{\frac{\Delta-J}{2}+\Df+1-d}}\,,
\ee
where the cross-ratios are now given by 
\be
u'=\frac{-(e+x+y)^2}{x^2y^2}\,,\qquad v'=\frac{(e+x)^2(e+y)^2}{x^2y^2}\,.
\ee
Equation \eqref{neat B integral} is an \emph{exact} rewriting of the integral \eqref{functional action},
where the $s$-channel block has been replaced by elementary factors at the cost of extra integrations.
Note that both square brackets and cross-ratios are positive because all vectors are future-directed timelike or null.

The key feature of \eqref{neat B integral} is that, for large twist,
the last factor localizes the integral to small $x,y\sim \frac{1}{\Delta}$.
More precisely, if we take $\Delta\sim J$ (but still $\Delta-J\gg 1$) and $x,y$ small, we find the simple limit:
\be\begin{aligned}
 \frac{\big[{-}n{\cdot}(e+x+y)\big]^{2-d-J}}{\big[{-}(e+x+y)^2\big]^{\frac{\Delta-J}{2}+\Df+1-d}}
 &\to 
 e^{(\Delta-J)e{\cdot}(x+y) + J n{\cdot}(x+y)} \nn\\
 &= e^{-\Delta(x^0+y^0)+J\vec{e}_z\.(\vec{x}+\vec{y})} \equiv e^{p\.x+p\.y}\,,
 \end{aligned} \label{heavy limit I}
\ee
where
\be
\begin{aligned}
p &= (\De, J \vec e_z)\,.
\end{aligned}\label{exponential from power}\ee
Thus:
\be
 \lim_{\Delta,J\gg 1} I_{\De,J}^\Omega = 
 \frac{(-1)^J}{b_{\Delta,J}}\int_{x,y>0} \frac{d^dx d^dy}{(x^2y^2)^{\tildeDf}}
  \Omega(u',v') e^{x\.p+y\.p}\,. \label{I impact parameter}
\ee
The integral of a kernel against a heavy block is the Laplace transform of the kernel!

Equation \eqref{I impact parameter} is closely related to the impact parameter representation of refs.~\cite{Cornalba:2006xm,Cornalba:2007zb,Penedones:2007ns}.
Note however that here we are not transforming correlators --- instead we are transforming functionals that act on correlators.
The action localizes to the Regge limit $w,\wb\to -\infty$ with angle of approach
\be \eta \approx \frac{-x{\cdot}\bar{y}}{|x||y|}\,,  \label{angle of approach}
\ee
where $\bar{y}\equiv -y-2e (y\.e)$ denotes the spatially-reflected vector.
The integral then depends on $\Delta$ and $J$ only through the hyperbolic angle between $p$ and $\bar{p}$:
\be \label{xi impact}
 \frac{-p{\cdot}\bar{p}}{|p||\bar{p}|} = \frac{\Delta^2+J^2}{\Delta^2-J^2}\approx \etaAdS\,.
\ee

Our task is now clear: as explained in subsection \ref{ssec:what does it mean}, 
to probe the flat space limit of a holographic theory,
we must find physical sum rules $\Omega(u',v')$ such that the integral \eqref{I impact parameter} is localized
as much as possible around $\etaAdS\to 1$. This will be achieved using the functional
$C_{k,\nu}$ in \eqref{eq:magicmoment}, which uses harmonic analysis to inject momentum $\nu$
conjugate to $\etaAdS$, thereby providing a crucial link between AdS and CFT quantum numbers.

\subsection{Regge moments of heavy blocks in the Regge limit}
\label{sec:reggemomentsheavy2}

To the leading order in the Regge limit, the physical functional $C_{k,\nu}$ is proportional
to the Regge moment $\widehat{\Pi}_{k,\nu}$ (see \eqref{eq:leadingreggemomentexpansionofc})
defined in (\ref{eq:reggemomentdefinition}) and (\ref{eq:inversetransforms}), which fits the template in~\eqref{functional action}.
The kernel $\Omega$ corresponding to $\widehat \Pi_{k,\nu}$ is found by computing the Jacobian between $dr\,d\eta$ and $dw\,d\wb$:
\be \label{Pi as I integral}
\frac{\widehat{\Pi}_{k,\nu}[G^s_{\De,J}]}{2\sin^2(\pi \frac{\tau-2\De_\f}{2})} =I^\Omega_{\Delta,J} \quad \mbox{ with }\quad
\Omega(u',v') = \frac{4^d (\rho\bar{\rho})^{\frac{k+d-1}{2}}\cP_{\frac{2-d}{2}+i\nu}\p{\frac{\rho+\bar{\rho}}{2\sqrt{\rho\bar{\rho}}}}}
{(1-\rho\bar{\rho})^{d-2} (1-\rho^2)(1-\bar{\rho}^2)}\,.
\ee
The $2\sin^2(\cdots)$ factor, with $\tau=\De-J$, accounts for the double-discontinuity of the block.
Let us first focus on the leading term as $\tau\to\infty$.
It can be computed by substituting into \eqref{I impact parameter} and taking small $x,y\to 0$
 (where $|x||y|\to4\sqrt{\rho\bar{\rho}}$) which gives the Laplace transform
\begin{align}
\lim_{\De,J\gg 1}\frac{\widehat{\Pi}_{k,\nu}[G^s_{\De,J}]}{{2\sin^2(\pi \frac{\tau-2\De_\f}{2})}} &=
\frac{(-1)^{J}4^{1-k}}{b_{\De,J}}\int\limits_{x,y>0} d^d x d^d y\,(|x||y|)^{k-d-1+2 \De_\f} \cP_{\frac{2-d}{2}+i\nu}\p{\frac{-x\.\bar{y}}{|x||y|}} e^{x\.p+y\. p}\,,
 \label{eq:reggemomentspacetimedeltaspace}
\end{align}
where $p=(\De,J\vec e_z)$.

This integral can be done readily because the Fourier-Laplace transform of a 
Gegenbauer function (times a power) is again a Gegenbauer (times a power) \cite{Penedones:2007ns,Kulaxizi:2017ixa,Li:2017lmh}:
\be
 \int_{x>0} d^dx |x|^{a-d} \cP_{\frac{2-d}{2}+i\nu}\p{\frac{-x\.y}{|x||y|}}e^{x\.p}
 = 2^{a-1} \pi^{\frac{d-2}{2}} \gamma_a(\nu)\times |p|^{-a}\cP_{\frac{2-d}{2}+i\nu}\p{\frac{-p\.y}{|p||y|}}\,, \label{Fourier identity}
\ee
where $\gamma_a(\nu)$ the product of $\G$-functions defined in (\ref{eq:gammafactors}). 
The proportionality is guaranteed by rotational and scale symmetry of the transform,
and a simple derivation of (\ref{Fourier identity}) using the ``split representation'' is presented in appendix \ref{app:Fourier Gegen}.
Using this result twice, we essentially replace $x$ and $y$ with $p$ and $\bar p$ in \eqref{eq:reggemomentspacetimedeltaspace}:
\be
\lim_{\De,J\gg 1}\frac{\widehat{\Pi}_{k,\nu}[G^s_{\De,J}]}{{2\sin^2(\pi \frac{\tau-2\De_\f}{2})}} = 4^{2\Df-1}\pi^{d-2}\frac{(-1)^{J}}{b_{\De,J}}
\times \gamma_{2\Df+k-1}(\nu)^2 \times \frac{\cP_{\frac{2-d}{2}+i\nu}\p{ \frac{-p\.\bar p}{|p||\bar p|}}}{(|p||\bar p|)^{k+2\De_\f-1}}\,.
\label{eq:finalreggemomentofablock}
\ee

Equation (\ref{eq:finalreggemomentofablock}) is a crucial result: up to factors that depend only on
$(\Delta,J)$, it shows (compare with \eqref{harmonic Pi}) that the Regge angle-of-approach $\eta$
is related to the impact-parameter $\etaAdS$
by multiplication by $\gamma_{2\Df+k-1}(\nu)^2$ in AdS momentum space.
Our ability to localize in $\etaAdS$ is tantamount to our ability to invert that factor. This is precisely what we do in the definition of $ C_{k,\nu}$ (\ref{eq:magicmoment}).

The factors in the above explain the normalization in \eqref{eq:positivedensityfactors}.
Using the expression for the MFT coefficients (\ref{eq:mftope}),
the formula simplifies to:
\be
\lim_{\De,J\gg 1} \widehat{\Pi}_{k,\nu}[\De,J] = \frac{\gamma_{2\Df+k-1}(\nu)^2}{a_{\Df}}
\frac{2\cP_{\frac{2-d}{2}+i\nu}(\etaAdS)}{m^{2k}}\,.
\ee
It is  straightforward to check that this agrees with (\ref{eq:bknuinreggelimit}) quoted previously,
using the Regge moment expansion (\ref{eq:leadingreggemomentexpansionofc}).

Interestingly, we can formally define functionals that are $\de$-function localized in $\etaAdS$ to leading order in the Regge limit by inverting the harmonic decomposition of $ C_{k,\nu}$:
\begin{align}
\label{eq:psidefinition}
\Psi_{k,\gamma} &\equiv \int_0^\oo \frac{d\nu}{2\pi} \gegenbauermeasure(\nu) \cP_{\frac{2-d}{2}-i\nu}(\gamma)  C_{k,\nu}\,.
\end{align}
They satisfy
\begin{align}
\label{eq:psiknuinreggelimit}
\lim_{\De,J\gg 1}\Psi_{k,\gamma}[\De,J] &=
 \frac{2}{m^{2k}}
\x \frac{\de(\gamma-\etaAdS)}{2^{d-2}(\gamma^2-1)^{\frac {d-3}{2}}}\,.
\end{align}
Thus, $\Psi_{k,\gamma}$ gives information about the density of heavy states with
fixed $\etaAdS\approx\frac{\De^2+J^2}{\De^2-J^2}$ in the heavy limit. We expect that (\ref{eq:psiknuinreggelimit}) is valid only for $\cosh^{-1}\g \gg 1/m$, since it involves an integral over large $\nu$, and subleading $1/m^2$ corrections can be enhanced at large $\nu$, as we will see in the next section.
We will see an application of (\ref{eq:psiknuinreggelimit}) in section~\ref{sec:eikonal}.
There is an additional technical caveat that makes $\Psi_{k,\gamma}$ a distribution which must be paired with an appropriate test function: it would impossible to perfectly ``unsmear'' and localize to infinite accuracy.
Yet, the space of test functions, described precisely in appendix~\ref{sec:impactparam},
contains narrowly-peaked functions.

\subsection{Higher orders in the large-$m$ expansion}
\label{ssec:regge higher orders}

\begin{figure}
\centering{
\begin{subfigure}[t]{0.4\textwidth}\centering{\def\svgwidth{6cm}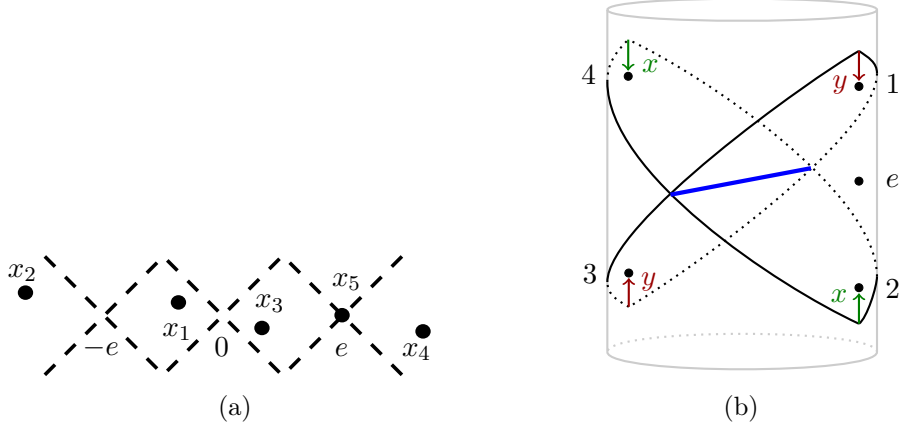\caption{}}\end{subfigure}
\quad
\begin{subfigure}[t]{0.4\textwidth}\centering
\begin{tikzpicture}[scale=0.75]
    \draw[looseness=0.14,thick,black!20!white] (0.94,7.03) to[out=-90,in=-90] (5.72,7.03) to[out=90,in=90] (0.94,7.03);
    \draw[looseness=0.22,thick,black!20!white] (0.94,0.96) to[out=-90,in=-90] (5.72,0.96);
    \draw[looseness=0.23,dotted,thick,black!20!white] (5.72,0.96) to[out=90,in=90] (0.94,0.96); 
    \draw[thick,black!20!white] (0.94,0.96) -- (0.94,7.03);
    \draw[thick,black!20!white] (5.72,0.96) -- (5.72,7.03);
    \draw[looseness=0.5,thick] (5.4,1.46) to[out=35,in=-90] (5.72,2.3);
    \draw[looseness=0.63,thick,dotted] (5.72,2.3) to[out=90,in=-25]  (1.31,6.5);
    \draw[looseness=0.67,thick] (5.4,1.46) to[out=155,in=-90] (0.94,5.75);
    \draw[looseness=1,thick,dotted] (0.94,5.75) to[out=90,in=-135] (1.31,6.5);
    \draw[looseness=1,thick,dotted] (1.32, 1.76) to[out=150,in=-90] (0.94,2.2);
    \draw[looseness=0.44,thick] (0.94,2.2) to[out=90,in=-150] (5.4,6.3);
    \draw[looseness=1,thick] (5.4,6.3) to[out=-35,in=90] (5.72,5.9);
    \draw[looseness=0.48,thick,dotted] (5.72,5.9) to[out=-90,in=30] (1.32, 1.76);
    \draw[fill] (1.32, 2.36) circle (2pt);
    \draw[fill] (5.4,2.10) circle (2pt);
    \draw[fill] (1.31,5.85) circle (2pt);
    \draw[fill] (5.4,5.67) circle (2pt);
    \draw[fill] (5.4,3.99) circle (2pt);
    \draw[blue,ultra thick] (2.06,3.75) -- (4.55,4.22);
    \node [left] at (0.94,2.35) {$3$};
    \node [right] at (5.7,5.72) {$1$};
    \node [right] at (5.69,2.09) {$2$};
    \node [left] at (0.92,5.88) {$4$};
    \draw [red!60!black,->, thick] (1.32, 1.76) -- (1.32, 2.27);
    \draw [red!60!black,->, thick] (5.4,6.3) -- (5.4,5.77);
    \draw [green!50!black,->, thick] (5.4,1.46) -- (5.4,2.0);
    \draw [green!50!black,->, thick] (1.31,6.5) -- (1.31,5.95);
    \node [red!60!black,left] at (1.99, 2.21) {$y$};
    \node [red!60!black,right] at (4.75,5.7) {$y$};
    \node [green!50!black,right] at (4.72,1.95) {$x$};
    \node [green!50!black,left] at (2.03,6.03) {$x$};
    \node [right] at (5.7,4.0) {$e$};
\end{tikzpicture}
\caption{}
\end{subfigure}}
\caption{
(a) Symmetrical frame \eqref{eq:otherframe} with $-x_1=x_3=y$, $-x_2=x_4=\frac{x}{-x^2}$.
It simplifies subleading terms in the Regge limit compared with Figure~\ref{fig:fivepoints}.
(b) The same coordinates on the cylinder, with $x,y\to 0$ interpreted as displacements away from
the Regge limit (compare with~\cite{Cornalba:2007fs}). 
In holographic theories, injecting a large momentum $\nu$ into $x,y$
effectively flows a large transverse momentum through the bulk AdS${}_{d-1}$
shown as a thick blue line. This is key to probing small-impact parameter scattering.
(The shown coordinates are on a different Poincar\'e patch compared with (a); the double commutator in the right kinematics is proportional to block computed on the left.)
\label{fig:cylinder}}
\end{figure}

It will be important to understand the structure of higher order terms in the large-$m^2$ expansion of the action of $C_{k,\nu}$. For one thing, these can be enhanced at large-$\nu$.
To study these corrections, let us choose a slightly different conformal frame from (\ref{eq:firstconformalframe}) with nicer symmetry properties (Figure~\ref{fig:cylinder}):
\begin{align}
\label{eq:otherframe}
(x_1,x_2,x_3,x_4;x_5) &= \p{-y,\frac{-x}{-x^2},y,\frac{x}{-x^2}; e}, \qquad x,y>0\,.
\end{align}
An advantage of this frame is that 
the radial coordinates defined in \eqref{eq:definitionofreta} are extremely simple:
\be
r=|x||y|\,,\qquad  \eta = -\frac{x\.y}{|x||y|}\,,
\ee
where $|x|=(-x^2)^{1/2}$ and $|y|=(-y^2)^{1/2}$.

We perform the Faddeev-Popov procedure for the frame (\ref{eq:otherframe}) in appendix~\ref{app:heavy}, 
and derive the following exact expression for the Regge moment of an $s$-channel block:

\be \label{moment maintext}
\frac{\widehat{\Pi}_{k,\nu}[G^s_{\De,J}]}{2\sin^2(\pi \frac{\tau-2\De_\f}{2})} = 4^{2\Df}\frac{(-1)^{J}}{b_{\Delta,J}}
\int d^dx d^dy\ \big(|x||y|\big)^{k+d-1} \cP_{\frac{2-d}{2}+i\nu}\p{\frac{{-}x\.y}{|x||y|}}
\mathcal{T}(x,y)\,, \ee
where the product of three-point functions $\mathcal{T}(x,y)$ is now given by 
\be\begin{aligned}
\mathcal{T}(x,y)&=
\ \frac{1}{(x^2y^2)^{\tl\De_\f}}
\frac{(1 + 2x\.y +x^2y^2)^{d-1-2\Df}}{(e-x)^2(e-y)^2} \left( \frac{(e-x)^2(e-y)^2}{(e+x)^2(e+y)^2}\right)^{\frac{\Delta+J}{2}}
\times \frac{[-n{\cdot}\mathcal{V}_e]^{J}}{[n{\cdot}\mathcal{V}_{-e}]^{J+d-2}}\,,
\\
 \mathcal{V}_e &= \left((1-2e\.x)(1-2e\.y)-x^2y^2\right)e + (e+y)^2x  + (e+x)^2\bar{y}\,,
\end{aligned}\ee
where $\bar{y}=-y-2e(y\.e)$ denotes the space-reflected vector. In the integral (\ref{moment maintext}), both $x$ and $y$ range over the diamond $0<x,y<e$.

To expand in the Regge limit, we simply rescale $(x,y)\mapsto (x/m,y/m)$ and express
$(\Delta,J)$ in terms of $(m,\etaAdS)$ (see \eqref{eq:impactparameter}).
The integrand then becomes a simple exponential times a tower of $1/m$ corrections,
multiplying polynomials in $x$ and $y$.
The convenient feature of the symmetrical frame is that only even powers appear!
The polynomials can be integrated straightforwardly by taking derivatives with respect to $p$.
The procedure can be carried systematically to rather high order.

Results for the Regge moments $\widehat{\Pi}_{k,\nu}$ can be immediately translated into results for
the physical sum rules $C_{k,\nu}$ using the series in appendix \ref{sec:functionalsexpandreggemoments}.
For illustration we record the first subleading correction:
\begin{align} \label{regge subleading}
 &\frac{m^{2k}}{2} C_{k,\nu}[\De,J] = \nn\\
 &\quad
 \cP(\etaAdS)
  + \frac{1}{6m^2}\Bigg\{
 \cP'(\etaAdS)\left[(\etaAdS^2-1)(\nu^2-3a^2+6a(2\Df-d)-2)-(d-4)(d-2)\right]
\nonumber\\
&\hspace{33mm}+\etaAdS\cP(\etaAdS)
\Big[(\nu^2-a(a+d-2))(\tfrac{7d}2+3a-1-6\Df)-2(\Df-d-1)_3
\nonumber\\
&\hspace{60mm}-2(a+\tfrac{d-2}{2})(1-a^2-a\tfrac{d-2}{2})\Big]\Bigg\} +O\p{\frac{1}{m^4}}\,,
\end{align}
where $a=2\De_\f+k-d/2$, and
we omitted the subscript on $\cP_{\frac{2-d}{2}+i\nu}$ for readability.
This formula is tested numerically in Figure~\ref{fig:convergence_regge} below,
and a similar series to order $1/m^8$ is attached in an ancillary file.
The expansion has the following important features:
\begin{itemize}
\item The dependence on $\etaAdS$ occurs only through
the harmonic function $\cP(\etaAdS)$ and its derivatives,
with coefficients that are polynomial in $\etaAdS$.
The coefficient of $\cP^{(n)}/m^{2j}$ does not grow faster than $(\etaAdS)^{j+n}$ at large $\etaAdS$.
\item The $\nu$-dependence at each order in $1/m$ is an entire function of $\nu$.
When expanded around $\etaAdS\to 1$, the coefficient of $1/m^{2j}$ does not grow no faster than
$\nu^{2j}$ at large $\nu$.
\end{itemize}
The first property ensures that the $1/m$ series remains applicable in the Regge limit no matter the spin of the heavy operator:  the combination $\frac{\etaAdS}{m^2}\approx \frac{\Delta^2+J^2}{(\Delta+J)^2(\Delta-J)^2}$ is always small as long as the twist is large.

The second property ensures that the regimes of $\nu\to 0$ and $\nu\sim m$
are smoothly connected by a series in $\nu/m$,
as long as impact parameters are not large (in AdS units), $\etaAdS-1\ll 1$.

Let us elaborate on this second property, focusing
on the bulk-point limit, where $m\to\infty$ with spin $J$ held fixed.
In this limit, $\etaAdS-1=\frac{2\mathcal{J}^2+(d-2)^2/2}{m^2}\to 0$, where $\mathcal{J}^2=J(J+d-2)$,
and the action of $ C_{k,\nu}$ on heavy blocks with fixed $\nu$ and $J$ has the form
\begin{align}
\label{eq:largemu}
\frac{m^{2k}}{2}
C_{k,\nu}[\De,J] &=
1 + \frac{Q_{k,2}(\nu)}{m^2} + \frac{Q_{k,4}(\nu)}{m^4} + \dots 
&&
(\textrm{$m\gg 1$; fixed $\nu,J$})\,,
\end{align}
where each $Q_{k,n}(\nu)$ is a polynomial of degree $n$ in $\nu^2$.
For example,\footnote{We expect that it should be possible to understand the $\De_\f$-dependence in (\ref{eq:qtwok}) by comparing to dispersion relations for massive scalars in flat space. To derive it, one could generalize the saddle point in section~\ref{sec:heavymomentbulkpoint} to include large $\De_\f$. Note that $\De_\f$-dependence does not appear in the leading large-$\nu$ terms of $Q_{k,n}(\nu)$, which are our main focus. }
\begin{align}
\label{eq:qtwok}
 Q_{k,2}(\nu) &= 
\nu ^2 \left(-\frac{2 \mathcal{J}^2}{d-1}+\frac{k+1}{2}\right)-\frac{(d-2)^2 \mathcal{J}^2}{2 (d-1)}+2 (d+k-1) \Delta _{\phi } (\Delta _{\phi }-d)
\nn\\
&\quad +\frac{1}{24} \left(8d^3 +3d^2(k+1)-4d(3k^2-3k+8)-4(k^2+2)(k-3)\right).
\end{align}
To make contact with flat space physics, the regime of large AdS momentum $\nu\sim m$ is particularly significant.  In this regime, one finds a series in $\nu/m$ by keeping the leading term of each $Q_{k,m}$,
and we find, for example:
\begin{align}
Q_{k,2}(\nu) &= \nu^2  \left(-\frac{2 \mathcal{J}^2}{d-1}+\frac{k+1}{2}\right) + O(1)\,, \nn\\
Q_{k,4}(\nu) &= \nu ^4 \left(\frac{2 \mathcal{J}^4}{d^2-1}-\frac{(d (k+3)+k-1)\mathcal{J}^2 }{d^2-1}+\frac{(k+2)^2}{8} \right) + O(\nu^2)\,,
\label{eq:QPolynomials}
\end{align}
where $\mathcal{J}^2 = J(J+d-2)$.

Amazingly, \eqref{eq:QPolynomials} agrees precisely with
the forward-limit expansion of the flat-space expressions \eqref{S heavy}!  This is the first hint that the
physical functional $C_{k,\nu}$ indeed provides a direct link to flat space physics.

We will now explain a shortcut to compute the leading terms at large $\nu$,
bypassing the cumbersome Regge limit expansion.

\subsection{The bulk point limit and spacelike scattering}
\label{sec:heavymomentbulkpoint}

We now study the action of dispersive functionals in the bulk-point limit of small $\frac{2J}{\De}$. We will be interested in finding a formula that works for both small $\nu$ and $\nu \sim m$, as that will allow us to mimic the construction of positive functionals in flat space by integrating over $\nu$.
This requires us re-sum the leading terms at large $\nu$ of the polynomials $Q_{k,2n}(\nu)$.  We will do so by identifying a saddle-point of integrals like \eqref{moment maintext}.

We begin by studying $\widehat \Pi_{k,\nu}[\De,J]$ at large $\nu$; after we identity the saddle point,
it will be straightforward to modify the result to find $C_{k,\nu}[\De,J]$ at large $\nu$. We start from formula (\ref{moment maintext}) for the $\widehat{\Pi}_{k,\nu}$ moment of a heavy block
and plug in the split representation (\ref{eq:integralgegenbauer}) for the Gegenbauer function $\cP_{\frac{2-d}{2}+i\nu}(\cdots)$ as an integral over a null vector $z$. Using $\SO(d-1)$ invariance, we can trade the integral over $z$ for an integral over $n$, and fix $z=(1,1,0,\dots,0)$. The integral over $n$ can then be done using (\ref{eq:integralgegenbauer}), leaving 
\begin{align}
\label{eq:sodminusoneswapperoo}
\frac{\widehat{\Pi}_{k,\nu}[G^s_{\De,J}]}{2\sin^2(\pi \frac{\tau'-2\De_\f}{2})}
 &= \frac{4^{2\Df}(-1)^{J}}{b_{\Delta,J}} 
\int d^dx d^dy\  \frac{\big(|x||y|\big)^{k+d-1}}{(x^2y^2)^{\tl\De_\f}}
\frac{(1 + 2x\.y +x^2y^2)^{d-1-2\Df}}{(e-x)^2(e-y)^2} \nn\\
&
\x \p{\frac{x^-}{|x|}}^{\frac{2-d}{2}+i\nu} \p{\frac{y^-}{|y|}}^{\frac{2-d}{2}-i\nu}\left( \frac{(e-x)^2(e-y)^2}{(e+x)^2(e+y)^2}\right)^{\frac{\Delta+J}{2}}
 \frac{|\cV_e|^{J}}{|\cV_{-e}|^{J+d-2}} \cP_{J} \p{\frac{\cV_e\.\cV_{-e}}{|\cV_e||\cV_{-e}|}},
\end{align}
where we've defined lightcone coordinates $x=(x^+,x^-,\vec x)$ with $x^- = -z\.x$ (similarly $y^-=-z\.y$).
The Gegenbauer function $\cP_J$ is as expected for the spin-$J$
index contraction between two three-point functions in the conventional shadow representation \cite{Polyakov:1974gs}.

\begin{figure}
\centering
\includegraphics[width=\textwidth]{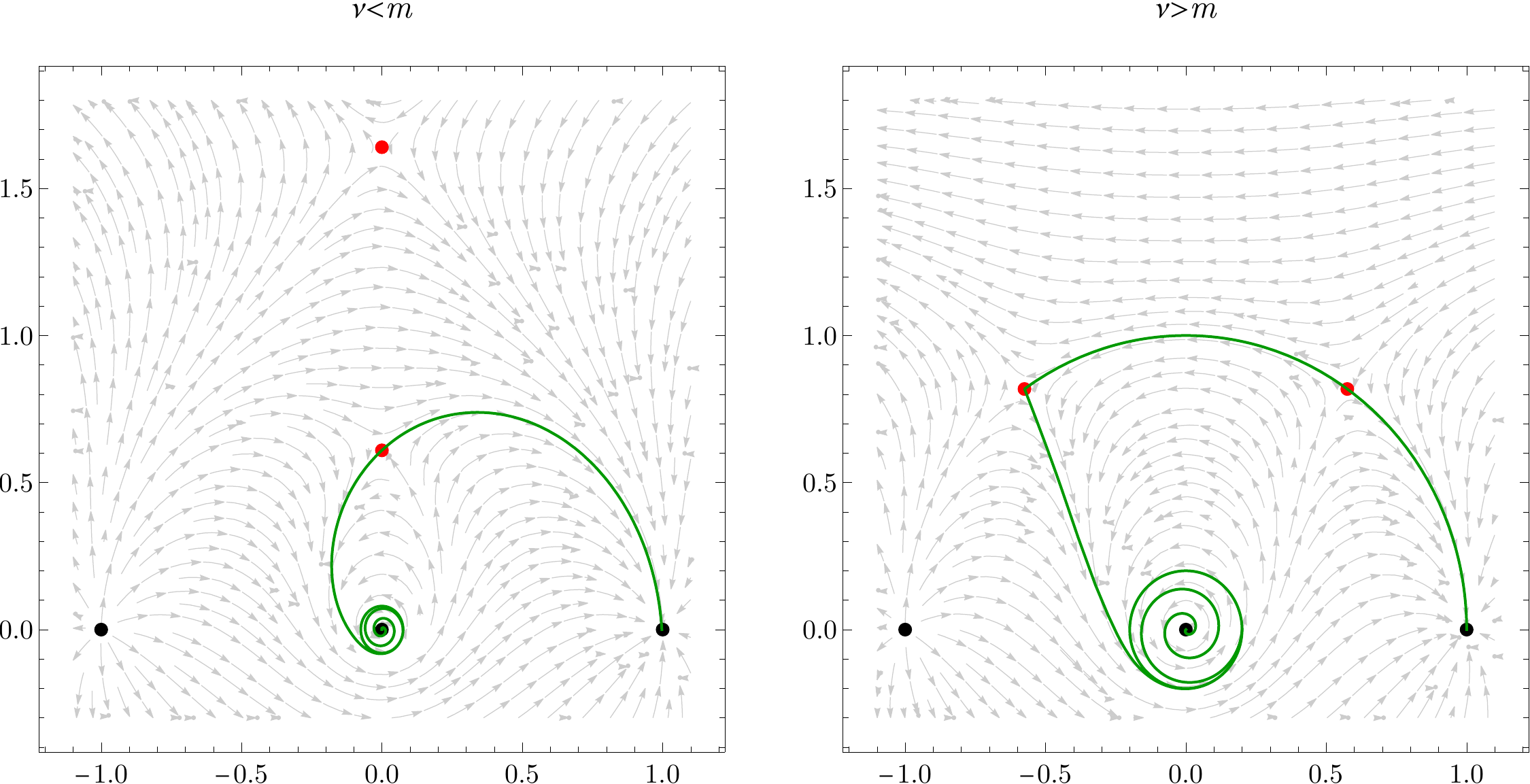}
\caption{Steepest-descent flows for the absolute value of the function $(x^-)^{\frac{i\nu}{2}} (\tfrac{1-x^-}{1+x^-})^{\frac{m}{2}}$, for $\nu<m$ and $\nu>m$ respectively. The two saddle points $p^-_{\bullet,-}$ and $p^-_{\bullet,+}$ are shown in red, and the branch points $-1,0,1$ are in black. The desired integration contour in (\ref{eq:examplexminusintegral}) runs between $0$ and $1$. In green, we show deformations of this contour to pass through saddle points. For $\nu<m$, the contour is homologous to a steepest descent contour (Lefschetz thimble) through $p^-_{\bullet,-}$ (the lower saddle point).  Near the origin, we choose the contour to spiral many times around 0, so that the function becomes arbitrarily small due to the factor $(x^-)^{i\nu/2}\propto e^{-\nu \arg x^-/2}$. (This is similar to contours described in \cite{Witten:2013pra}.) In the case $\nu>m$, the dominant saddle is the rightmost one. One of the flows from that saddle ends at the subdominant left saddle. From there, the contour approaches the origin and spirals around 0 in a similar way to the case $\nu<m$.
\label{fig:saddleflows}
}
\end{figure}

If $\De$ were large but not $\nu$, this integral would reduce to a Laplace transform similar to \eqref{I impact parameter}. Instead, when both $\De$ and $\nu$ are large (and $\De\sim m$), 
the integral gets pushed away from the $x,y\to 0$ limit and develops saddle points.
The most important $x$-dependent factors in (\ref{eq:sodminusoneswapperoo}) are
\begin{align}
\label{eq:mostimportantfactors}
\p{\frac {x^-}{|x|}}^{i\nu} \p{\frac{(e-x)^2}{(e+x)^2}}^\frac{m}{2}.
\end{align}
These factors lead to four saddle points for the $x$ integral at (in lightcone coordinates)
\begin{align}
\label{eq:bulkpointsaddle}
p_{\pm,\pm} &=  \p{-i\frac{m\pm \sqrt{m^2-\nu^2}}{\nu}, i \frac{m\pm\sqrt{m^2-\nu^2}}{\nu},\vec 0}.
\end{align}
Similarly, there are four saddle points $q_{\pm,\pm}$ for the $y$-integral, given by replacing $\nu\to-\nu$ in (\ref{eq:bulkpointsaddle}). To find a saddle-point approximation for our integral, we would like to express the integration contour as a linear combination (in an appropriate relative homology group) of steepest-descent flows from saddle points.\footnote{See \cite{Witten:2010cx} for a pedagogical introduction to these methods.} This is easiest to analyze in $d=2$ dimensions, where the function (\ref{eq:mostimportantfactors}) factorizes. For example, the $x^-$ integral takes the form
\be
\label{eq:examplexminusintegral}
\int_0^1 dx^- (x^-)^{\frac{i\nu}{2}} \p{\frac{1-x^-}{1+x^-}}^{\frac{m}{2}} \x (\cdots)\,.
\ee
By plotting steepest-descent flows, we find that for $\nu<m$, the saddle point $p^{-}_{\bullet,-}$ dominates the integral over $x^-$ (here $\bullet$ can be either $+$ or $-$ since this sign choice does not affect the $-$ component in (\ref{eq:bulkpointsaddle})), see Figure~\ref{fig:saddleflows}. This conclusion holds for the other variables as well: for $\nu<m$, the saddle points $p_{-,-}$ and $q_{-,-}$ dominate the $x$ and $y$-integrals, respectively.
These are the saddles which approach the origin as $\nu\to0$.
Plugging in their values and computing the Gaussian determinant in general $d$, we find
\begin{align}
\lim_{\De,\nu \gg 1}\frac{\widehat{\Pi}_{k,\nu}[G^s_{\De,J}]}{2\sin^2(\pi \frac{\tau-2\De_\f}{2})}
 &= \frac{(-1)^{J}\g_{2\De_\f+k-1}(\nu)^2}{b_{\De,J}m^{4 \De_\f-2}}\frac{4^{2 \De_\f +k -2}  \pi ^{d-2} \mathcal{P}_{J}\p{1-\frac{2 \nu ^2}{m ^2}}}{
 (m^2-\nu ^2)(m+\sqrt{m ^2-\nu ^2})^{2k-2} 
 }\nn\\
 &\qquad\qquad\qquad\qquad\qquad\qquad\qquad\qquad\qquad\qquad\qquad (\nu<m)\,,
 \label{eq:bulkpointresult}
\end{align}
where with some foresight we have re-expressed the value at the saddle point in terms of $\g_{2\De_\f+k-1}(\nu)^2$.

When $\nu =m$, the saddle points $p_{\pm,\pm}$ collide (and similarly for $q$),
and then separate again for $\nu>m$. After this collision, a single saddle still dominates each of the $x$- and $y$-integrals, but the values at the saddles change, so that the formula (\ref{eq:bulkpointresult}) is no longer valid. Thus, in order to apply (\ref{eq:bulkpointresult}), we must consider functionals with compact support in $\nu$ space: $\nu\in[0,\De_\mathrm{gap}]$.  As stated before, $\nu$ will play an analogous role to the flat-space transverse momentum $p$. Fortunately, in our flat space bounds, we chose to consider functionals with compact support $p\in[0,M]$. Thus, formula (\ref{eq:bulkpointresult}) is perfectly adapted to uplift flat-space bounds to the CFT case. It would be nice to better understand the relationship between flat-space and CFT functionals outside the ranges $\nu\in[0,\De_\mathrm{gap}]$ and $p\in [0,M]$.

To compute the action of $ C_{k,\nu}$ (instead of $\widehat \Pi_{k,\nu}$) in the limit $m,\nu\gg 1$, we need to account for any additional factors in the kernel defined by \eqref{eq:magicmoment}, evaluated on the saddle-point 
\begin{align}
r_*=\frac{\nu^2 }{(m+\sqrt{m ^2-\nu ^2})^2}\,,\qquad \eta_*=1\,. \label{eq:saddle}
\end{align}
The fact that the integral is dominated by $\eta\approx 1$ is physically intuitive because of the large  momentum $\nu$ in AdS space.  Fortunately, the kernel for $C_{k,\nu}$ simplifies dramatically at large $\nu$, and one finds
(see \eqref{eq:largenukernelexpression}) that the above simply gets multiplied by:
\begin{align}
\label{eq:extrafactors}
\frac{ a_{\De_\f}}{\g_{2\De_\f+k-1}(\nu)^2 } \frac{(1-r^4)}{((1+r^2)^2-4 r^2 \eta^2)^{\frac{k+1}{2}}}\,,
\end{align}
evaluated at the saddle-point  \eqref{eq:saddle}.
The result, after rewriting the prefactor in terms of $p_{\De,J}^\mathrm{MFT}$ and $n_J$ in the limit $\De\gg 1$, is
\begin{align}
\label{eq:bulkpointbknuagain}
\lim_{\De,\nu\gg 1} 
C_{k,\nu}[\De,J]
&= 
 \frac{2 m ^2-\nu ^2}{m ^2-\nu ^2}\frac{\cP_{J}\big(1-\frac{2 \nu ^2}{m ^2}\big)}{(m ^2 (m ^2-\nu ^2))^{k/2}}\,.
\end{align}
This gives (\ref{eq:bulkpointbknu}).
One can expand in $\nu/m$ to reproduce the terms of degree $\nu^{2n}$ in each polynomial $Q_{k,2n}(\nu)$, as recorded in \eqref{eq:QPolynomials}.

Equation~\eqref{eq:bulkpointbknuagain} holds for large $\nu$ and fixed spin $J$.
As visible from the structure of the $Q_{k,j}$ polynomials introduced in \eqref{eq:largemu}, 
finite-spin corrections involve the ratio $J^2/m^2$ but \emph{not} $J^2/\nu^2$:
the result remains valid as long as the AdS impact parameter
is small compared to the AdS radius ($\etaAdS-1\ll 1$).

\begin{figure}
\centering{\def\svgwidth{6cm}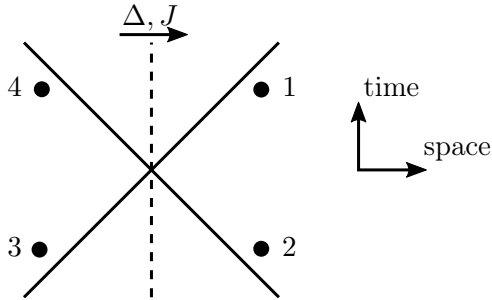}
\caption{Interpretation of the complex saddle point \eqref{eq:bulkpointsaddle} and spacelike scattering.
We start with the commutator squared $\<[3,4][1,2]\>\sim\dDisc_s[\cG]$
in the Rindler positive kinematics of Figure~\ref{fig:cylinder}(b).
The $s$-channel OPE inserts a basis of intermediate states along the dashed line,
which propagate horizontally along a spacelike direction.
This explains why a real exponent appears in the block \eqref{I impact parameter}.
Integrating the block against the oscillatory function $\mathcal{P}_{\frac{2-d}{2}+i\nu}(\eta)$ with $\nu$ large
creates the complex saddle point which effectively interchanges space and time,
thus realizing a bulk-focused scattering process $34\to 12$.
\label{fig:spacelike}}
\end{figure}

An interpretation of the saddle point is presented in Figure~\ref{fig:spacelike}: the
fact that \eqref{eq:bulkpointsaddle} is pure imaginary implies an interchange of (one direction of) space
and time, in effect a double Wick rotation.  This allows to reach the bulk scattering process $34\to 12$
from kinematics where $1$ and $3$ are spacelike!
Qualitatively similar complex coordinates were used
in \cite{Afkhami-Jeddi:2016ntf,Afkhami-Jeddi:2017rmx} to localize in the bulk down to
distances parametrically $\sim \frac{1}{\Delta_{\rm gap}}$,
but there is a crucial difference: since we start with an integral over real spacetime, we know \emph{exactly}
when the saddle point is valid, namely for $|\nu|<\Delta_{\rm gap}$.
This will enable us in the next section to find sharp bounds involving $\Delta_{\rm gap}$. It is remarkable that one can access moments of the $S$-matrix while remaining in spacelike kinematics. We refer to this phenomenon as {\it spacelike scattering}. We refer the reader to \cite{Paulos:2016fap,Hijano:2019qmi,Komatsu:2020sag} for interesting recent work on recovering the flat-space S-matrix from CFT correlators.

Interestingly, \eqref{eq:bulkpointsaddle}  coincides precisely with the saddle point found in the flat-space limit
of light-light-heavy three-point functions \cite{Li:2021snj}. 
This makes us confident that the dictionary between CFT and
flat-space will straightforwardly extend to spinning operators.

\subsection{The $C_{k,\nu}$ sum rule in Mellin space: numerical validations}

A remarkable feature of dispersive sum rules is that they are natural in all spaces:
as emphasized in~\cite{Caron-Huot:2020adz},
the position space dispersion relation is the same as the natural Mellin space one.
As was further explained in \cite{Meltzer:2021bmb}, it is also a standard dispersion relation in momentum space. It is thus easy to convert between spaces.
Here we discuss the $C_{k,\nu}$ sum rule from the Mellin perspective;
we show in appendix \ref{app:superconvergence} that it is also simply related to Lorentzian inversion sum rules.

The Mellin representation for identical operators takes the form
\be
\cG(u,v) = \iint\!\!\frac{d\mS\,d\mT}{(4\pi i)^2}\,u^{\tfrac{\mS}{2}-\Df}v^{\tfrac{\mT}{2}-\Df}
\Gamma\!\left(\Df-\tfrac{\mS}{2}\right)^2\Gamma\!\left(\Df-\tfrac{\mT}{2}\right)^2\Gamma\!\left(\Df-\tfrac{\mU}{2}\right)^2 M(\mS,\mT)\,.
\label{eq:mellinRep2}
\ee
Here $\mS$, $\mT$, $\mU$ are the Mellin variables, which are constrained to satisfy $\mS+\mT+\mU=4\Df$. (We use non-italicized letters  $\mS,\mT,\mU$ to distinguish Mellin variables  from flat-space Mandelstam variables $s,t,u$.)
The Mellin amplitude is a meromorphic function with poles encoding the spectrum, and
bounded in the Regge limit ($|M/\mS^2|\to 0$ as $\mS\to\infty$ with fixed $\mU$).
Dispersive sum rules in Mellin space are defined as\footnote{The normalization used here differs from that of
$\widehat{B}_{2,\mT}$ in ref.~\cite{Caron-Huot:2020adz} by a factor $(\tfrac{\mT}{2}-\Df)$. \simon{check this.}}
\be
\widehat{B}_{k,\mT}[M(\mS,\mT)]
=(-1)^{\frac{k}{2}} \oint_{|\mS'|=\infty} \frac{d\mS'}{2\pi i} \frac{1}{2\Df-\mS'} \frac{M(\mS',\mT'=\mT+2\Df-\mS')}{
(\Df-\tfrac{\mS'}{2})_{\frac{k}{2}}(\Df-\tfrac{\mT'}{2})_{\frac{k}{2}}}\,,
\label{eq:BMellinDefinition}
\ee
which converges (to zero) for $k\geq 2$.
The denominators play the role of subtractions and improve large-$\mS$ convergence;
as discussed in~\cite{Penedones:2019tng,Caron-Huot:2020adz}, it is physically
natural to put the subtraction points at double-twist locations, as this simply
cancels naive zeros caused by the gamma-functions in the representation \eqref{eq:mellinRep2}.
The $B_{k,v}$ sum rules we used above are Mellin conjugate to the $\widehat{B}_{k,\mT}$ (see equation~(4.85) of \cite{Caron-Huot:2020adz}):
\be
 B_{k,v} = \Gamma(\tfrac{k}{2})\int \frac{d\mT}{4\pi i} v^{\tfrac{\mT}{2}-\Df}
\Gamma\big(\tfrac{\mT}{2}\big)^2
\Gamma\big(\Df-\tfrac{\mT}{2}\big)\Gamma\big(\Df-\tfrac{\mT}{2}+\tfrac{k}{2}\big)
\widehat{B}_{k,\mT}\,. \label{Bkv from Bkt}
\ee
On the other hand, the action of $\widehat{B}_{k,t}$ on a heavy block can be calculated
directly in Mellin space.  The Mellin amplitude has poles
$\sim\frac{p_{\Delta,J}\mathcal{Q}^n_{\Delta,J}(\mT)}{\mS-\Delta+J-2n}$ for each descendant $n$ of an
exchanged conformal primary, where $\mathcal{Q}$ is a Mack polynomial and $n\geq 0$ is an integer.
Therefore, we can write the heavy contribution to the sum rule as
\be
 \widehat{B}_{k,\mT}[G]\Big|_{\rm heavy} = \sum_{\tau > \Delta_{\rm gap}} p_{\Delta,J} 
 \widehat{B}_{k,\mT}[G^s_{\Delta,J}]\,,
\ee
where
\be
\widehat{B}_{k,\mT}[G^s_{\Delta,J}] = (-1)^{\frac{k}{2}}\sum\limits_{n=0}^\infty
\frac{\mathcal{Q}^n_{\Delta,J}(\mT+2\Df-\tau-2n)}{
(\tfrac{2\Df-\tau}{2}-n)_{\frac{k}{2}}
(\tfrac{\tau-\mT}{2}+n)_{\frac{k}{2}}} \left[\frac{1}{2\Df-\tau-2n}+ \frac{1}{\mT-\tau-2n}\right]\,.
\label{Bkt action}
\ee
In the square bracket we added the $\mS$ and $\mT$ poles, assuming symmetry of the correlator,
as before. Eqs.~\eqref{Bkv from Bkt} and \eqref{Bkt action} together
provide a numerically stable formula to evaluate the action of $B_{k,v}$ on conformal blocks \cite{Caron-Huot:2020adz}. We will now deduce a similar Mellin-space representation directly for the $C_{k,\nu}$
functional.

\begin{figure}
\centering{
\includegraphics[width=0.75\textwidth]{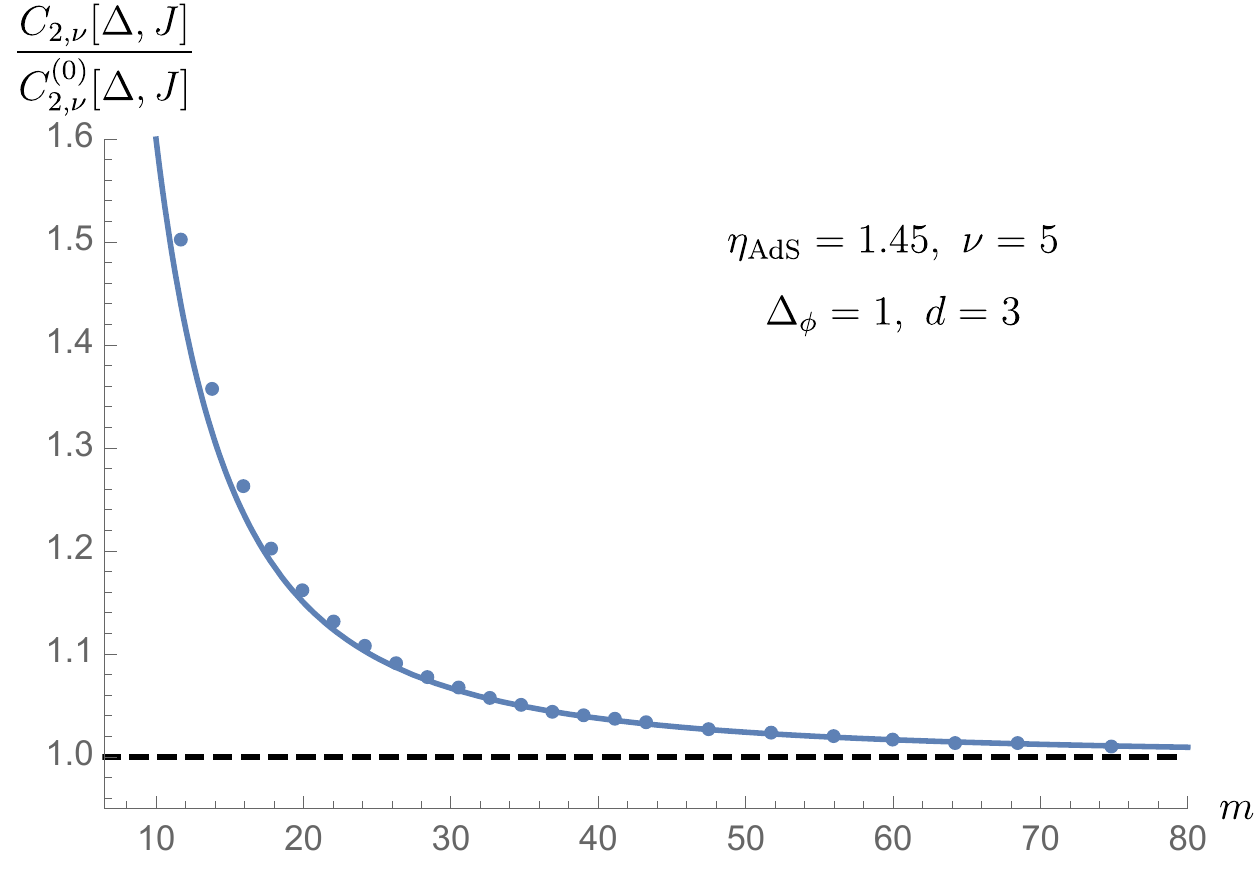}
\caption{Convergence in the Regge limit: $m\to\infty$ with $\nu$ and $\etaAdS$ fixed.
The dots show numerical evaluation of the Mellin representation \eqref{eq:CfromBMellin}, divided
by the leading order formula $C^{(0)}_{2,\nu}[\Delta,J]=\frac{2}{m^4}\mathcal{P}_{\frac{2-d}{2}+i\nu}(\etaAdS)$, so as to asymptote to 1.  The solid line shows the leading $1/m^2$ correction from \eqref{regge subleading}, which is seen to track the exact result
even at moderate $m\approx 10$.
}
\label{fig:convergence_regge}}
\end{figure}

\begin{figure}
\centering{
\includegraphics[width=0.8\textwidth]{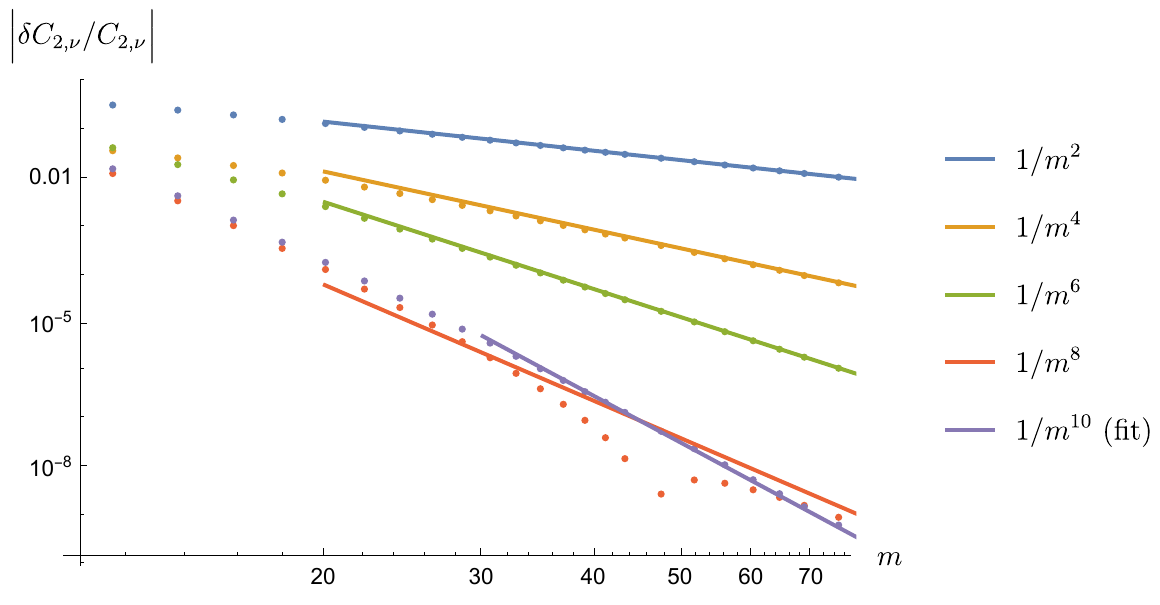}
\caption{Residual errors in the Regge limit, for the functional
in Figure~\ref{fig:convergence_regge} after subtracting various partial sums of the $1/m^2$ series from high-precision Mellin-space numerics.
Including the $1/m^8$ term yields $10^{-9}$ relative accuracy at $m=70$, nontrivially validating both the analytic and numerical formulas.
Although the coefficient of $1/m^8$ is anomalously small in this particular example,
the residual is well fitted by a $1/m^{10}$ power law.
}
\label{fig:convergence_regge_residual}}
\end{figure}

We compute the two transforms in \eqref{eq:magicmoment} one after the other.
First, the transform from $v$ to the angle $\eta$ gives simply a power of $\eta$:
\be
\frac{(-1)^{\frac{k}{2}}\pi \Gamma(k)}{2^{3k-2}\Gamma\big(\tfrac{k}{2}\big)^2}
\int_{\eta^2}^\oo \frac{dv\ v^{\tfrac{\mT}{2}-\Df}} {\big[(v-\eta^2)^{\frac{k+1}{2}}\big]_+} 
 = \frac{\pi^{\frac{3}{2}}\Gamma\big(\Df+\tfrac{k-\mT-1}{2}\big)}{2^{2k-1}
 \Gamma\big(\tfrac{k}{2}\big)
 \Gamma\big(\Df-\tfrac{\mT}{2}\big)} \eta^{\mT-2\Df-k+1} \,.
\ee
Second, the $\eta$-integral then reduces to a moment of the Gegenbauer function:
\be
  \int_1^\oo [d\eta] \cP_{J}(\eta)\eta^{-X} = 
  \frac{2^{X+d-4} \Gamma(\tfrac{d-1}{2})}{\sqrt{\pi}\Gamma(X)}
  \Gamma(\tfrac{X+J}{2}) \Gamma(\tfrac{X+2-d-J}{2})
\ee
with $X=2\Df+k-1-\mT$.
Combining these with \eqref{eq:magicmoment} and \eqref{Bkv from Bkt} we find:
\be
 C_{k,\nu} = \frac{2^{d-2-2k}\pi^{\frac{3}{2}}\Gamma(\tfrac{d-1}{2})
 a_{\De_\f}}{\g_{2\De_\f+k-1}(\nu)^2}
 \int \frac{d\mT}{4\pi i}
 \Gamma\big(\tfrac{\mT}{2}\big)^2 \gamma_{2\Df+k-1-\mT}(\nu)
\widehat{B}_{k,\mT}\,.
\label{eq:CfromBMellin}
\ee
We used \eqref{eq:CfromBMellin} with \eqref{Bkt action} (together with standard expressions for the Mack polynomials $\mathcal{Q}$ \cite{Costa:2012cb})
to evaluate numerically the action of $C_{k,\nu}$ functionals.

\begin{figure}
\centering{
\includegraphics[width=0.9\textwidth]{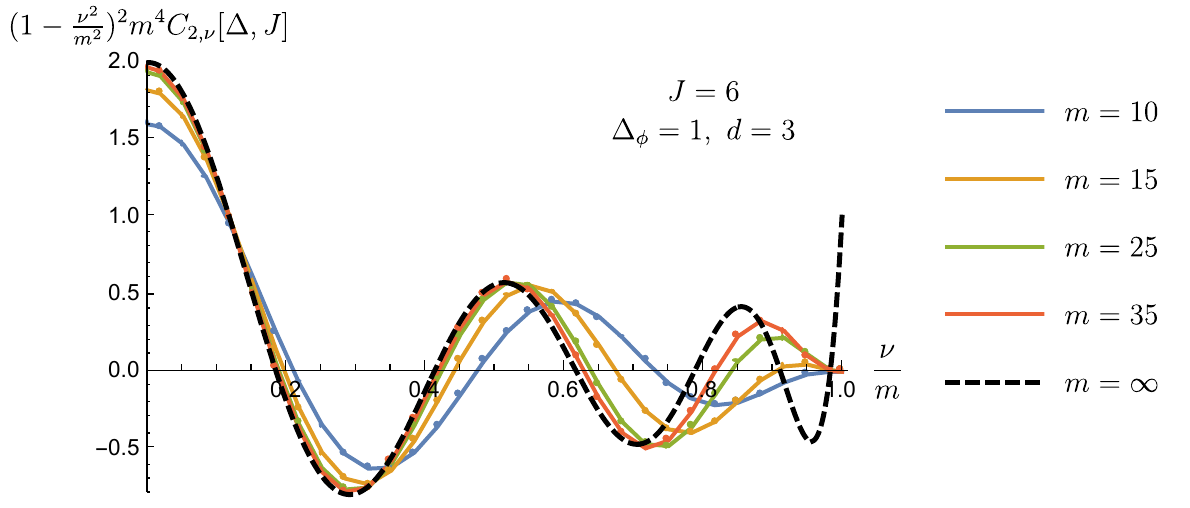}
\caption{
Convergence of functional action in
the bulk-point limit: $m\to\infty$ with $J$ and $\nu/m$ fixed.
The dots, connected to guide the eye,
show the exact action computed using the Mellin representation; the dashed line shows the saddle-point result from \eqref{eq:bulkpointbknuagain}.
The slower convergence near $\frac{\nu}{m}=1$ is compatible with the observed collision of saddle points. 
}
\label{fig:convergence_bulkpoint}}
\end{figure}

While the Mellin amplitude is often regarded (with reason) as a CFT \emph{analog} to the S-matrix, it is important to stress that sum rules with prescribed Mellin-$\mT$ are \emph{not}
directly the ones with simple flat-space limit. Rather, the sum rules with definite
transverse momentum $\nu$ are the $\mT$- integrals in \eqref{eq:CfromBMellin}.

Comparison between numerical and analytical results in the Regge and bulk point limits are reported in Figures~\ref{fig:convergence_regge}, \ref{fig:convergence_regge_residual}, and \ref{fig:convergence_bulkpoint}. 
For heavy operators $m\gg 1$, we find that the Mellin numerics require including many descendants $n$, however the dependence on $n$ is fairly smooth and can be accurately modelled using polynomial interpolation. For example, to obtain the $10^{-9}$-accurate data points in Figure~\ref{fig:convergence_regge_residual}, we evaluated the $\mT$ integral to high accuracy at
2000 values of $n$ between 0 and $10^6$; reliable results to lower accuracy require much fewer $n$-evaluations.
At large $\nu$,
the Mellin integrand becomes highly oscillatory but remains numerically stable since it is only one-dimensional.

These comparisons confirm that the main results of this section: the $1/m^2$ series
in \eqref{regge subleading} and bulk-point saddle point in \eqref{eq:bulkpointbknuagain},
indeed accurately approximate the action on heavy blocks of a well-defined functional $C_{k,\nu}$.


\section{Light contributions to holographic sum rules}\label{sec:lightAction}

\subsection{Light contributions and Regge moments}
\label{ssec:lightResidue}
\newcommand\cK{\mathcal{K}}
\newcommand\arcbelowright{\rotatebox[origin=c]{180}{\reflectbox{$\curvearrowright$}}}
\newcommand\arcaboveright{\rotatebox[origin=c]{180}{\reflectbox{$\arcbelowright$}}}
\newcommand\cA{\mathcal{A}}
\newcommand\cM{\mathcal{M}}

The heavy side of a holographic sum rule
\begin{align}
\left.\w\right|_\mathrm{heavy} &= -\left.\w\right|_\mathrm{light}
\end{align}
is determined by the Regge moments of the functional $\w$. A surprising and useful simplification is that $\left.\w\right|_{\mathrm{light}}$ is also entirely determined by the Regge moments of $\w$!  This allows us to discuss heavy and light contributions to the sum rule (\ref{eq:heavylightsplit}) in the same language.

The contribution of light states to a holographic sum rule is
\begin{align}
\left.\w\right|_\mathrm{light} = \sum_{\tau\leq \De_\mathrm{gap}} p_{\De,J} \w[G^s_{\De,J}] + \textrm{subtractions}\,,
\end{align}
where ``subtractions'' denotes possible additional terms such as $(-1)^{\frac k 2 - 1}$ in (\ref{Bkv sum rule}) arising from subtracting known functions to ensure convergence of $\omega$.
We assume that the light OPE data agrees with a tree-level AdS effective field theory (EFT) with a derivative expansion.
Naively, to compute $\w|_\mathrm{light}$, we must sum the contributions of light single-trace operators, as well as any double-trace trajectories for which $\w$ is nonzero. In particular, this requires computing OPE data coming from exchanges and contact diagrams.

However, there is an efficient shortcut to computing $\w|_\mathrm{light}$ that bypasses these steps and gives a result that depends only on the Regge moments of $\w$. Suppose $\omega$ is a dispersive functional that vanishes on all double-trace families with $\tau> 2\De_\f+2p$ for some $p\geq 0$. We assume that the OPE data of double-traces with $\tau\leq 2\De_\f+2p$ is well-approximated by a bulk EFT with a truncated derivative expansion.  We denote the four-point function computed using this EFT by $\cG^\mathrm{EFT}$. The important properties of $\cG^\mathrm{EFT}$ are that it has approximately the same low-lying double-trace OPE data as $\cG$, and that it is crossing-symmetric. Generically, $\cG^\mathrm{EFT}$ will grow with spin $n$ for some $n>1$ in the Regge limit, and hence cannot itself be a physical correlator.

Suppose that $\w$ has the Regge moment expansion
\begin{align}
\w &\sim \Pi_{k,\eta} + O(\Pi_{n+1})\,.
\end{align}
That is, $\w$ is a pure spin-$k$ Regge moment, up to moments with Regge spin greater than $n$ (the Regge spin of $\cG^\mathrm{EFT}$).
Then for a tree-level bulk EFT, $\w|_\mathrm{light}$ can be computed from the following contour integral:
\ba
\left.-\w\right|_\mathrm{light} &= \frac{1}{4}\oint dr\, r^{k-2} \cG^\mathrm{EFT}_+(r,\eta) = \frac{2\pi i}{4} \left.\GG_+^\mathrm{EFT}(r,\eta)\right|_{r^{1-k}} \qquad \textrm{(tree level)}\,.
\label{eq:eftformula}
\ea
The $r$-contour encircles the $u$-channel Regge limit $r=0$. It simply picks out the $r^{1-k}$ term in the Laurent expansion of the integrand around $r=0$. The $+$ subscript on $\cG^\mathrm{EFT}_+$ indicates that we should approach the Regge limit by taking $w,\bar w$ in the upper half plane (and analytically continuing from there as we go around the $r$-contour).  In order for (\ref{eq:eftformula}) to make sense, $\cG^\mathrm{EFT}_+$ must be single-valued near $r=0$, with a Laurent expansion containing only integer powers of $r$. This is indeed the case for tree-level bulk EFTs. More generally, when loops are included, $\omega|_\mathrm{light}$ can still be computed from its Regge moments via a contour integral close to the Regge limit; however the integral doesn't necessarily localize to a residue.

The intuitive reason for (\ref{eq:eftformula}) is as follows. The failure of $\w|_\mathrm{light}$ to vanish comes from the fact that $\cG^\mathrm{EFT}$ grows in the Regge limit with spin $n>k$. Thus, the value of $\w|_\mathrm{light}$ should come from an ``arc at infinity'' that is sensitive to the growing terms in the Regge limit. We make this intuition precise and prove (\ref{eq:eftformula}) in appendix~\ref{app:omegalight}.

More generally, suppose that $\w$ is a physical functional with Regge expansion
\begin{align}
\w &\sim \sum_{k=2}^\oo\int_1^\oo d\eta\, a_k(\eta) \Pi_{k,\eta}\,.
\end{align}
By linearity, we have
\begin{align}
\label{eq:moregeneralformula}
\left.-\w\right|_\mathrm{light} &= \frac{1}{4}\sum_{k=2}^n \int_1^\oo d\eta \oint dr\, r^{k-2} a_k(\eta) \cG_+^\mathrm{EFT}(r,\eta)\,.
\end{align}
The sum truncates at $k=n$, where $n$ is the Regge spin of $\cG^\mathrm{EFT}$, since the highest power of $1/r$ appearing in the Laurent expansion of $\cG^\mathrm{EFT}$ is $r^{1-n}$.

As a sanity check on (\ref{eq:moregeneralformula}), note that if $n<2$, then $\w|_\mathrm{light}$ vanishes. This is a general consequence of the fact that $\w$ is $s\leftrightarrow t$ antisymmetric and has at least spin-2 Regge decay, as explained in \cite{Caron-Huot:2020adz}. As an example, scalar exchange diagrams and $\phi^4$-type contact interactions do not contribute to $\w|_\mathrm{light}$.

Since $\w|_\mathrm{light}$ is entirely determined by the expansion of $\w$ in Regge moments, it is convenient to introduce the following notation
\begin{align}
\label{eq:abusenotation}
&\left.\Pi_{k,\eta}\right|_\mathrm{light} \equiv \left.\w\right|_\mathrm{light},\quad
\textrm{where $\w$ is any physical functional such that\ }
\w \sim \Pi_{k,\eta} + O(\Pi_{n+1})\,.
\end{align}
Here, $n$ is the Regge spin of $\cG^\mathrm{EFT}$. Note that $\Pi_{k,\eta}|_\mathrm{light}$ is {\it not\/} equal to a sum over light operators of OPE coefficients times $\Pi_{k,\eta}[G^s_{\De,J}]$ (since the expansion in Regge moments does not necessarily converge on light operators). In this sense, (\ref{eq:abusenotation}) is an abuse of notation --- we hope it does not cause confusion.

The fact that $\w|_\mathrm{light}$ is controlled by an expansion in the Regge limit is true in any space --- in particular it is true in Mellin space. In the next section we use this fact to efficiently compute the contribution of contact diagrams. We return to the position-space formula (\ref{eq:moregeneralformula}) in section~\ref{eq:lightpositionspace} to aid in interpreting the Mellin results.

\subsection{Light contributions from Mellin space}\label{ssec:mellin}

Our main goal is to prove bounds on couplings of effective field theories in AdS. In order to do that, we must first choose a parametrization of the EFT couplings. We will parametrize them using their Mellin representation. We will then compute the light contributions $C_{k,\nu}|_\mathrm{light}$ for specific EFT correlators. The Mellin representation for identical operators takes the form \eqref{eq:mellinRep2}. A basis for contact interactions consists of the following symmetric polynomials
\be
M_{a,b}(\mS,\mT) = c_{2a+3b}(\mS^2+\mT^2+\mU^2)^a(\mS\mT\mU)^b\,,
\ee
where
\be
c_{n} = \frac{2^{n-3}\Gamma(2\Df-\tfrac{d}{2}+n)}{\pi^{\frac{d}{2}}\Gamma(\Df)^2\Gamma(\Df-\tfrac{d-2}{2})^2}\,.
\label{eq:cNormalization}
\ee
The normalization is chosen so that the AdS bulk interaction which gives rise to $M_{a,b}(\mS,\mT)$ would give rise to the following S-matrix when used in flat space
\be
\cM_\mathrm{flat}(s,t) = \cM_{a,b}(s,t) \equiv (s^2+t^2+u^2)^a(stu)^b + \text{subleading}\,,
\ee
where the corrections are subleading in the high energy limit $s,t,u\rightarrow\infty$ with ratios fixed. Note that following \cite{Penedones:2010ue}, the normalization \eqref{eq:cNormalization} is equivalent to the following relationship between $M(\mS,\mT)$ and $\cM(s,t)$
\be
\label{eq:joaotransform}
M(\mS,\mT) = \frac{1}{8\pi^{\frac{d}{2}}\Gamma(\Df)^2\Gamma(\Df-\tfrac{d-2}{2})^2}
\int\limits_{0}^{\infty}d\beta \beta^{2\Df-\frac{d}{2}-1}e^{-\beta}\cM(2\beta \mS,2\beta \mT)\,.
\ee

Using the labelling scheme introduced in flat space in subsection \ref{ssec:SMatrixReview},
let us parametrize the Mellin amplitude of the EFT as follows
\ba
M_{\text{EFT}}&= M_{\text{non-growing}}+8\pi G M_{\text{gravity}}+\\
&\quad+g_{2}\,M_{1,0}+g_3\,M_{0,1}+g_4 M_{2,0}+g_{5} M_{1,1} + g_{6} M_{3,0}+g'_{6}M_{0,2}+g_{7} M_{2,1}+\ldots\,.
\label{eq:MellinEFT}
\ea
Here $M_{\text{non-growing}}$ consists of terms which do not grow in the Regge limit. In the present context, it receives contributions only from the exchange of light scalars and from the scalar contact interaction $M_{0,0}$. $M_{\text{gravity}}$ is the sum of graviton exchanges in the three channels.\footnote{An explicit formula for the graviton exchange in Mellin space was found in \cite{Costa:2014kfa}.} The second line is an infinite tower of higher-derivative contact interaction $g_{2a+3b}M_{a,b}$, where $g_n$ denotes the strength of an interaction with $2n$ derivatives.

We would now like to compute the contribution of each term in the EFT expansion \eqref{eq:MellinEFT} to
$C_{k,\nu}|_\mathrm{light}$. This can be done by combining the Mellin representation of $C_{k,\nu}$ in \eqref{eq:CfromBMellin} with the definition of $\widehat{B}_{k,\mT}$ in \eqref{eq:BMellinDefinition}. We start by calculating $\widehat{B}_{k,\mT}[M_{a,b}]$ directly as the residue at $\mS' = \infty$ in \eqref{eq:BMellinDefinition}. This gives $\widehat{B}_{k,\mT}[M_{a,b}]$ as a polynomial in $\mT$. $C_{k,\nu}[M_{a,b}]$ can then be found by doing the $\mT$ integral in \eqref{eq:CfromBMellin} using the formula
\be
\int\limits_{-i\infty}^{+i\infty}\!\!\frac{dz}{2\pi i}\Gamma(z-a)\Gamma(z-b)\Gamma(c-z)\Gamma(d-z) = 
\frac{\Gamma (c-a) \Gamma (d-a) \Gamma (c-b) \Gamma (d-b)}{\Gamma (c+d-a-b)}
\,,
\ee
valid for $\mathrm{Re}(a,b)<\mathrm{Re}(c,d)$, with the contour passing to the right of the poles at $z=a,b$ and to the left of those at $z=c,d$.

The simplest nonvanishing case is $C_{2,\nu}[M_{1,0}]$, for which the above procedure gives the simple answer
\be
- C_{2,\nu}|_{g_2} = 2\,,
\label{eq:C2g2}
\ee
which exactly agrees with the $2g_2$ term in the flat-space sum rule ${\cal C}_{2,u}$ in \eqref{S light}.
We can repeat the exercise for the $stu$ and $(s^2+t^2+u^2)^2$ couplings, finding
\ba
- C_{2,\nu}|_{g_3} &= \nu ^2-\left(2 \Df -\tfrac{d-4}{2}\right) \left(2 \Df+\tfrac{d-4}{2} \right)\\
- C_{2,\nu}|_{g_4} &= 8 \nu ^4 +4 \left(d^2-4 d \Df -8 d+8 \Df +16\right)\nu ^2-\left(2 \Df-\tfrac{d-4}{2}\right)\times\\
&\qquad\times\left[16 (3 d-16) \Df ^2-4 d (d-4) \Df +(d-4)^3-192 \Df ^3\right]\\
- C_{4,\nu}|_{g_4} &= 4\,.
\label{eq:C2C4Light}
\ea
These results agree again with the flat-space formulas for ${\cal C}_{2,u}$ at the leading order at large $\nu$ as expected. Let us record several more general results for the leading moments ($k=2a+2b$)
\ba
- C_{2a,\nu}[M_{a,0}] &= 2^a\\
- C_{2a+2,\nu}[M_{a,1}] &=2^a \left[\nu ^2-\left(2 \Df -\tfrac{d-4a-4}{2}\right) \left(2 \Df+\tfrac{d-4a-4}{2} \right)\right]\,.
\ea
It is possible to check that in general $ C_{k,\nu}[M_{a,b}]$ is a polynomial in $\nu^2$ of degree at most $2a+3b-k$, and that the leading behavior at large $\nu$ is
\be
- C_{k,\nu}[M_{a,b}] = \frac{2^a\left(\tfrac{k-2b+2}{2}\right)_{\frac{2a+2b-k}{2}}}{\left(\frac{2a+2b-k}{2}\right)!}
\nu^{2(2a+3b-k)} + O(\nu^{2(2a+3b-k-1)})\,.
\label{eq:BknuContactsLeading}
\ee
In particular, the $\nu^{2(2a+3b-k)}$ term is absent unless $b\leq k/2$. Note that he leading term \eqref{eq:BknuContactsLeading} comes from the maximal power of $\mT$ in $\widehat{B}_{k,\mT}[M_{a,b}]$, which allows us to find the closed form. Note that $C_{k,\nu}$ receives contributions only from anomalous dimension on the first $k/2$ double-trace trajectories, as well as anomalous OPEs on the leading trajectory. It follows that $C_{k,\nu}$ identically vanishes on all contacts which have a zero at $\mS=2\Df,2\Df+2,\ldots,2\Df+k-2$ and a double zero at $\mS = 2\Df$.

It remains to evaluate the contribution of the graviton exchange $M_{\text{gravity}}$ to the light moments. The graviton exchange grows as spin two in the Regge limit and therefore contributes only to the $k=2$ moments. Using the explicit form of $M_{\text{gravity}}$ given in \cite{Costa:2014kfa}, we find
\be
M_{\text{gravity}}(\mS,\mU) \sim -8\pi G\sum\limits_{m=0}^{\infty}
\frac{\Df ^2 \sin ^2\!\left[\tfrac{\pi}{2}(d-2\Df)\right] \Gamma \big(\tfrac{d-2\Df}{2}+m\big)^2}{4 \pi ^{\frac{d}{2}+2} m! \Gamma \left(\frac{d}{2}+m+1\right)}\frac{\mS^2}{\mU-(d-2+2 m)}\,,
\label{eq:MGravityRegge}
\ee
as $\mS\rightarrow\infty$ at fixed $\mU$. Here $G$ is measured in units of the AdS radius. The contribution to $C_{2,\nu}$ of each term with fixed $m$ in \eqref{eq:MGravityRegge} can be evaluated in terms of a hypergeometric function ${}_3F_2$. The resulting sum over $m$ takes a remarkably simple form (as we checked numerically\footnote{Alternatively, one can derive \ref{eq:BknuGravity} analytically using (\ref{eq:eftformula}) and computing the Regge limit of the graviton exchange diagram using its $u$-channel partial wave expansion continued to the Regge limit.})
\be
-\left. C_{2,\nu}\right|_\mathrm{\text{gravity}} = \frac{8\pi G}{\nu^2 + (d/2)^2}\,.
\label{eq:BknuGravity}
\ee
This also agrees precisely with the flat-space formula \eqref{S light} at large $\nu$:
\be
\left. -{\cal C}_{2,u}\right|_\mathrm{\text{gravity}} = \frac{8\pi G}{-u}
\ee
under $u\to -\nu^2$.

\subsection{Light contributions from position space}
\label{eq:lightpositionspace}

Remarkably, both (\ref{eq:BknuContactsLeading}) and (\ref{eq:BknuGravity}) are consistent with a simple formula in terms of the flat-space amplitude
\begin{align}
-\left. C_{k,\nu}\right|_{\mathrm{light}} &= 
\mathrm{Res}_{s=0} \left[\frac{2s-\nu^2}{s(s-\nu^2)}\frac{\cM_\mathrm{flat}(s,-\nu^2)}{[s(s-\nu^2)]^{k/2}}\right]\x (1+O(1/\nu^2))\,.
\label{eq:BknuFlatSpaceLimit}
\end{align}
The physical reason is clear: either in Mellin space or position space, $C_{k,\nu}|_\mathrm{light}$ comes from an expansion around the Regge limit. However, the nontrivial transform (\ref{eq:joaotransform}) relating the Mellin amplitude and $\mathcal{M}_\mathrm{flat}$ somewhat obscures the simplicity of (\ref{eq:BknuFlatSpaceLimit}). Next, we describe a direct position-space computation of the large-$\nu$ limit of contact diagram contributions, showing that (\ref{eq:BknuFlatSpaceLimit}) is controlled by a saddle point analogous to the one we encountered in the bulk point limit in section~\ref{sec:heavymomentbulkpoint}.

We begin with a general contact diagram
\begin{align}
\cG(X_i) &= \int_\mathrm{AdS} d^{d+1}Q \left.\prod_{i=1}^4 i\cD \frac{\cC_{\De_\f}}{(-2 X_i\.Q_i + i\e)^{\De_\f}}\right|_{Q_i=Q},
\end{align}
where we have written the integral in the embedding space formalism, where the metric is
\begin{align}
X\.X &= -X^+ X^- + X_\mu X^\mu\,.
\end{align}
The $X_i$, satisfying $X_i^2=0$ are boundary points, and $Q$ satisfying $Q^2=-1$ is a bulk point. $\cD$ is a differential operator in the $Q_i$, coming from the bulk Lagrangian. The constant $\cC_{\De_\f}$ is given by
\begin{align}
\cC_{\De_\f} &= \frac{\Gamma (\Delta_\phi )}{2 \pi ^{d/2} \Gamma(\De_\f-\frac{d-2}{2})}\,.
\end{align}
We take the boundary points to be in the causal configuration $2<1$ and $3<4$, with the remaining pairs of points spacelike separated.

The $\widehat{\Pi}_{k,\nu}$ moment of the diagram is a conformally-invariant integral
\begin{align}
\left.-\widehat{\Pi}_{k,\nu}\right|_{\mathrm{light}} &=
-2^{2d-2}\vol \SO(d-2) \oint_{\substack{2<1 \\ 3<4}} \frac{d^d x_1 d^d x_2 d^d x_3 d^d x_4}{\vol \tl \SO(d,2)} 
\frac{\Omega(u',v') \cG(X_i)}{(-2X_1\.X_3)^{\tl \De_\f}(-2 X_2\.X_4)^{\tl \De_\f}}\,,
\label{eq:conformalcontourintegral}
\end{align}
where $ \Omega(u',v')$ is the kernel defining the functional. The symbol $\oint$ indicates that we must compute the light moment using a contour integral in cross-ratio space
\begin{align}
\label{eq:contourforr}
\int dr\,r^{k-2}\, \mathrm{dDisc} \cG &\to - \oint dr\, r^{k-2}\,\cG\,.
\end{align}
We explain how to implement this prescription shortly.

To compute (\ref{eq:conformalcontourintegral}), let us first gauge-fix the conformal symmetry in the same way as in our computation of heavy moments in the bulk-point limit in section~\ref{sec:heavymomentbulkpoint}. Because of the causal configuration, we cannot place all of the points in the same Minkowski patch. The correct embedding coordinates for our problem are
\begin{align}
\label{eq:choiceforboundarypoints}
X_1 &= -(1,y^2,-y) \nn\\
X_2 &= (x^2,1,x) \nn\\
X_3 &= (1,y^2,y) \nn\\
X_4 &= -(x^2,1,-x)\,.
\end{align}
After choosing the boundary positions (\ref{eq:choiceforboundarypoints}), the remaining conformal symmetries are a Lorentz symmetry $\SO(d-1,1)$ and dilatation symmetry. The Lorentz group $\SO(d-1,1)$ acts as the group of isometries on a transverse hyperboloid $H_{d-1}$ in AdS. We can use it to set the transverse position of $Q$ to the center of $H_{d-1}$. In embedding coordinates, this corresponds to
\begin{align}
Q &= (Q^+,Q^-,Q^\mu) = (\xi^+,\xi^-,e\sqrt{1-\xi^+\xi^-})=Q_0\,,
\end{align}
where $e$ is a unit vector in the time direction. The $\SO(d-1)$ of rotations around $e$ remains unfixed.
Meanwhile, dilatation symmetry acts on the boundary points by $(x,y)\to (\l x,\l^{-1} y)$.
We fix it by inserting 
\begin{align}
\frac 1 2\de\p{\frac{|x|}{|y|}-1}.
\end{align}
Finally, to implement the contour prescription (\ref{eq:contourforr}), we insert a $\de$-function $\de(r-|x||y|)$ and perform the contour integral over $r$ at the end. Overall, we find
\begin{align}
\label{eq:gaugefixedlightintegral}
\left.-\widehat{\Pi}_{k,\nu}\right|_{\mathrm{light}}&= \oint dr \int \frac{d^d x\, d^d y}{\vol S^{d-2}}\frac{d\xi^+ d\xi^-(1-\xi^+ \xi^-)^{\frac{d-3}{2}}}{2} \de(r-|x||y|)\frac 1 2\de\p{\frac{|x|}{|y|}-1} \frac{2^{5d-2}r^{k+d-1}}{(16 r^2)^{\tl \De_\f}}\nn\\
&\quad\quad \x  \cP_{\frac{2-d}{2}+i\nu}\p{-\frac{x\.y}{|x||y|}} \left.\prod_{i=1}^4 i\cD \frac{1}{(-2 X_i\.Q_i+i\e)^{\De_\f}}\right|_{Q_i=Q_0}.
\end{align}
Our goal is to compute this integral at large $\nu$, where it will be dominated by flat-space physics.

The next step is to rewrite the boundary to bulk propagators using a Laplace transform
\begin{align}
\frac{1}{(-2X_i\.Q_i+i\e)^{\De_\f}} &= \frac{e^{-i\frac{\pi}{2} \De_\f}}{2^{\De_\f} \G(\De_\f)}\int_0^\oo \frac{d\w_i}{\w_i} \w_i^{\De_\f} e^{-iQ_i\.\w_i X_i}\,.
\end{align}
We can now define
\begin{align}
i\cM_\mathrm{AdS} &\equiv \p{e^{-i \sum_i Q_0 \w_i X_i}}^{-1} \p{i\cD e^{-i\sum_i Q_i\.\w_i X_i}}_{Q_i=Q_0}.
\end{align}
This is a potentially complicated function of the boundary positions and depends on our conventions for defining bulk contact interactions. 
However, in the flat space limit, the exponential $e^{-i\sum_i Q_i\.\w_i X_i}$ becomes a plane wave with momentum 
\begin{align}
p_i^A &\sim \w_i (X_i^A - (X_i\.Q_0)Q_0^A)\,.
\end{align}
$\cM_\mathrm{AdS}$ then becomes simply the flat space amplitude evaluated on the momenta $p_i$
\begin{align}
\label{eq:flatspacelimitofM}
\cM_\mathrm{AdS} &= \cM_\mathrm{flat}(s,u) + \textrm{subleading at large $\nu$}\,,
\end{align}
where $s=-(p_1+p_2)^2$ and $u=-(p_1+p_3)^2$. 

Now to compute (\ref{eq:gaugefixedlightintegral}), we first rescale
\begin{align}
x \to \frac{x}{\w_2+\w_4},\quad y\to \frac{y}{\w_1+\w_3}
\end{align}
and use the $\de$ functions to solve for $\w_3$ and $\w_4$. We then insert the split representation of the Gegenbauer (\ref{eq:integralgegenbauer}), using $\SO(d-1)$ symmetry to set $z=(1,1,0,\dots,0)$, so that we can make the replacement
\begin{align}
\cP_{\frac{2-d}{2}+i\nu}\p{-\frac{x\.y}{|x||y|}} &\to 
\p{\frac{x^-}{|x|}}^{\frac{2-d}{2}+i\nu} \p{\frac{y^-}{|y|}}^{\frac{2-d}{2}-i\nu},
\end{align}
where we use lightcone coordinates $x^\pm=x^0\pm x^1$.

In the large $\nu$ limit, the remaining integral over $x,y,\xi^\pm,\w_1,\w_2$ can be done by saddle point. The important factors are
\begin{align}
\label{eq:importantfactors}
\p{\frac{x^-}{|x|}}^{i\nu} \p{\frac{y^-}{|y|}}^{-i\nu} e^{-i \sum_i Q_0\. \w_i X_i}\,,
\end{align}
where $x$ and $y$ have been rescaled and $\w_3,\w_4$ solved for as before. Explicitly, the argument of the exponential becomes
\begin{align}
-i \sum_i Q_0\. \w_i X_i&\to -\frac{i \xi ^- \left(-r | x| -| y| +2 r^{1/2}(r \omega _2+\omega _1)\right)}{2 \sqrt{r}}-\frac{i \xi ^+ \left(r | y| +| x| -2 r^{1/2}(r \omega _1+ \omega _2)\right)}{2 \sqrt{r}}\nn\\
&\quad-i ( e\cdot x + e\cdot y )\sqrt{1-\xi ^- \xi ^+}\,.
\end{align}
Extremizing (\ref{eq:importantfactors}), we find a saddle at
\begin{align}
(x^-,x^+,y^-,y^+,\xi^-,\xi^+,\w_1,\w_2) &= \p{e^{i\pi} \nu,\nu,\nu,e^{i\pi}\nu,0,0,\frac{i\nu}{2\sqrt{r}},\frac{i\nu}{2\sqrt{r}}},
\end{align}
with vanishing transverse coordinates for $x$ and $y$. We have written the phases $e^{i\pi}=-1$ explicitly to indicate the direction one must analytically continue the corresponding variables to reach the saddle.

In evaluating the gaussian determinant, we must be careful: when deforming the contour to pass through the saddle point, we must rotate all the $x$ and $y$ coordinates by an overall factor of $i$, resulting in an extra phase $e^{i \pi d}$. 

The result after the saddle point integral is
\begin{align}
\left.-\widehat{\Pi}_{k,\nu}\right|_{\mathrm{light}} &= \frac{\g_{2\De_\f+k-1}(\nu)^2 }{ (\nu/2)^{2k} a_{\De_\f}}
 \oint \frac{dr}{2\pi i}\,\frac{r^{k-1}}{1-r^2} \mathcal{M}_\mathrm{flat}\p{s=-\frac{(1-r)^2}{4 r}\nu^2,u=-\nu ^2}\nn\\
&\quad \x (1+O(1/\nu^2)\,.
\end{align}
To compute the result for $ C_{k,\nu}$, we must insert the factors (\ref{eq:extrafactors}) into the integrand. We then change variables from $r$ to $s$, expressing the result as a contour integral around $s=\oo$. Finally, bringing the contour in from infinity, we use crossing symmetry in the form $s\to \nu^2-s$ to express the answer purely as a residue at $s=0$ up to an overall factor of 2. The result is (\ref{eq:BknuFlatSpaceLimit}).
Note that even though we have derived \eqref{eq:BknuFlatSpaceLimit} for contact Witten diagrams, it works for the contribution of gravity (\ref{eq:BknuGravity}) as well. 

\subsection{Interlude: gravity in impact parameter space and eikonalization}
\label{sec:eikonal}

Having computed both the heavy and light contributions to $C_{k,\nu}$, we can now consider some simple consequences of the $C_{k,\nu}$ sum rules. The contribution of gravity to $C_{2,\nu}$ is special because it is delocalized in impact parameter space under the transform (\ref{eq:psidefinition}), which goes from $C_{2,\nu}$ to $\Psi_{2,\g}$. By contrast, contact diagrams contribute polynomials in $\nu$, which become distributions localized near $\g=1$. Thus, in this section, we will consider the consequences of the $\Psi_{2,\g}$ sum rule 
\begin{align}
\left.-\Psi_{2,\g}\right|_{\mathrm{light}} &= \<\Psi_{2,\g}[\De,J]\>\,,
\end{align}
for $\gamma>1$. This discussion is outside the main thread of this paper, so an impatient reader should feel free to skip to section~\ref{sec:holo}.

To obtain $\Psi_{2,\g}|_\mathrm{gravity}$, we use (\ref{eq:BknuGravity}) together with the identity
\begin{align}
\label{eq:gravitonpropagator}
\int_0^\oo \frac{d\nu}{2\pi}  \gegenbauermeasure(\nu)\frac{\cP_{\frac{2-d}{2}-i\nu}(\gamma)}{\nu^2+(d/2)^2} = \frac{\vol S^{d-2}}{2^{d-2}}h_{d-1}(\gamma-1)\,,
\end{align}
where $h_{\De_0}(u)$ is the bulk-to-bulk propagator in AdS${}_{d-1}$ for a field of dimension $\De_0$ with chordal distance $u=\eta-1$:
\begin{align}
h_{\De_0}(u) &= \frac{\Gamma (\De_0 )  }{2\pi ^{\frac{d-2}{2}} \Gamma \left(\frac{2-d}{2}+\De_0 +1\right)} (2u)^{-\De_0 } {}_2F_1\left(\De_0 ,\De_0-\frac{d-3}{2} ;2 \De_0-d +3;-\frac{2}{u}\right)\,.
\end{align}
The function $h_{d-1}(u)$ is precisely the propagator for a graviton in the transverse $H_{d-1}$ space. 
Overall, we find
\begin{align}
-\left.\Psi_{2,\gamma}\right|_\mathrm{light} 
&=  \frac{8\pi G\vol S^{d-2}}{2^{d-2}} h_{d-1}(\gamma-1) + \textrm{contacts}\,,
\end{align}
where ``contacts'' are distributional terms near $\g=1$.

Via (\ref{eq:psiknuinreggelimit}), this gives a universal prediction for the $k=2$ moment of the double-discontinuity of heavy states in the Regge limit in any holographic CFT:
\begin{align}
\label{eq:universalreggegravity}
8\pi G \frac{\vol S^{d-2}}{2^{d-2}} h_{d-1}(\gamma-1) &= \left\<
\frac{2}{m^4} \frac{\de(\gamma-\eta_\mathrm{AdS})}{2^{d-2} (\gamma^2 -1)^{\frac{d-3}{2}}}
\right\> \x(1 + O(1/\De_\mathrm{gap})) \nn\\
&\qquad\qquad\qquad\qquad\qquad\qquad   (\cosh^{-1}\g \gg 1/\De_\mathrm{gap})\,,
\end{align}
where the expectation value on the right-hand side is defined by (\ref{eq:cftpositivemeasure}). The condition $\cosh^{-1}\g\gg 1/\De_\mathrm{gap}$ comes from the fact that we expect large-$\nu$ corrections to the heavy density to become important at smaller impact parameters.
 
Positivity of the right-hand side implies that $G>0$, so gravity must be attractive.
One may also try to turn the equation around and use low-energy gravity to
constrain light-light-heavy couplings at a given impact parameter $\etaAdS$.
How constraining is that?

It turns out that the sum rules are rather easy to fulfill, and can be satisfied by very different physical models. 
Perhaps the simplest ``UV completion'' is the just eikonal exponentiation of the tree-level phase shift.
The anomalous dimensions of double-trace states due to tree-level graviton exchange in the Regge regime are \cite{Cornalba:2006xm,Cornalba:2007zb} 
\begin{align}
\label{eq:treelevelgraviton}
\Gamma_\mathrm{tree}(\De,J) &= 
 8G\De^2 \frac{h_{d-1}(\etaAdS-1)}{\etaAdS+1}\,.
\end{align}
Of course the full anomalous dimension gets corrected at loop level. The eikonal approximation amounts to trusting the result (\ref{eq:treelevelgraviton}) even when $\G_\mathrm{tree}(\De,J)$ is $O(1)$, which is generally expected
at large enough impact parameter where nonlinear curvature effects and tidal excitations can be ignored.
Plugging (\ref{eq:treelevelgraviton}) into the right-hand side of (\ref{eq:universalreggegravity}), expanding in the Regge limit and performing the integral over $J$ (including a factor of $1/2$ because only even spins appear, and another factor of $1/2$ because tree-level twists of double-trace families are spaced by 2), we find
\begin{align}
\frac{\vol S^{d-2}  }{2^{d-2}} \int \frac{d\De}{4} \frac{16(\eta +1)}{\pi  \Delta ^3} 2\sin ^2\left(\frac{\pi \Gamma_\mathrm{tree}(\Delta,\etaAdS)}{2}\right)\,.
\end{align}
The $2\sin^2(\dots)$ factor represents the double-discontinuity of the exchanged double-twist operators.
Naively expanding the integrand in small $\Gamma_\mathrm{tree}$ would result in a divergent integral $\sim\int d\De\, \De$. Instead, in accordance with the eikonal approximation, we do not expand the exponentials in small $\Gamma_\mathrm{tree}$,
and the large phases regulate the integral. We obtain the left-hand side of (\ref{eq:universalreggegravity}).
It is a nice consistency check of our calculations that the eikonal model satisfies the sum rules.\footnote{Note that the $k=2$ moment of the double-discontinuity determines the stress-tensor pole in the Lorentzian inversion formula. Our calculation shows that this pole can be produced by eikonalization (among other mechanisms).}

However, this success cannot by any means be used to justify the eikonal approximation.
Very different models of the heavy sector also satisfy the sum rules.
For example, in tree-level string theory, we expect (\ref{eq:universalreggegravity}) to be saturated by stringy states in the weakly coupled regime.


\section{Holographic sum rules and local physics in AdS}
\label{sec:holo}

We now combine the results of the previous sections to derive bounds on couplings of EFTs in AdS. Recall that our principal tools in flat space are the dispersive sum rules $\cC_{k,u}$
and their improved versions $\cC^\imp_{k,u}$. These sum rules relate EFT couplings to massive states. If a sum rule gets nonnegative contributions from states with $m\geq M$ and $J=0,2,\dots$, we say that it is ``heavy-positive'', and it leads to nontrivial bounds on EFT couplings.

The central observation of this section is:
\begin{itemize}
\item[$(\star)$] Generic heavy-positive flat-space sum rules can be directly translated to give heavy-positive CFT sum rules, via the dictionary $\cC_{k,u} \to C_{k,\nu}$ and $u\to -\nu^2$, up to possible relative corrections suppressed by positive powers of $1/\Delta_{\rm gap}$.
\end{itemize}
We will show this by translating various classes of flat space sum rules to CFT, with increasing levels of sophistication.

Each CFT sum rule obtained in this way places bounds on low-energy EFTs in AdS.
The interpretation of these individual bounds depends subtly on conventions for defining bulk coupling constants, but we believe that the interpretation of the infinite set is clear and transparent.

To make our physical set up clear, let us elaborate here on this point, beginning with non-gravitational bounds
(that is, bounds on CFTs without a stress tensor or equivalently $c_T=\infty$).
In flat space, there exists an infinite series of functionals which prove lower and upper bounds on EFT couplings, in term of a cutoff scale $M$.
The corresponding heavy-positive CFT sum rules will prove
an infinite set of bounds on the CFT interactions defined in \eqref{eq:MellinEFT}:
\def\rAdS{R_{\rm AdS}}
\begin{subequations}\label{bounds}
\begin{align} 
 0 &\leq g_2\,,  \\
 -\#g_2 +O(\sum g/\rAdS^2) &\leq g_3\Delta_{\rm gap}^2 \leq
 \#g_2 + O(1/\rAdS^2)\,, \\
 0 &\leq g_4\Delta_{\rm gap}^4 \leq \#g_2 + O(1/\rAdS^2)\,, \\
 -\#g_2+O(\sum g/\rAdS^2) &\leq g_5\Delta_{\rm gap}^6 \leq
  \#g_2+ O(1/\rAdS^2)\,,\\
&\cdots\nonumber
\end{align}
\end{subequations}
Crucially, such two-sided inequalities exist for \emph{all} EFT parameters
beyond $g_2$ \cite{Caron-Huot:2020cmc,Tolley:2020gtv}.
The O(1) constants $\#$ depend on spacetime dimension but do not grow with derivative order
(see Table 3 of \cite{Caron-Huot:2020cmc}).

The error terms in eqs.~\eqref{bounds} are of two different types.
The first type of error, denoted $O(1/\rAdS^2)$, originates from the mentioned ``small deformations'' that restore positivity.  These are needed because the optimal functionals in flat space typically have zeros,
so there is a danger that $1/m^2$ corrections in going from $\cC_{k,u}$ to $C_{k,\nu}$ cause negative dips.
Conceptually, this issue is easily solved by starting with non-optimal flat space functionals, which are strictly positive everywhere. For large enough $\Delta_{\rm gap}<m$, we are then guaranteed that $1/m^2$ corrections will not spoil positivity, provided they do not grow in some kinematic limits, which we will check.
As exemplified in subsection \ref{sec:derivativesofforward}, this ``de-extremalization'' step only changes result by a relative $1/\Delta_{\rm gap}^2$.

The second type of error, denoted $O(\sum g/\rAdS^2)$, involves the use of IR crossing symmetry or ``null constraints''.
For example, the lower bound on $g_3$ from \cite{Caron-Huot:2020cmc,Tolley:2020gtv} exploits that
two distinct IR experiments can measure the same EFT coupling $g_4$ (see \eqref{w abc} below) to bound couplings to heavy spinning states.  The difficulty is that the straightforward uplift of a flat space null constraint is not a CFT null constraint.  Rather, it measures infinitely many EFT couplings, albeit with suppressed coefficients:
\begin{align} \label{null constraint CFT}
-\# g_2 +O(\sum g/\rAdS^2) \equiv 
-\# g_2 + \#' \frac{g_3\Delta_{\rm gap}^2}{\Delta_{\rm gap}^2}
+\#''\frac{g_4\Delta_{\rm gap}^4}{\Delta_{\rm gap}^2}
+\#'''\frac{g_5\Delta_{\rm gap}^6}{\Delta_{\rm gap}^2} + \ldots\,.
\end{align}
The various coefficients $\#'$ are computable for example
from the subleading terms in \eqref{eq:BknuContactsLeading} and they may have any signs.
The key point is that all terms are parametrically suppressed by $1/\Delta_{\rm gap}^2$ compared to the first,
if one assumes the scaling implied by eqs.~\eqref{bounds}.  In other words,
by substituting into \eqref{null constraint CFT} the subsequent functionals \eqref{bounds}
which prove upper and lower bounds on $g_3,g_4,g_5\ldots$,
one gets a better functional in which the error is reduced term-by-term in $k$ and
order-by-order in $1/\Delta_{\rm gap}^2$, and the first term is left to dominate.

In the following, we thus make the following technical assumption: that the
\emph{infinite} collection of bounds \eqref{bounds} implies
\be \label{technical assumption}
O(\sum g/\rAdS^2) = O(1/\rAdS^2)\,.
\ee
This seems reasonable, since the pattern of the bounds suggests that
neglected terms get smaller and smaller with $k$; this is further discussed in \ref{sec:lightandadsbounds}.
It seems utterly implausible to us that a counter-example to \eqref{technical assumption} could be found.

The inclusion of gravity does not significantly affect the above:
as long as the EFT contains only particles of spin $2$ or less, all higher-subtracted sum rules
smoothly expand around the forward limit and only spin-2 sum rules require
the improvement $\cC^{\rm imp}_{2,u}$ from section~\ref{ssec:SMatrixReview}.
Thus, although $g_2$ can be negative with gravity \cite{Caron-Huot:2021rmr},
one can still show that $g_4\geq 0$ and that
all ratios of the form $g_k M^{2(k-4)}/g_4$ with $k\geq 4$ admit two-sided bounds of the form in \eqref{bounds}.

With this technical assumption, the calculations in this section establish effective field theory scaling,
in the same sense that it is established in flat space.
That is, any EFT bound which can be proved in a flat space assuming
weak interactions between particles of spin $J\leq 2$ at energies less than $M$,
yields a corresponding bound in a CFT with a large higher-spin gap.
For example, if the CFT contains a real scalar operator with dimension $\Df\ll \Delta_{\rm gap}$,
and a stress tensor, the flat-space results of \cite{Caron-Huot:2021rmr} uplift to yield bounds on
the relative strength of all scalar self-interactions:
\begin{subequations}\label{bounds1}
\begin{align} 
 0 &\leq \tilde{g}_2 \equiv g_2 +\#/(c_T \Delta_{\rm gap}^2)\,, \\
 -\# \tilde{g}_2(1+O(1/\rAdS^2)) &\leq g_3\Delta_{\rm gap}^2 \leq
 \# \tilde{g}_2(1+O(1/\rAdS^2))\,, \\
 0 &\leq g_4\Delta_{\rm gap}^4 \leq \# \tilde{g}_2(1+O(1/\rAdS^2))\,, \\
 -\# \tilde{g}_2(1+O(1/\rAdS^2)) &\leq g_5\Delta_{\rm gap}^6 \leq
  \# \tilde{g}_2(1+O(1/\rAdS^2))\,.\\
&\cdots \nonumber
\end{align}
\end{subequations}
The optimal finite constants $\#$ in CFT are thus \emph{at least as good} as in flat space.
The numerical offset between $g_2$ and $\tilde{g}_2$ on the first line, as well as the optimal value of
other coefficients, can depend on interactions between the scalar and light spin-2 particles
besides the graviton, if they exist.

With our physical setup and assumptions hopefully clear, we now describe the CFT translates of various sum rules.

\subsubsection*{Notation and conventions}

In the following, we use a mixture of flat-space and CFT notation, depending on what is being emphasized in the discussion. We use $M$ to refer to the flat-space cutoff and $\De_\mathrm{gap}$ to refer to the CFT twist gap. In units where $R_\mathrm{AdS}=1$, they are equal: $M=\De_\mathrm{gap}$.

We use $\beta$ to refer to the AdS impact parameter. Recall that $\beta\approx \log\frac{\De+J}{\De-J}$. In the flat-space region $\beta\ll 1$, this becomes $\beta\approx \frac{2J}{\De}\approx \frac{2J}{m}$. In the context of flat space physics, we sometimes write $b=\frac{2J}{m}$.

The contribution of gravity in flat space is naturally expressed in terms of Newton's constant $G$. In CFT, it is natural to use the stress-tensor two point coefficient $c_T$. They are related by
\begin{align}
8 \pi  G \frac{d\,\Gamma\!\left(\frac{d}{2}\right)^3}{4\pi ^{\frac{d}{2}}\,\Gamma (d+2)} &= \frac{c^{\text{free}}_T}{c_T},
\end{align}
where $c^{\text{free}}_T$ is the stress tensor two-point coefficient for the free boson.\footnote{The value of $c_T$ depends on the convention for the two-point function, but $c^{\text{free}}_T/c_T$ is convention-independent. In the convention of \cite{Dolan:2000ut}, we have $c^{\text{free}}_T = d/(d-1)$.}

Finally, we set $R_\mathrm{AdS}=1$ throughout. However, we use the notation $O(1/R_\mathrm{AdS}^2)$ to indicate a quantity that vanishes in the flat-space limit if the $R_\mathrm{AdS}$-dependence were restored. For example, $O(1/R_\mathrm{AdS}^2)$ can be read as $O(1/\De_\mathrm{gap}^2)$ or $O(1/M^2)$, depending on the context. We hope this does not cause confusion.

\subsection{Positive functionals from the forward limit}
\label{sec:positiveforwardlimit}

Perhaps the simplest heavy-positive flat space sum rules are the forward limits of $\cC_{k,u}$:
\begin{align}
\cC_{k,u=0}[m,J] &= \frac{2}{m^{2k}}\,.
\end{align}
In non-gravitational EFTs, we expect $\cC_{k,0}$ to converge for $k=2,4,\dots$, while in gravitational EFTs, $\cC_{k,0}$ converges for $k=4,6,\dots$.\footnote{$C_{2,0}$ may ``converge'' in a technical sense in CFT since, unlike in flat space, the forward limit of the graviton pole contribution \eqref{eq:BknuGravity} is finite $\sim \frac{8\pi G \rAdS^2}{(d/2)^2}$, where we restored $\rAdS$.
We nonetheless treat it as divergent since it diverges quadratically
in the flat space limit, and the resulting bound $\frac{8\pi G \rAdS^2}{(d/2)^2} + 2 g_2 \geq 0$
does not possess the expected EFT scaling with $M$.
We obtain bounds on $g_2$ with the expected $M$-scaling in section~\ref{sec:implocfun}.
}

By analogy, in CFT we should consider the functionals $C_{k,\nu=0}$. We claim that these are indeed heavy-positive, assuming all single-trace operators satisfy $\tau \geq \De_\mathrm{gap}\equiv M$ and $M$ is sufficiently large. To see this, note that since $\nu=0\ll \De$ for heavy operators, the Regge limit formula (\ref{eq:bknuinreggelimit}) applies, giving
\begin{align} \label{Ck forward}
C_{k,0}[\De,J] &=
\frac{2\cP_{\frac{2-d}{2}}(\cosh\beta)}{m^{2k}}\x\p{1+ O\p{\frac {1}{M^2}}}\,,
\end{align}
where the heavy density $C_{k,0}[\De,J]$ is defined by (\ref{eq:definitionofheavydensity}).
Furthermore, we  see by inspection that 
\begin{align} \label{cP pos}
\cP_{\frac{2-d}{2}}(\cosh\beta) &= {}_2F_1 \p{\frac{d-2}{2},\frac{d-2}{2},\frac{d-1}{2},\frac{1-\cosh\beta}{2}}
\end{align}
is positive for all impact parameter $\beta$.
(Recall from \eqref{eq:impactparameter} that $\cosh\beta\approx \frac{\Delta^2+J^2}{\Delta^2-J^2}$.)

\def\gnice{g^{\rm clean}}

It follows that $C_{k,0}$ gives a bound
\begin{align} \label{Ck0 bound}
0 \leq -\left.C_{k,0}\right|_{\mathrm{light}} &= 2^{k/2}g_k + O(\sum g/\rAdS^2) \qquad\quad\mbox{($k$ even)}\,,
\end{align}
where the error is an infinite sum of couplings, each suppressed by an additional $1/\rAdS^2$ compared with the expected dimensional analysis scaling.
Alternatively, in this case it would be reasonable to {\it define} the bulk coupling $2^{k/2} g_k$ as the forward limit of the $C_{k,\nu}$ sum rule:
\be
 0 \leq -\left.C_{k,0}\right|_{\mathrm{light}} \equiv 2^{k/2}\gnice_k\,.
\ee
Assuming that the coefficients in the EFT series in \eqref{eq:MellinEFT} each satisfy the anticipated bounds,
the difference between $g_k$ and $\gnice_k$ is indeed negligible.
This is the technical assumption stated in \eqref{technical assumption}.
However, since we are on our way to justifying this assumption, we stay clear of circular reasoning
by distinguishing between the Mellin couplings $g$ and the ``easy to bound'' couplings $\gnice$.

Since the dependence on $\etaAdS$ is the same for all $k$,
the collection of $2^{k/2}\gnice_k$ for even $k$ satisfy the
same properties as moments of $\frac{1}{m^{2k}}$ in a positive distribution
\cite{Adams:2006sv,Kundu:2021qpi,Arkani-Hamed:2020blm}.
In particular, the following two-sided bounds hold:
\be
 0\leq 2^{k/2}\gnice_k \Delta_{\rm gap}^{2k-4}\leq 2\gnice_2 \times (1+O(1/\Delta_{\rm gap}^2))\qquad\quad\mbox{($k\geq 4$ even)}\,.
\ee
The sole source of error here is the fact that the optimal combination of 
$C_{2,0}$ and $C_{k,0}$ which proves the flat space version of this bound, vanishes at threshold $m=\Delta_{\rm gap}$. One then needs to give oneself a little bit of breathing room to deal with $1/m^2$ effects in
\eqref{Ck forward}, for example by slightly increasing the coefficient of $C_{2,0}$.

As discussed below \eqref{regge subleading}, $1/m^2$ corrections
are uniform in $\beta$ (the expansion parameter is $\frac{\cosh\beta}{m^2}$, which is less than
$\frac{1}{2\Delta_{\rm gap}^2}$ if we have a twist gap $\Delta_{\rm gap}$).
Therefore, a small finite shift suffices at once to obtain positivity on all heavy states.

In this example, positivity of $C_{k,0}$ in the bulk point limit $\beta\ll 1$ was actually guaranteed by positivity in flat space, since $C_{k,\nu}$ in the bulk point limit (\ref{eq:bulkpointbknu}) agrees with $\cC_{k,u}$ in flat space (up to $1/m^2$ corrections). However, positivity in the Regge regime $\beta\sim O(1)$ was not guaranteed and depended on the behavior of the function $\cP_{\frac{2-d}{2}}(\cosh\beta)$. This is a general feature of functionals obtained from flat space. Most of the work in showing positivity involves controlling the Regge regime, given that the bulk point regime is automatic.

\subsection{Derivatives around the forward limit and general EFT scaling}
\label{sec:derivativesofforward}

Additional heavy-positive sum rules can be obtained via derivatives around the forward limit. As an example, let us explore the simplest flat-space sum rules that yield two-sided bounds on $g_3/g_2$ in non-gravitational theories.  The minimal analysis  in flat space  (we follow section 3.1 of \cite{Caron-Huot:2020cmc})
uses three sum rules:
\begin{align} \label{w abc}
\w^{(a)}_{\rm flat} \equiv \tfrac12 \cC_{2,0},\quad
\w^{(b)}_{\rm flat} \equiv -M^2\partial_{u} \cC_{2,u}\big|_{u=0},\quad
\w^{(c)}_{\rm flat} \equiv M^4(d^2-1)\left[\tfrac14(\partial_u)^2\cC_{2,u} - \cC_{4,u}\right]_{u=0}\,.
\end{align}
These are directly uplifted to CFT by setting $u\mapsto -\nu^2$:
\begin{align}
\w^{(a)} \equiv \tfrac12 C_{2,0},\quad
\w^{(b)} \equiv \Delta_{\rm gap}^2\partial_{\nu^2} C_{2,\nu}\big|_{\nu=0},\quad
\w^{(c)} \equiv \Delta_{\rm gap}^4(d^2-1)\left[\tfrac14(\partial_{\nu^2})^2C_{2,\nu} - C_{4,\nu}\right]_{\nu=0}\,.
\end{align}
We normalized the three sum rules
so that they are homogeneous in $M=\Delta_{\rm gap}\rAdS$ and $\rAdS$.
The combinations of interest in the flat space limit
will have relative coefficients of order unity, and we say that such combinations
have manifest dimensional analysis scaling.

The flat space sum rules $\w_{\rm flat}^{(i)}$,
to tree-level accuracy in the EFT, measure  respectively $g_2$, $M^2g_3$ and 0 --- the third is a ``null constraint'' which compares two IR measurements of $g_4$.
Their CFT uplifts measure infinite combinations of the couplings defined in \eqref{eq:MellinEFT}.
Like we did in the preceding subsection,
we can conveniently use a sum rule to \emph{define} a coupling $\gnice_3$:
\begin{align} \label{omega abc light}
 -\w^{(a)}\Big|_{\rm light} \equiv \gnice_2\,,\qquad
 -\w^{(b)}\Big|_{\rm light} \equiv \Delta_{\rm gap}^2\gnice_3\,,\qquad
 -\w^{(c)}\Big|_{\rm light} = O(\sum g/\rAdS^2)\,.
\end{align}
The null constraint $\omega^{(c)}|_\mathrm{light}$, remains an infinite sum of couplings of the form
discussed in \eqref{null constraint CFT}, where each coefficient is suppressed by an extra factor $1/\rAdS^2$ that we anticipate will turn into $1/\Delta_{\rm gap}^2$
upon using other bounds, as argued below \eqref{technical assumption}.

In the bulk point limit, the heavy actions of these functionals are, from (\ref{eq:bulkpointbknu}):
\be \label{omega abc limits}
\left(\begin{array}{c}
\w^{(a)}\\
\w^{(b)}\\
\w^{(c)}
\end{array}\right)[\Delta,J] \to \frac{1}{m^4}
\left(\begin{array}{c} 1 \\ \frac{M^2}{m^2}\left(3-\frac{4\cJ^2}{d-1}\right) \\ \frac{M^4}{m^4}\cJ^2(2\cJ^2-5d-1)
\end{array}\right)\times \left(1+O\p{\frac{1}{m^2},\frac{J^2} {m^2}}\right)\,,
\ee
with $\cJ^2\equiv J(J+d-2)$ the Casimir. In this approximation,
it is easy to see that the functional $3\w^{(a)}-\w^{(b)}$ is nonnegative on all states with $m\geq M$.
It vanishes at threshold for $J=0$.
This could potentially cause problem if the $1/m^2$ correction has the wrong sign.
However, the sign of that correction won't be important for our story because we can always simply
increase the coefficient of $\w^{(a)}$ by a small amount:
\be
 \w^{(\e)}\equiv (3+\e)\omega^{(a)} - \omega^{(b)}\geq 0 \mbox{ on all heavy states }\ \ \Rightarrow\ \ 
 \Delta_{\rm gap}^2\gnice_3 \leq (3+\e) \gnice_2\,. \label{g3 upper bound}
\ee
Because we started with a sum rule with manifest dimensional analysis scaling,
the $\e\sim \frac{1}{\Delta_{\rm gap}^2}$ correction has a negligible impact: this strategy allows us to prove
a small perturbation of the flat space bound.

\begin{figure}
\centering{\includegraphics[width=11cm]{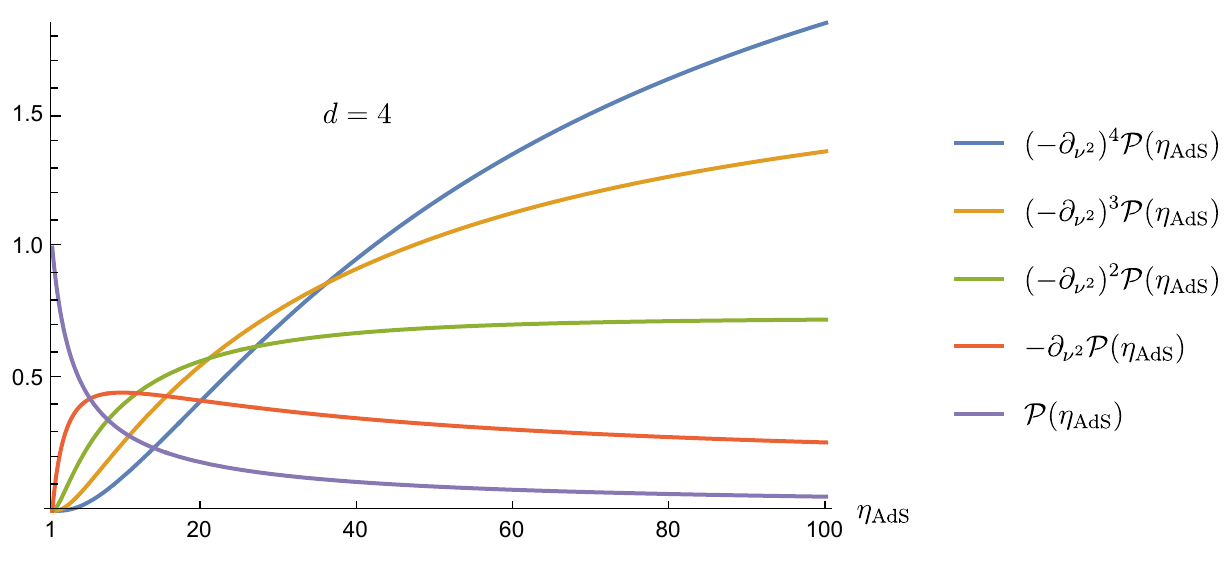}}
\caption{Derivatives around zero momentum transfer
$(-\partial_{\nu^2})^n\mathcal{P}_{\frac{2-d}{2}+i\nu}(\cosh\beta)\big|_{\nu=0}$
are positive for all impact parameters, even beyond the AdS radius ($\beta\sim 1$).  At small impact parameters, they reproduce the flat space behavior $\sim \beta^{2n}$.
\label{fig:gegenbauer derivatives}}
\end{figure}

To complete the proof that the left-hand-side of \eqref{g3 upper bound} is positive, we must
consider those heavy states that are outside the reach of \eqref{omega abc limits}: those with $J/m$ not small.
We use the Regge limit formula in \eqref{eq:bknuinreggelimit}, reproduced here for convenience:
\be
 C_{k,\nu}[\Delta,J]\to \frac{2}{m^{2k}} \cP_{\frac{2-d}{2}+i\nu}(\cosh\beta)\qquad (m\gg 1,\nu\sim1, \beta \sim 1)\,.
\ee
In this limit, $\partial_{\nu^2}$ is no longer parametrically scales like $1/\Delta_{\rm gap}^2$:
the combination \eqref{g3 upper bound} is thus dominated by the term with the largest number of derivatives,
\begin{align}
\w^{(\e)}[\Delta,J] &\to -\Delta_{\rm gap}^2\partial_{\nu^2}C_{2,\nu}[\Delta,J]\big|_{\nu=0}
\\
 &= 2\frac{\Delta_{\rm gap}^2}{m^4}\left[(-\partial_{\nu^2})\cP_{\frac{2-d}{2}+i\nu}(\cosh\beta)\right]_{\nu=0}\,.
\label{g3 upper bound Regge}
\end{align}
Importantly, the $\beta\to 0$ limit of the Regge limit overlaps with the bulk point limit.
Substituting $\cosh\beta=1+\frac{2\cJ^2+(d-2)^2/2}{m^2}$ into \eqref{g3 upper bound Regge} and expanding
at large $m$, we indeed reproduce the $J\to\infty$ limit of \eqref{omega abc limits}:
\be
 \w^{(\e)}[\Delta,J] \to \frac{M^2}{m^6}\frac{4\cJ^2}{d-1} \qquad (m,J\gg1,\ \frac{J}{m}\ll 1)\,.
\ee
This implies that the expression in \eqref{g3 upper bound Regge} is automatically positive at $\beta\to 0$. Interestingly, it never changes sign!
It is plotted in Figure~\ref{fig:gegenbauer derivatives},
and this completes the proof of the upper bound in \eqref{g3 upper bound}.

Physically, $\beta\ll 1$ means that we are in the flat space limit where the impact parameter
is small compared to $\rAdS$. Meanwhile, $J\gg 1$ or $\beta\gg \frac{1}{M}$ means that the impact parameter is large compared to the energy scale of the scattering process.
The strategy of establishing positivity by exploiting the overlap between these regimes works in many examples. We will apply it in section \ref{sec:implocfun} to show positivity of functionals constructed using the algorithm of \cite{Caron-Huot:2021rmr}.

For now, let us study one more example involving derivatives around the forward limit: the lower bound on $g_3/g_2$.  By adding an $\e$-correction to the flat space functional from \cite{Caron-Huot:2020cmc},
we can construct a CFT functional which is strictly positive on all states with $m>\Delta_{\rm gap}$
in the bulk point limit:
\begin{align}
{\w'}^{(\e)} &= \left(\kappa_{d+1}+\e,\ 1, \frac{(d-1)\kappa_{d+1}-5d-3}{2d(d^2-1)}\right)\cdot
\left(\begin{array}{c}
\w^{(a)}\\ \w^{(b)}\\ \w^{(c)}
\end{array}\right)\,, \label{g3 lower bound}
\end{align}
where again $\e\sim \frac{1}{\Delta_{\rm gap}^2}$, and $\kappa_{d+1}$ is a constant given in \cite{Caron-Huot:2020cmc}
which ensures that ${\w'}^{(\e)}[\Delta,J]>0$ for all heavy states in the bulk point limit.

In the Regge regime, the heavy density is dominated by the maximal derivative term, $(-\partial_{\nu^2})^2C_{2,0}$:
\begin{align}
\label{eq:lowerboundreggeterm}
{\w'}^{(\e)}[\De,J] &= \frac{M^4}{m^4}\frac{(d-1)\kappa_{d+1}-5d-3}{4d}\left[(-\ptl_{\nu^2})^2 \cP_{\frac{2-d}{2}+i\nu}(\cosh\beta)|_{\nu=0} + O\p{\frac 1 {m^2}}\right]\,.
\end{align}
Equation (\ref{eq:lowerboundreggeterm}) is a good approximation for $1/M\ll \beta$.
The agreement between the bulk point and Regge formulas in the overlap regime $1/M\ll \beta \ll 1$ is readily verified using that:
\begin{align} \label{overlap comparison}
\left.(-\partial_{\nu^2})^2 \cP_{J}\p{1-\frac{2\nu^2}{m^2}}\right|_{\nu=0} &\approx \frac{4\cJ^4}{m^4(d^2-1)}
\approx \left.(-\partial_{\nu^2})^2 \cP_{\frac{2-d}{2}+i\nu}(\cosh\beta)\right|_{\nu=0}\,.
\end{align}
Therefore, to complete the proof of positivity of \eqref{g3 lower bound} in CFT, it suffices to show that the bracket in \eqref{eq:lowerboundreggeterm} never changes sign.
In appendix~\ref{app:gegenbauerformulas} we prove at once
that all derivatives $(-\partial_{\nu^2})^n\cP$ around $\nu=0$ are positive; for illustration, a few cases are plotted in Figure~\ref{fig:gegenbauer derivatives}.
Equating the light and heavy contributions in eqs.~\eqref{omega abc light} and \eqref{g3 lower bound},
and including the previous upper bound, we find the following two-sided bound :
\be
-\kappa_{d-1}\gnice_2 + O(\e,\sum g/\rAdS^2)  \leq \Delta_{\rm gap}^2\gnice_3
\leq 3\gnice_2 + O(\e)\,. \label{lower bound nice}
\ee
The error includes the ``de-extremalization'' step to keep the functional positive in the bulk-point limit,
as well as the fact that null constraints have become infinite sums of couplings each with small coefficients.

The arguments in these examples generalize to any functional $\omega_\mathrm{flat}$ constructed from a finite number of derivatives around the forward limit:
the dominant behavior at $\sim\rAdS$ impacts is entirely determined by the $J\to\infty$ limit
of the bulk-point limit.
Thus, positivity for all spins in the bulk-point regime implies positivity at all impact parameters.
In particular, all flat-space sum rules constructed in \cite{Caron-Huot:2020cmc}
give rise to heavy-positive CFT functionals via this construction,
thus giving two-sided bounds on
all higher-derivative contact interactions that can appear in four-point correlators.

The error on the left of \eqref{lower bound nice} is \emph{expected} to be small according to dimensional analysis scaling, which is what we are trying to establish here.
As stressed above \eqref{technical assumption}, bounds of this form, taken individually, do not justify dimensional analysis scaling: they only confirm that it is self-consistent.
However, as also argued below \eqref{technical assumption}, we believe that the \emph{infinite} collection of these bounds leaves dimensional analysis scaling as the only plausible option.


\subsection{Impact parameter localized functionals}
\label{sec:implocfun}

In flat space, expanding dispersive sum rules around $u=0$ fails in the presence of gravity. Instead, a working strategy is to integrate the $\cC_{2,-p^2}^\imp$ sum rule against a suitably-chosen wavefunction $f(p)$ to create a functional that is localized in impact parameter space \cite{Caron-Huot:2021rmr}. In this section, we mimic this construction in CFT. We show that a positive flat-space sum rule obtained by integrating $\cC_{2,-p^2}^\imp$ can be translated to give a positive CFT functional that proves the same bounds in the limit $M\to \oo$.

We can construct an improved version of $C_{2,\nu}$ by replacing $\cC_{k,u}\to C_{k,\nu}$ and $u\to -\nu^2$ in the definition \eqref{C2 improved def flat} of $\cC^\imp_{2,u}$:
\begin{align}
C_{2,\nu}^\mathrm{imp} &=
C_{2,\nu} - \sum_{n=2}^{k_{\rm max}/2} \p{n\,\nu^{4n-4} C_{2n,0} + \nu^{4n-2} \left.\ptl_{\nu^2}C_{2n,\nu}\right|_{\nu=0}}
\nn\\&=
 C^{\rm unimproved}_{2,\nu}+ C_{2,\nu}^{\rm improvement}\,.
\label{eq:definitionofk}
\end{align}
Note that we have truncated the sum over $k=2n$, since formulas from the preceding sections
are not valid for $k\sim \Delta_{\rm gap}$; this is expected to have a negligible impact, as discussed in section \ref{sec:lightandadsbounds}.

Consider now a flat-space sum rule
\begin{align}
\omega_\mathrm{flat} &\equiv \int_0^M \frac{dp}{M} f\p{\frac{p}{M}} \cC^\imp_{2,-p^2}\,,
\end{align}
and the analogous CFT sum rule
\begin{align}
\omega_\AdS &\equiv \int_0^M \frac{d\nu}{M} f\p{\frac{\nu}{M}} C^\imp_{2,\nu}\,. \label{omega AdS f}
\end{align}
Suppose that $\omega_\mathrm{flat}[m,J]$ is heavy-positive. We will argue that, for sufficiently large $M$, $\omega_\AdS[\De,J]$, or perhaps a small correction to it, is heavy-positive as well.
Furthermore, the contributions of light states to $\omega_\mathrm{flat}$ and $\omega_\AdS$ agree as $M\to \oo$, so $\omega_\AdS$ proves the same bounds on EFT couplings as $\omega_\mathrm{flat}$ in that limit.

By construction, $\omega_\mathrm{flat}[m,J]$ and $\omega_\AdS[\De,J]$ agree up to $1/M^2$ corrections in the bulk point regime of fixed $J$; hence $\omega_\AdS[\De,J]$ is automatically positive in that regime for sufficiently large $M$.  Our task is to show that $\omega_\AdS[\De,J]$ is additionally positive outside the bulk point regime.

Our strategy will use the following observation about the functional $\omega_\mathrm{flat}$:
at large impact parameters (but still within the AdS radius), $\frac{1}{M}\ll \beta\ll 1$,
the ``unimproved'' and ``improvement'' terms dominate in different limits, and they are
thus separately positive.  Our task will be to ensure that separate positivity continues to hold at larger $\beta$.

For fixed $m$ and large impact parameter (or large $J$),
the improvement term dominates for the simple reason that
forward derivatives of Gegenbauer polynomials grow with spin; on the other hand, at large spin,
the $\cC_{2,u}$ (or $C_{2,\nu}$) contribution tends to decay because it is an integral over highly oscillatory
Gegenbauer polynomials.
The later thus only matters where the former vanishes.  This happens as $m\to\infty$ because the improvement
is built from higher-subtracted sum rules.
To summarize, the key regimes are:
\be\begin{aligned}
\beta\gg \frac{1}{M},\ m\ {\rm fixed}:\ & C^{\rm improvement} \mbox{ dominates}\,, \\
\beta\ {\rm fixed},\ m\gg M:\ & C^{\rm unimproved} \hspace{0.06in}\mbox{ dominates}\,.
\label{schematic limits}
\end{aligned}\ee
Together, these regions cover all cases not already covered by the bulk-point regime $\beta\sim \frac{1}{M}$, see Figure~\ref{fig:wadsregions}.

\tikzmath{\oneOverM=1.3;\M=1.5;\betaLLOne=5;\mMax=6;\betaMax=9;\mGGM=3.0;\mMinBetaMax=\M+0.3;\tGapAngle=183;}

\begin{figure}
\centering
\begin{tikzpicture}
\draw[thick,->] (0,0) -- (0,\mMax);
\draw[thick,->] (0,0) -- (\betaMax,0);
\draw[gray,dashed] (\oneOverM,0) -- (\oneOverM,\mMax);
\draw[gray,dashed] (\betaLLOne,0) -- (\betaLLOne,\mMax);
\draw[thick] (0,\M) to[out=2,in=\tGapAngle] (\betaMax,1.8);
\node [below] at (\oneOverM,0) {$\beta\sim 1/M$};
\node [below] at (\betaLLOne,0) {$\beta \ll 1$};
\node [above] at (0,\mMax) {$m$};
\node [right] at (\betaMax,0) {$\beta$};
\node [right] at (\betaMax,\mMinBetaMax) {$\tau>\Delta_\mathrm{gap}$};
\draw [very thick,blue] (\oneOverM,\mMax) -- (\oneOverM,\M+0.03) to[out=2,in=\tGapAngle] (\betaMax,\mMinBetaMax);
\draw [very thick,red] (\betaMax,\mGGM) -- (0,\mGGM) -- (0,\mMax);
\draw [fill=red,opacity=0.2,red] (0,\mMax) -- (0,\M) -- (0.8,\M+0.02) to[out=31,in=-148] (\betaMax-0.2,\mMax);
\draw [fill=blue,opacity=0.2,blue] (\betaMax,\mMinBetaMax) to[out=\tGapAngle,in=2] (0,\M) -- (0,\M+0.4) to[out=21,in=-148] (\betaMax-0.25,\mMax) -- (\betaMax,\mMax);
\node [left] at (0,\M) {$M$};
\node [left] at (0,\mGGM) {$m\gg M$};
\path [pattern=north east lines, pattern color=black, opacity=0.4] (0,\M) -- (0,\mMax) -- (\betaLLOne,\mMax) -- (\betaLLOne,1.66) to[out=185,in=2] (0,\M);
\node [red!50!black] at (6.3,5.4) {$\omega_\mathrm{AdS}^\mathrm{unimproved}$};
\node [blue!70!black] at (7.5,4.0) {$\omega_\mathrm{AdS}^\mathrm{improvement}$};
\draw [red!18!white,fill=red!18!white,opacity=0.7] (2.15,3.65) -- (2.95,3.65) -- (2.95,4.13) -- (2.15,4.13)-- cycle;
\node [black!80!white] at (2.57,3.9) {flat};
\end{tikzpicture}
\caption{\label{fig:wadsregions}A cartoon of the regions used in the proof of positivity of $\omega_\mathrm{AdS}$. In the blue-shaded region $\omega_\mathrm{AdS}^\mathrm{improvement}$ dominates; in the red-shaded region $\omega_\mathrm{AdS}^\mathrm{unimproved}$ dominates; in the purple overlap they are comparable. The gray-hatched region is the bulk-point regime, where positivity follows from agreement with $\omega_\mathrm{flat}$. To establish positivity of $\omega_\mathrm{AdS}$, we will show positivity of $\omega_\mathrm{AdS}^\mathrm{improvement}$ in the region $\beta \gg 1/M$ bounded by the solid blue lines, and positivity of $\omega_\mathrm{AdS}^\mathrm{unimproved}$ in Regge region $m\gg M$ bounded by the solid red lines.
}
\end{figure}

For reasons that will become clear shortly, to prove positivity of $\omega_\AdS$, we will need to assume
that $f(p)$ is sufficiently regular near $p\to 1$:
\begin{align} \label{eq:thingswemustassume}
f(p) \sim (1-p)^l + \ldots\quad(p\to 1)\,, \qquad \mbox{with }l>n+1-\frac{d}{2}\,,
\end{align}
where $n$ is an exponent controlling the $p\to 0$ limit: $f(p)\sim p^n$.
It is straightforward in practice to find flat-space functionals obeying these conditions, see for example~(\ref{g2_g3_functionals}).\footnote{Based on experience with numerics, we expect that the conditions (\ref{eq:thingswemustassume}) do not affect the optimal bounds that would be obtained by scanning over an infinite-dimensional space of functions $f(p)$.}
In addition, we assume that $\omega_\mathrm{flat}[m,J]$ is everywhere strictly positive. This can be ensured by performing a small deformation of a nonnegative functional. The example functionals~(\ref{g2_g3_functionals}) are strictly positive.

\subsubsection{Positivity of the improvement term}

We begin with the contribution to \eqref{omega AdS f} from $C_{2,\nu}^{\rm improvement}$.  Since this involves
derivatives around the forward limit, the analysis is similar to that in subsection \ref{sec:derivativesofforward} and relies
on the matching region $\frac{1}{M}\ll \beta\ll 1$.

The contribution of improvement terms in flat space is
\begin{align}
\omega_\mathrm{flat}^{\rm improvement}[m,J] &= A(m) \cP_J'(1)+ B(m)\cP_J(1)\,,
\label{eq:bulkpointexpression}
\end{align}
where
\begin{align} \label{AB}
A(m) &\equiv \int_0^1 dp f(p) \frac{4M^6p^6}{m^6(m^4-M^4p^4)}\,,\nn\\
B(m) &\equiv -\int_0^1 dp f(p) \frac{M^4p^4(4m^2-3M^2p^2)}{m^6(m^2-M^2p^2)^2}\,,
\end{align}
and
\begin{align}
\cP_J(1)=1\,,\quad \cP_J'(1)=\frac{\cJ^2}{d-1},\quad\cJ^2=J(J+d-2)\,.
\end{align}
At large $J$ with fixed $m$, the term proportional to $\cP_J'(1)$ overwhelms all other terms in $\omega_\mathrm{flat}$. Since $\omega_\mathrm{flat}$ must be positive at large $J$, it follows that $A(m)\geq 0$ for all $m\geq M$. By a small deformation of $f(p)$, we can assume that $A(m)$ is strictly positive.

Since we already showed in section~\ref{sec:derivativesofforward} that the extension of $\cP_J'(1)$ into AdS is positive at all impact parameters,
we must only show is that it retains its dominance over $\cP_J(1)$.
In fact, already in flat space, there always exists a minimum impact parameter $b_0\sim \frac{1}{M}$ beyond which $A(m)\cP_J'(1)$ uniformly dominates over $B(m)\cP(1)$ for all $m$.
To see this, note that the dimensionless ratio $\frac{-M^2B(m)}{m^2 A(m)}$ is bounded from above for all $m\geq M$,
as can be argued by studying its limits as $m\to M$ and $m\to \oo$, and using that $A(m)$ is strictly positive.
Thus, the following quantity exists:
\begin{align}
\label{eq:defofbnought}
M^2b_0^2 &\equiv \max_{m>M} \max\p{0,-4(d-1)\frac{M^2 B(m)}{m^2 A(m)}}\,.
\end{align}
In flat space, the improvement contribution is strictly positive for all $b> b_0$.

\begin{figure}
\centering
\includegraphics[width=0.52\textwidth]{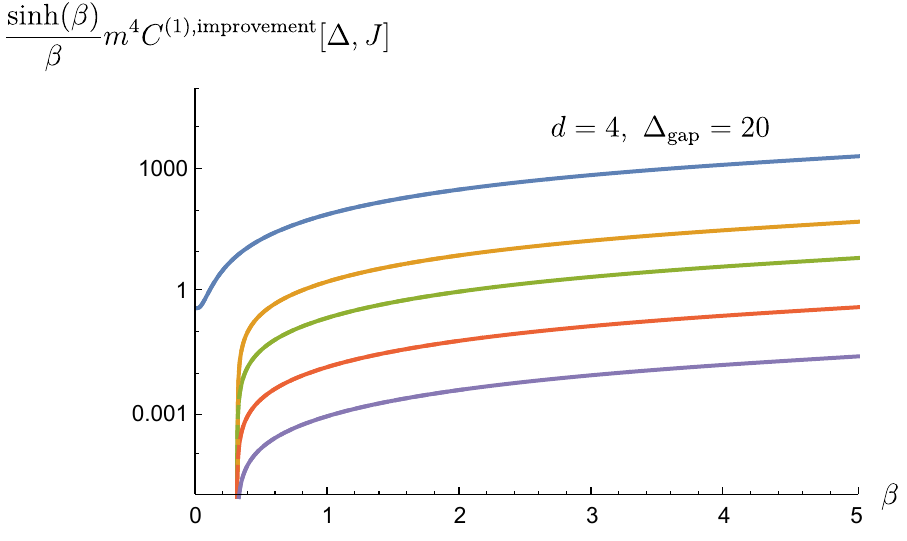}
\includegraphics[width=0.47\textwidth]{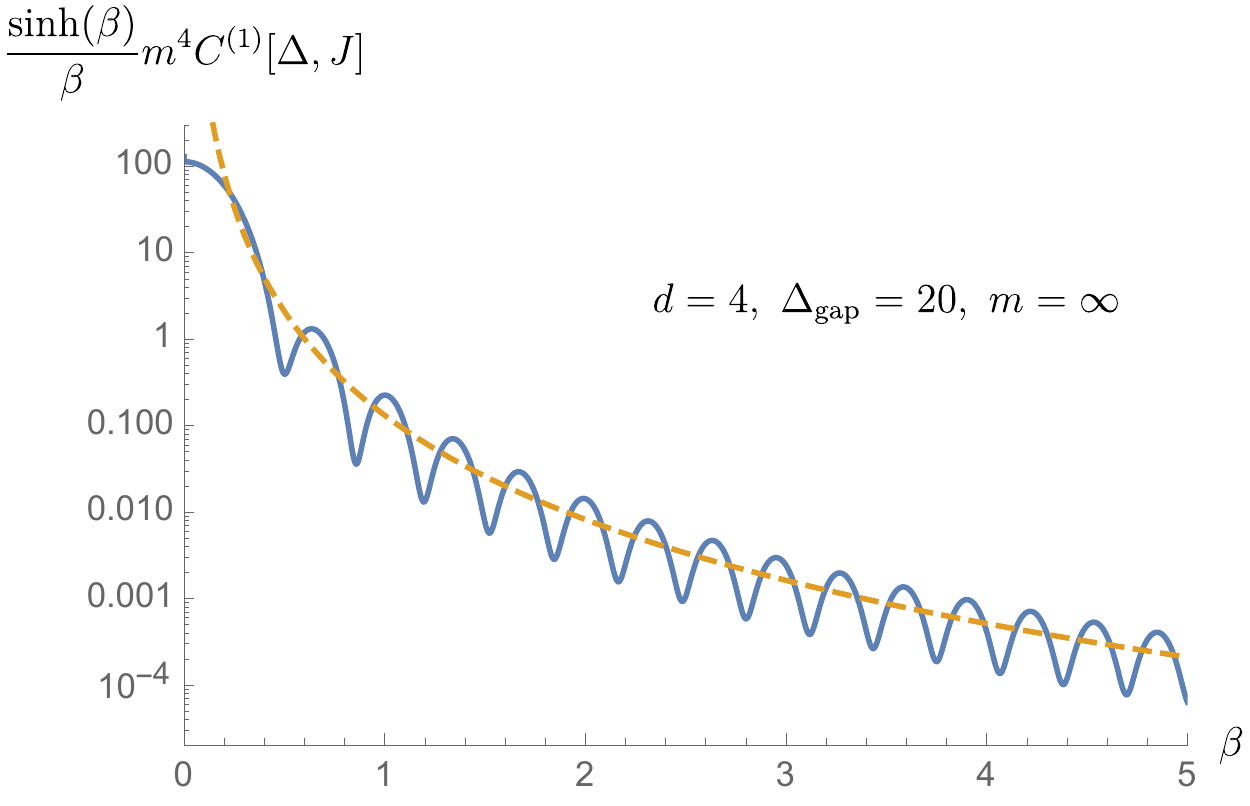}
\caption{Contributions from the improvement (from top to bottom: $m/\Delta_{\rm gap}=1.01, 3,5,10,20$) and unimproved (with $m=\infty$) terms for the AdS version of the flat space functional ${\cal C}^{(1)}$ in \eqref{g2_g3_functionals}. At large impact parameters $\beta\gg 1/\Delta_{\rm gap}$,
the total functional $\omega^{(1)}_{\rm AdS}[\Delta,J]$,
is well approximated by the sum of the two panels
and is therefore positive (and qualitatively similar to Figure~\ref{fig:functionals}).
The dashed line on the right panel shows the $\mathcal{Q}_3$ contribution from formula \eqref{eq:expansionofQ}.
The width of the peak in (b) is proportional to $1/\Delta_{\rm gap}$.
\label{fig:positivity 4d}
}
\end{figure}

Now consider the Regge limit of $m\gg 1$ with fixed $\beta=\cosh^{-1}\etaAdS$. Using (\ref{eq:bknuinreggelimit}) and (\ref{eq:definitionofk}) we have
\begin{align}
\label{eq:reggeexpression}
\omega_\mathrm{AdS}^{\rm improvement}[\Delta,J] &=
A(m)(-\tfrac {m^2}{2}\ptl_{\nu^2}) \cP_{\frac{2-d}{2}+i\nu}(\cosh \beta)|_{\nu=0} +
B(m)\cP_{\frac{2-d}{2}}(\cosh \beta) + O(1/m^2)\,,
\end{align}
where $A(m)$ and $B(m)$ are the same as in flat space \eqref{AB}.
Similarly to \eqref{overlap comparison}, this smoothly matches onto \eqref{eq:bulkpointexpression} in the overlapping regime $\frac{1}{M}\ll \beta\ll 1$,
which always contains the minimum impact parameter $b_0$ at which the $A(m)$ term takes over.
The proof is completed by noting, by inspection, that the ratio
\begin{align}
\label{eq:theratiothingy}
\frac{(-\tfrac 12\ptl_{\nu^2})\cP_{\frac{2-d}{2}+i\nu}(\cosh \beta)|_{\nu=0}}{\cP_{\frac{2-d}{2}}(\cosh\beta)}
\end{align}
is a monotonically increasing function of $\beta$.\footnote{For example, in $d=4$ equation (\ref{eq:theratiothingy}) is equal to $\beta^2/24$.} Thus, once established, dominance of $A(m)$ can't be lost, 
and $\omega_\mathrm{AdS}^{\rm improvement}[\Delta,J]$ is positive for all $\beta>b_0$.

Figure \ref{fig:positivity 4d} shows this behavior, and also illustrates the fact that
while the improvement terms grow with spin, the unimproved contribution decays.
In the region $b<b_0$ where improvement is not positive on its own, positivity of the total functional
was already guaranteed by positivity in flat space.

\subsubsection{Positivity of the unimproved contribution in the Regge limit $m\to\infty$}
\label{sec:posunimprovedRegge}

We now analyze positivity of the unimproved contribution. We focus on the limit $m\to\infty$ where it is relevant
according to \eqref{schematic limits}.  Although positivity will \emph{not} always automatically follow from that in flat space,
we will find that it can always be ensured by a small deformation of the original functional $f(p)$.

We begin with bulk-point/flat-space expressions. Given $J\sim \frac{\beta m}{2} \gg 1$, we can use the impact parameter approximation (\ref{eq:impactparamlimitofgeg}) for $\cP_J(1-2\nu^2/m^2)$.
Writing $\nu=Mp$, we have
\begin{align}
\label{eq:theintegralc2}
\lim_{m\to\infty} m^4 \omega_{\rm flat}[m,J] &=
2\int_0^1 dp f(p) \tl J(M\beta p) \qquad (1/M\ll \beta\ll 1)\,,
\end{align}
where $\tl J(x)$ is the Bessel function defined in (\ref{eq:impactparamlimitofgeg}).
Here we do not distinguish $\omega$ from $\omega^{\rm unimproved}$, because the improvement contributions vanish
in this limit.

In the large-$\beta M$ limit, $\tl J(M \beta p)$ oscillates rapidly with frequency $M\beta\gg 1$. These oscillations exponentially damp contributions from the interior of the integration region $p\in(0,1)$. The integral is  then dominated by the endpoints $p=0$ and $p=1$. To determine their contributions, we expand the non-oscillatory part of the integrand around each endpoint and extend the integral to infinity. For example, if $f(p)\sim f_0 p^n$, the contribution near $p=0$ is
\begin{align}
\label{eq:nonoscicont}
m^4\omega_{\rm flat}[m,J]\big|_{p=0} &\to 2\int_0^\oo dp\, f_0\, p^n \tl J(M\beta p) = 2\frac{2^n \Gamma (\frac{d-1}{2}) \Gamma (\frac{n+1}{2})}{\Gamma (\frac{d-n-2}{2})}\frac{f_0}{(M\beta)^{n+1}}
\quad (1/M\ll \beta \ll 1)\,.
\end{align}
To determine the contribution near $p=1$, we use $f(p)\sim f_1(1-p)^l$ together with the large-$x$ expansion of $\tl J(x)$. Throwing away unimportant constants, we find
\begin{align}
\label{eq:oscillatoryconttoc2}
m^4\omega_{\rm flat}[m,J]\big|_{p=1}&\propto f_1\frac{\cos(\beta M - \frac{\pi (d+2l)}{4})}{(\beta M)^{\frac{d}{2}+l}} \x \p{1+O\p{\frac{1}{\beta M}}} 
\quad (1/M\ll \beta \ll 1)\,.
\end{align}
The oscillatory contribution (\ref{eq:oscillatoryconttoc2}) is subdominant to the non-oscillatory contribution (\ref{eq:nonoscicont}) in the large-$M$ limit, since we have assumed $l > n+1-\frac{d}{2}$ in (\ref{eq:thingswemustassume}).

\begin{figure}[ht!]
\centering
\includegraphics[width=0.75\textwidth]{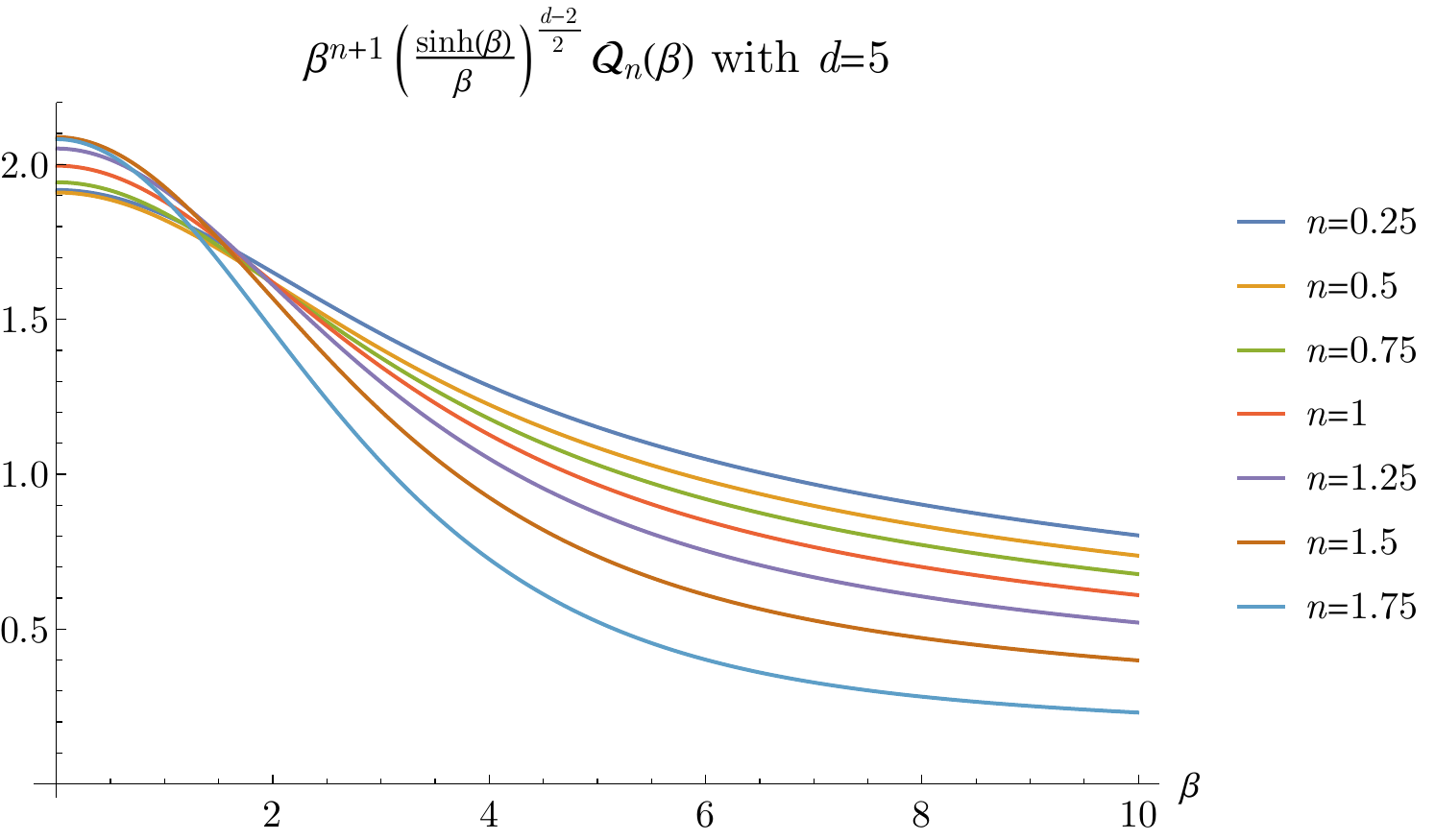}
\caption{The function $\beta^{n+1}(\frac{\sinh \b}{\b})^{\frac{d-2}{2}}\cQ_n(\beta)$ in $d=5$ for various values of $n$. In each case, it is everywhere positive and smoothly interpolates between the limits (\ref{eq:expansionofQ}).
\label{fig:qn}
 }
\end{figure}

The result (\ref{eq:nonoscicont}) is precisely the large impact parameter limit of the transverse Fourier transform $\widehat f(b)$ of $f(p)$, evaluated at $b=M\beta$. The function $\widehat f(b)$ is everywhere positive otherwise $\omega_\mathrm{flat}[m,J]$ would not be positive \cite{Caron-Huot:2021rmr}. This leaves two options:
\be
\mbox{either}\quad \frac{f_0}{\Gamma(\frac{d-n-2}{2})}>0
\quad\mbox{or}\quad n=d-2\,, \label{either or asymptotics}
\ee
where in the second case, a subleading term as $p\to 0$ must contribute with the correct sign
(its exponent should then be used for $n$ in \eqref{eq:thingswemustassume}). 
We have examples of flat-space functionals which are positive due to either of these mechanisms; for example,
the $d=4$ functional in \eqref{g2_g3_functionals} behaves like $f^{(1)}(p)\to 2280 p^2 - 5225 p^3+\ldots$ and it is the second term which
ensures positivity at large $b$.

The analysis of $\omega_{\rm AdS}[m,J]$ is similar: one simply substitutes $\tilde{J}(\beta p)\mapsto \cP_{\frac{2-d}{2}+i\nu}(\cosh\beta)$ following (\ref{eq:bknuinreggelimit}). Before proceeding, let us note that in certain spacetime dimensions
this substitution is trivial.  For example, in CFT${}_4$,
\be
 \tilde{J}(\beta p) = \frac{\sin(\beta p)}{\beta p} = \frac{\sinh(\beta)}{\beta} \cP_{-1+i\nu}(\cosh\beta) \qquad (d=4)\,.
\ee
Thus, in $d=4$ positivity at all $\beta$ is immediate.
In other cases, however, the limit turns out to be controlled by \emph{different} coefficients.

The functional $C_{2,\nu}[\Delta,J]$ is rapidly oscillating with $\nu$, so the integral over $\nu$
is again dominated by its endpoints at $\nu=0$ and $\nu=M$:
\begin{align}
\label{eq:cdeltajreggething}
\lim_{m\to\infty}m^4\omega_{\rm AdS}[\De,J] &\to 2\int_0^\oo \frac{d\nu}{M} f_0 \p{\frac{\nu}{M}}^n \cP_{\frac{2-d}{2}+i\nu}(\cosh\beta) + \textrm{osc.}  \qquad (1/M\ll \beta )\,.
\end{align}
The first term comes from the endpoint $\nu \sim 0$. To obtain it, we used $f(\nu/M)\sim f_0(\nu/M)^n$ and (\ref{eq:bknuinreggelimit}), which are both valid for $\nu \ll M$. The term ``osc.'' is an oscillatory contribution from the endpoint $\nu=M$. At large $\nu$, the Gegenbauer behaves like $e^{i\nu \beta}$ (see equation~\ref{eq:limitofgegenbauer}), which leads to an integral similar to (\ref{eq:oscillatoryconttoc2}) that is subleading to the first term in (\ref{eq:cdeltajreggething}), again assuming $l>n+1-\frac{d}{2}$.\footnote{Here, we had to assume $m\gg M>\nu$ in order to evaluate $\omega_\mathrm{AdS}[\De,J]$ using (\ref{eq:bknuinreggelimit}). In appendix~\ref{sec:oscregge}, we check that the oscillatory terms are not enhanced outside this regime, i.e.\ for $m\sim M$.}
Dropping the oscillatory terms, we thus have
\begin{align}
\label{eq:cintermsofq}
m^4\omega_{\rm AdS}[\De,J] &\to \frac{2f_0}{M^{n+1}} \cQ_n(\beta)\,, \qquad (1/M\ll \beta)\,,
\end{align}
where
\begin{align}
\label{eq:defofqn}
\cQ_n(\beta) &\equiv \int_0^\oo d\nu\, \nu^n \cP_{\frac{2-d}{2}+i\nu}(\cosh \b)\,.
\end{align}
We describe the function $\cQ_n(\beta)$ in more detail in appendix~\ref{app:qfunction}. There we show that it has the asymptotic behavior
\begin{align}
\label{eq:expansionofQ}
\p{\frac{\sinh \b}{\b}}^{\frac{d-2}{2}}\cQ_n(\beta)&\to
\begin{dcases}
\frac{2^n \Gamma (\frac{d-1}{2}) \Gamma (\frac{n+1}{2})}{\Gamma (\frac{d-n-2}{2})}
\frac{1}{\beta^{n+1}}
\p{1+
O(\beta^2)
}
 & \b \ll 1\,,
\\
\frac{
2^{\frac{d-2}{2}}
\Gamma (\frac{d-1}{2})\G(n)\sin(\tfrac{\pi n}{2})
}{
\sqrt{\pi } \Gamma (\frac{d-2}{2})
} 
\frac{1}{\beta^{n+\frac{d-2}{2}}}\p{1+ O(1/\b^2)}
& \b\gg 1\,.
\end{dcases}
\end{align}
An example plot of $\cQ_n(\beta)$ is shown in Figure~\ref{fig:qn}. We see that it smoothly interpolates between the two limits (\ref{eq:expansionofQ}) without crossing zero. By checking numerical examples, we find that $\cQ_n(\beta)$ is everywhere positive as a function of $\beta>0$ for $0\leq n\leq 2$ in $d\geq 4$, and for $0\leq n\leq 1$ in $d=3$.
Thus, if the leading term as $p\to 0$ is in this range, positivity in AdS follows from positivity in flat space.

\begin{figure}
\centering
\includegraphics[width=0.51\textwidth]{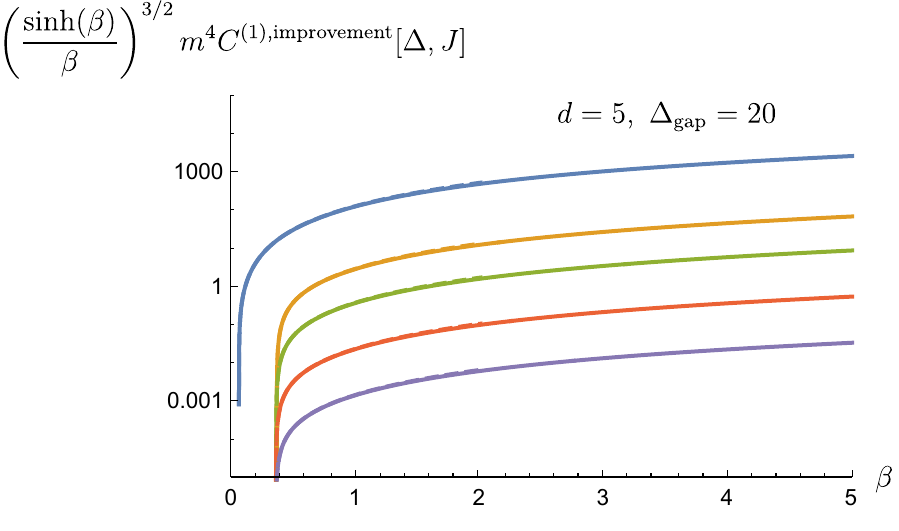}
\includegraphics[width=0.48\textwidth]{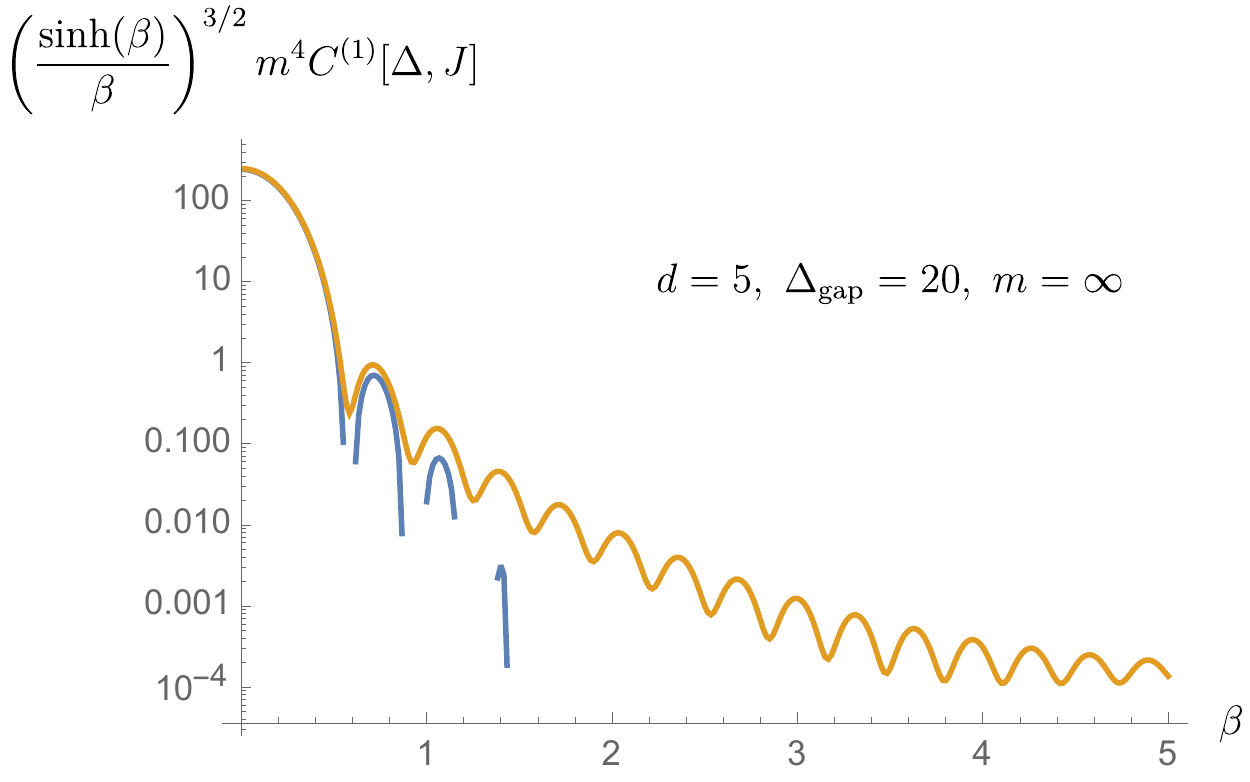}
\caption{Similar to Figure~\ref{fig:positivity 4d} but for the CFT${}_5$/AdS${}_6$ functionals
in eqs.~\eqref{f23 5d}. The left panel confirms that the improvement terms are again
positive on their own for $\beta>b_0$; dashed lines, almost identical,
show the flat space approximation of the functionals for $\beta<2$.
The discontinuous curve on the right panel shows that the Regge limit of the original functional
is not positive (negative regions show as missing parts in the logarithmic plot),
but becomes positive upon including the small perturbation \eqref{f23 5d shifted}.
\label{fig:positivity 5d}
}
\end{figure}

What if this is not that case?  As a concrete example, consider the following functionals, which are analogous
to \eqref{g2_g3_functionals} in $d=5$ ($D=6$):
\be\begin{aligned} \label{f23 5d}
f^{(1),d=5} &\equiv p^2(1-p^2)^2\left[ 30 + 12516 p - 18690 p^2 + 4345 p^3\right],\\
f^{(2),d=5} &\equiv p^2(1-p^2)^2\left[ 63 + 46690 p - 152082 p^2 + 115749 p^3\right]\,.
\end{aligned}\ee
In flat space, these are strictly positive and establish the two-sided bound
\be
-8.15\frac{g_2}{M^2} -28.8 \frac{8\pi G}{M^4} \leq g_3 \leq 3\frac{g_2}{M^2} +93.0\frac{8\pi G}{M^4} \qquad (D=6)\,.
\label{g2_g3_simple_bounds d=5}
\ee
The large-$b$ limit in flat space is controlled by the coefficient of $p^2$, which is positive. 
In AdS, however, the large-$\beta$ limit is controlled by the coefficient of $p^3$, which has the wrong sign:
the functionals $\omega_{\rm AdS}$ from \eqref{f23 5d} are \emph{not} positive!
However, since this problem occurs only for $\beta\sim 1$, where the Regge limit of the functional is very small in magnitude (the Regge limit of optimal functionals in flat space tend to be sharply localized around $b\sim 1/M$), we find that this can be easily fixed by adding a $p^{3/2}$ term:
\be \label{f23 5d shifted}
f^{(1),d=5} \mapsto f^{(1),d=5} + 5000\frac{p^{3/2}(1-p)^2}{M^{3/2}}\,,\qquad
f^{(2),d=5} \mapsto f^{(2),d=5} + 20000\frac{p^{3/2}(1-p)^2}{M^{3/2}}\,.
\ee
Despite the large-looking coefficients, these are actually small changes to eqs.~\eqref{f23 5d}.
The added terms have a negligible impact $\sim \frac{1}{M^{3/2}}$ on the bounds \eqref{g2_g3_simple_bounds d=5}.
They also do not spoil positivity elsewhere: they have a negligible impact on the bulk-point regime as well as on the improvement terms in \eqref{eq:reggeexpression}.
Positivity of the resulting functional is shown in Figure~\ref{fig:positivity 5d}.

We conclude that, for flat space functionals whose Regge limit is localized at small $b$, the large-$\beta$ regime in AdS is either automatically positive, or it can be corrected by a small perturbation.

\subsubsection{Bounds with maximal supersymmetry}

In this paper we have focussed on scalar correlators. Our arguments rely on a physical picture that we expect to go through straightforwardly  for spinning correlators, but as always, supersymmetry provides a shortcut.
With maximal supersymmetry, one can access   stress tensor four-point correlators in terms of scalar correlators with improved Regge behavior, in close analogy with  $2 \to 2$ graviton scattering in flat space \cite{Caron-Huot:2021rmr}.
Let us focus on the paradigmatic case of ${\cal N}=8$ $AdS_5$ gauged supergravity, which captures all tree-level interactions of the supergraviton multiplet of type IIB supergravity on $AdS_5 \times S^5$.
Superconformal Ward identities \cite{Eden:2000bk, Nirschl:2004pa} can be solved to express 
 the stress tensor four-point function in terms on an auxiliary scalar correlator. The corresponding 
 scattering amplitude (which is also simple in Mellin space \cite{Rastelli:2016nze, Binder:2019jwn}) is:
\be
{\cal M}_{\rm flat} =  \frac{ 8 \pi G}{stu} + \alpha + \dots
\ee
The  constant $\alpha$  corresponds to the leading ${\cal R}^4$ higher curvature
correction. We can now uplift to AdS$_5$ the flat space analysis on our previous paper~\cite{Caron-Huot:2021rmr} (section 3.8). Running the flat space numerics for $D=5$, we find a precise upper bound with the expected scaling,
\be
0\leq \frac{\alpha}{8 \pi G} \leq  \frac{10.9}{\Delta_{\rm gap}^6} + O(\frac{1}{\Delta_{\rm gap}^8}) \,.
\ee
In this case, we also have 
 a purely field-theoretic calculation~\cite{Gromov:2009zb} of the higher-spin gap,
 $\Delta_{\rm gap} \cong 2 \lambda^{1/4}$, where $\lambda = g_{\rm YM}^2 N \gg1$ is the 't~Hooft coupling of planar ${\cal N}=4$ SYM theory.
All in all, this establishes bulk locality  in the canonical example of  holographic duality, with a sharp bound that depends on the coupling $\lambda$.

\subsection{Comments on truncation errors}
\label{sec:lightandadsbounds}

Here we provide additional justification for the technical assumption \eqref{technical assumption},
that $O(\sum g/R_{\rm AdS}^2)=O(1/R_{\rm AdS}^2)$. The reader content with this
assumption can safely skip to the next subsection.

Each heavy-positive functional we have found places bounds on the bulk EFT, since the light contribution from states with $\tau \leq \De_\mathrm{gap}$ can be computed from the bulk EFT.
(From a purely CFT perspective, this is equivalent to including only the light state contribution to dDisc
in the Lorentzian inversion formula, then accounting for the heavy contribution by adding
a series of ambiguities with finite support in spin \cite{Caron-Huot:2017vep}.
The later are known to be in one-to-one correspondence with bulk local interactions \cite{Heemskerk:2009pn}.)

By design, the light contribution is given, to leading order as $R_\mathrm{AdS}\to \oo$, by a finite linear combination of bulk couplings (up to loop corrections).
However, unlike in flat space, there are additional finite-$R_\mathrm{AdS}$ corrections (even at tree-level),
and we would like to understand them.
These take the schematic form \eqref{null constraint CFT}:
\begin{align} \label{schematic gap}
\frac{1}{\Delta_{\rm gap}^2} \sum_{n,\bullet} c_{n,\bullet} g_{n,\bullet}\Delta_{\rm gap}^{2n-4}
\end{align}
where the coefficients $c_n$ are of order unity and the exponent of $\Delta_{\rm gap}$
conforms with the expectations of bulk dimensional analysis with $M=\Delta_{\rm gap}$;
the index $\bullet$ labels different contact interactions with the same scaling dimension.
As argued below \eqref{null constraint CFT}, since we have two-sided inequalities from flat space which
control \emph{each} EFT coefficient (to leading order in $1/\Delta_{\rm gap}$), we could substitute in those inequalities to eliminate any finite number of $g_{n,\bullet}$, at the cost of introducing new terms order $1/\Delta_{\rm gap}^4$. Here we would like to argue that this process seems convergent.

To deal with this in practice, we afford ourselves breathing room by lowering the cutoff slightly:
$M^2\mapsto (1-\epsilon)M^2$.  Since the coefficients in the flat space bounds for $g_{k,\bullet}$ do not grow
(the flat space EFT series is proven to converge for $|s|, |t|, |u|<M^2$ \cite{Caron-Huot:2020cmc}),
any contribution from a weight $k$ coupling will be weighted by $(1-\epsilon)^k$.  Now, the coefficients of $1/\Delta_{\rm gap}^{2j}g_{n,\bullet}$ in \eqref{schematic gap} could potentially
grow with $n$, but we only expect a power-law growth of the form $n^j$.

Intuitively, the reason our null constraints are not exact in CFT, and why \eqref{schematic gap} arises at all,
is that different sum rules measure coefficients at slightly values of Mellin variables
(ie. $\mS=2\Df$ versus $\mS=2\Df+2$ for subtracted sum rules).
These values are not very friendly to crossing symmetry:
Mellin-space dispersion relations may not converge at those values of $\mT$ \cite{Penedones:2019tng}.
Yet, the difference between all these evaluations are expected to be small,
they are effectively derivatives of monomials like $\mS^n$, whence the power dependence on $n$.

We thus estimate that
\be
\mbox{contribution from } \frac{g_{n,\bullet}}{\Delta_{\rm gap}^{2j}} \sim \frac{n^j (1-\epsilon)^n}{\Delta_{\rm gap}^{2j}} \quad
\mbox{at large $n$.}
\ee
Thus, even though we have formally infinitely many couplings to eliminate, the sum over all $n$ won't exceed
$\sim \frac{1}{(\epsilon \Delta_{\rm gap}^2)^j}$. As long as we choose $\epsilon$ sufficiently large
(so that $\epsilon \Delta_{\rm gap}^2\gg 1$) the procedure appears to converge.

The technical weakness of this argument is that EFT couplings $g_{n,\bullet}$ with $n$
too large may not be fundamentally defined in CFT.  For example, the $C_k$ sum rules
require evaluations at Mellin variables up to $\mS=2\Df+k-2$, so at least if using the tools in this paper,
it is hard to define couplings with $n\gtrsim \Delta_{\rm gap}$.
This is also why we introduced a cutoff $k_{\rm max}$ in the improved sum rules \eqref{eq:definitionofk}.\footnote{
The formulas of the preceding sections also become inaccurate at large $k$. 
For example, terms in the Regge series \eqref{regge subleading} go like $\frac{k^3}{m^2}$ at large $k$.}
On the other hand, given that, for sufficiently large $\epsilon$, eliminating those couplings changes our sum rules by a negligible amount, this would seem to be more of a technical annoyance rather than a real loophole to the idea of dimensional analysis scaling --- whence the technical assumption \eqref{technical assumption}.

A rigorous proof of \eqref{technical assumption}, for example
by bounding at once truncated tails as opposed to bounding individual terms,
is left to the future. This might be simplified by using exact CFT null constraints.

\subsection{Resolution of infrared divergences in CFT$_3$/AdS$_4$}

Dispersive bounds in $D{=}4$-dimensional flat space suffer from IR divergences in the presence of gravity.  To see why, consider integrating $\cC^\imp_{2,u}$ against a wavefunction $f(p)$, producing the sum rule
\begin{align}
\label{eq:dispersivesumruled4}
\int_0^1 dp f(p) \p{\frac{8\pi G}{(M^2 p^2)} + 2g_2 + g_3 M^2p^2} &= \int_0^1 dp f(p) \<\cC^\imp_{2,-M^2 p^2}[m,J]\>\,.
\end{align}
To place bounds on the left-hand side, we should search for $f(p)$ such that
\begin{align}
\label{eq:positivityonf}
\int_0^1 dp f(p) \cC^\imp_{2,-M^2 p^2}[m,J] &\geq 0 \qquad \forall m\geq M,J=0,2,\dots\,.
\end{align}
As usual, (\ref{eq:positivityonf}) implies that the transverse Fourier transform $\widehat f(b)$ is positive. The complication in $D{=}4$ dimensions is that positivity of $\widehat f(b)$ implies
\begin{align}
\label{eq:positivityatpzero}
0 <  \int d^2 \vec b\, \widehat f(|\vec b|) = \lim_{p\to 0} \frac{f(p)}{p}\,,
\end{align}
which means that the integral of $f(p)$ against the graviton pole in (\ref{eq:dispersivesumruled4}) is infinite. In other words, there is a fundamental tension between a finite action on graviton exchange and positivity in impact parameter space.
Of course, this simply reflects that the Regge limit of tree-level graviton-exchange in impact parameter space is infrared divergent.

To proceed, we must relax one of these two conditions.
The strategy in \cite{Caron-Huot:2021rmr} is to first allow the action on gravity to be divergent;
one can then find positive functionals, for example
\begin{align} \label{eq:examplefunctionalfourd}
f^{d=3}(p) &= p(1-p)^2 (1310+1540 p-10318 p^2-4697 p^3+15680 p^4)\,.
\end{align}
The leading term as $p\to 0$ is positive, in agreement with \eqref{eq:positivityatpzero}.
Another crucial property is that the first subleading term is negative:
\begin{align}
f^{d=3}(p) \to 1310 p - 1080 p^2 + \ldots \equiv a_1p + a_2p^2+\ldots\,.
\end{align}
This is necessary because we are in the second case of \eqref{either or asymptotics}: $n=d-2$,
so the large-$b$ flat-space asymptotics controlled by the subleading term:
\begin{align}
\widehat f(b) &= \int_0^1 dp\, f(p) J_0(b p)
\ \stackrel{b\gg 1}{\longrightarrow}\ 
\int_0^\oo dp\,(a_1 p + a_2 p^2 + \dots) J_0(b p) = -\frac{a_2}{b^3} + \dots\,.
\label{eq:largebcomputation}
\end{align}
The fact that the $a_1p$ term integrates to zero is an interesting quirk about Bessel functions.
Oscillatory contributions from $p\to 1$ vanish like $\sim 1/b^4$ thanks to the $(1-p)^2$ factor and can be ignored,
see \eqref{eq:oscillatoryconttoc2}.

Although $f(p)$ is positive, its action on gravity diverges. To get a meaningful bound,
ref.~\cite{Caron-Huot:2021rmr} then imposed a low-momentum cutoff $p_\mathrm{min}$ and considered
the regulated functional
\begin{align}
\w_{p_\mathrm{min}} &\equiv \int_{p_\mathrm{min}/M}^1 dp\, f(p) \cC_{2,-M^2 p^2}^\imp\,.
\end{align}
This is a well-defined functional with a finite
action on gravity (albeit logarithmically sensitive on $p_{\rm min}$),
but it violates positivity for $b$ larger than an IR cutoff $b_\mathrm{max}$.
We can determine $b_\mathrm{max}$ by studying the large impact parameter limit of $\omega_{p_\mathrm{min}}$:
\begin{align}
\int_{p_\mathrm{min}/M}^1 dp\,\p{a_1 p + a_2 p^2 + \dots} J_0(M b p) &= -\frac{a_1 p_\mathrm{min}^2}{2M^2} - \frac{a_2}{(Mb)^3} + \dots\,.
\label{pmin large b}
\end{align}
Locating where the two terms exchange dominance, one finds the curious scaling
$p_\mathrm{min}^2 = \frac{a_2}{a_1(Mb_\mathrm{max}^3)}$. Numerically, the resulting bound is:
\be \label{g2 bound 4D}
g_2 \geq -17.64\x\frac{8\pi G}{M^2}\log\frac{0.19M}{p_\mathrm{min}}\,.
\ee
This bound is only rigorous if one assumes additional information about the spectrum at large impact parameters.

We will now show that the same functional, in CFT, proves a rigorous bound with the more intuitive
scaling $p_{\rm min}\sim \frac{1}{R_{\rm AdS}}$.
By contrast with flat space, AdS possesses a built-in IR regulator, which automatically cures the IR problems described above! 

Consider the functional
\begin{align}
\w_\mathrm{AdS} &= \int_0^M \frac{d\nu}{M} f^{(d=3)}\p{\frac{\nu}{M}} C^\imp_{2,\nu}\,.
\end{align}
Finiteness of the action on gravity follows from the modified form of the graviton propagator in AdS:
\begin{align} \label{ads4 graviton}
-\w_\mathrm{AdS}|_\mathrm{light} &=\int_0^M\frac{d\nu}{M} f^{(d=3)}\p{\frac{\nu}{M}}
\left[\frac{8\pi G}{\nu^2+(d/2)^2} +2g_2 +g_3\nu^2\right]  \qquad (M\gg 1)\,,
\end{align}
where we used (\ref{eq:BknuGravity}) for the contribution of graviton exchange.
The zero-momentum logarithmic divergence has been cut-off at the inverse AdS radius.

We claim that $\w_\mathrm{AdS}$ is rigorously positive at large $M$. 
By construction, it is positive in the region of overlap between the bulk point and Regge limits $1/M\ll \beta \ll 1$.
For $1/M\ll \beta$ we split $\omega_\mathrm{AdS}$ into an improvement term and non-improvement term as before. The improvement term is positive, and the non-improvement term in the Regge regime gives
\begin{align}
\label{eq:twotermsimpactads}
\lim_{m\to\infty} m^4 \w_{\rm AdS}[\De,J] &\propto \frac{a_1\cQ_1(\beta)}{M^2} + \frac{a_2\cQ_2(\beta)}{M^3} + \dots\,,\qquad (1/M\ll \beta)\,,
\end{align}
where ``$\dots$'' represents subleading corrections at large $M$. 

Recall that in flat space, the term proportional to $a_1$ vanished at large impact parameter, due to the identity (\ref{eq:largebcomputation}). The consequence for AdS$_4$ is that $\cQ_1(\beta)$ is smaller than naively expected at small $\b$:
in the flat space regime, $\cQ_2$ dominates.
(Generically $\cQ_n(\beta)\sim1/\beta^{n+1}$ at small $\beta$, but $\cQ_1(\beta)$ approaches a constant as $\beta\to 0$.) 

\begin{figure}
\centering
\includegraphics[width=0.8\textwidth]{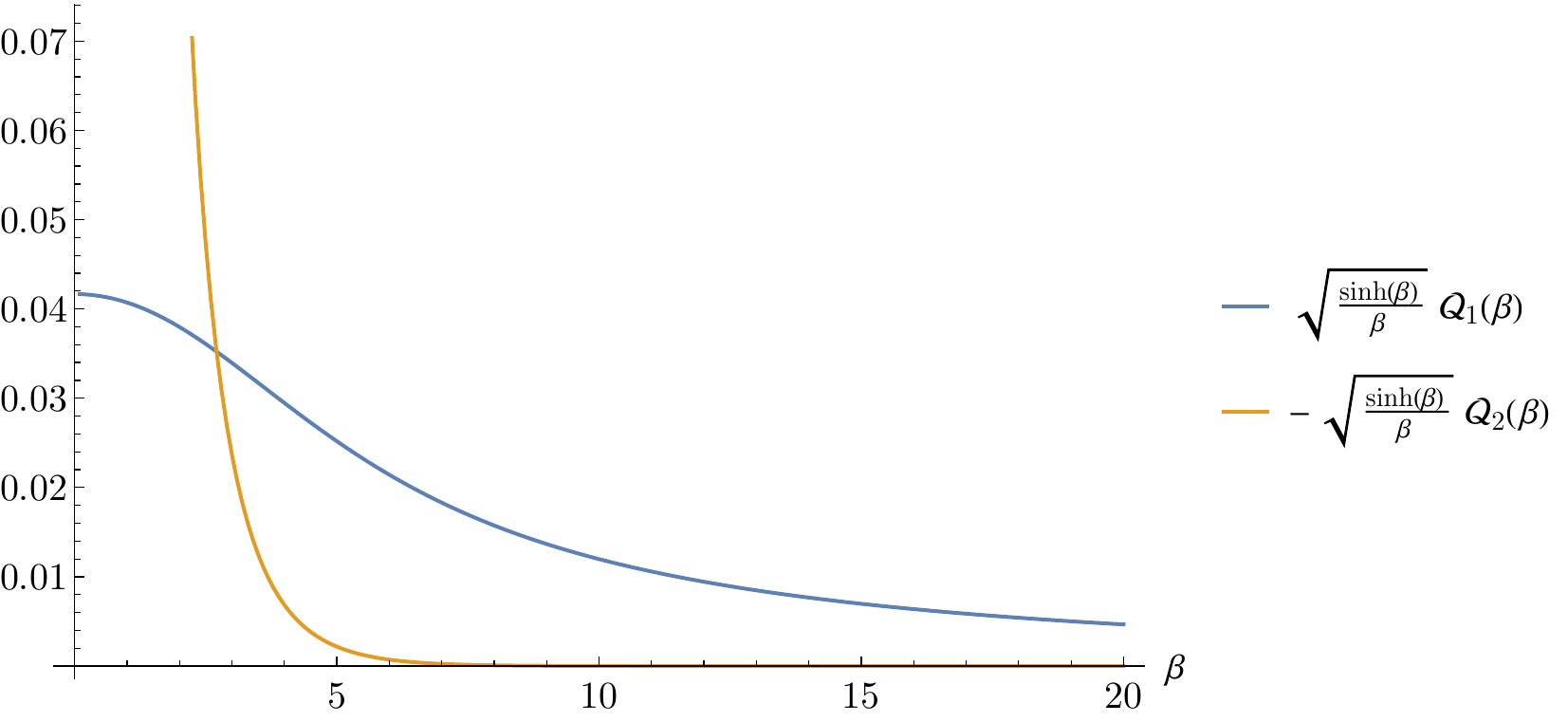}
\caption{The functions $\cQ_1(\beta)$ and $-\cQ_2(\beta)$, multiplied by a standard factor $(\sinh\b/\b)^{1/2}$. Both are everywhere positive. $-\cQ_2$ dominates \eqref{eq:twotermsimpactads}
in the flat space regime $\beta\ll 1$ but $\cQ_1$ takes over at large $\beta$.
\label{fig:q1q2plot}}
\end{figure}

However, importantly, $\cQ_1(\beta)$ is \emph{everywhere positive}.
This is the crucial distinction between AdS${}_4$ and the momentum cutoff in \eqref{pmin large b}: the latter goes negative
at large $b$, but the former stays positive!  The functional $\w_{\rm AdS}$ thus remains positive
even after $\cQ_1$ and $\cQ_2$ exchange dominance
(at large $\beta$, $\cQ_2(\beta)$ is exponentially smaller than $\cQ_1(\beta)$).
It follows that $\w_\mathrm{AdS}[m,J]$ is positive for $1/M\ll \beta$ at sufficiently large $M$.
We plot the functions $\cQ_1(\beta)$ and  $-\cQ_2(\beta)$ for $d=3$ in Figure~\ref{fig:q1q2plot}.

In conclusion, any positive functional in flat space, with logarithmically divergent action on gravity,
will uplift to an AdS functional where the graviton propagator is simply
given an effective mass as in \eqref{ads4 graviton}.
For example, the functional (\ref{eq:examplefunctionalfourd}) gives the bound \eqref{g2 bound 4D} with
$p_{\rm min}=\frac{3}{2}\frac{1}{R_\mathrm{AdS}}$: 
\begin{align} \label{eq:adsfourbound}
g_2 + O(1/\rAdS^2) &\geq -17.64 \frac{8\pi G}{M^2} \log(0.124\,M R_\mathrm{AdS}) \nn\\
 &= -4516 \frac{c_T^\textrm{free}}{\De_\mathrm{gap}^2 c_T} \log(0.124\, \Delta_{\rm gap})\,.
\end{align}
We also find other functionals giving lower and upper bounds on the coefficient of $stu$, $g_3$, for example
\be
-10.4g_2 -48.4 \frac{8\pi G}{M^2}\log(0.23\, M\rAdS)
\leq g_3 M^2 \leq 3g_2 + 207 \frac{8\pi G}{M^2}\log(0.11 M\rAdS)\,. \label{eq:ads4bounds 1}
\ee

The bounds (\ref{eq:adsfourbound}) and \eqref{eq:ads4bounds 1}
differ from the naive expectation from dimensional analysis by an additional logarithmic factor $\log(\Delta_{\rm gap})$. This factor appears because the IR divergences present in flat space reappear as one takes the AdS radius to infinity. If the flat-space divergences could be resolved using a purely flat-space mechanism, then it should be possible to remove the $\log$ term, obtaining stronger bounds.

\subsection{Bounds on anomalous dimensions}

In the above, we have derived constraints on low-energy physics in AdS. The constraints were expressed as bounds on various polynomial terms in the Mellin representation of the low-energy correlator. In this subsection, we will express these bounds in terms of physical observables, namely anomalous dimensions of light double-trace operators, denoted $\gamma_{n,\ell}$.

Recall that all the anomalous dimensions are small and of order $G$. We can measure $G$ from the coefficient of the stress tensor conformal block
\be
\cG(z,\zb) = f_{\phi\phi T_{\mu\nu}}^2 G_{d,2}(z,\zb) + \ldots\,,\qquad
f_{\phi\phi T_{\mu\nu}}^2 = 8 \pi  G \frac{\Df ^2\,\Gamma\!\left(\frac{d+2}{2}\right)^3}{2 \pi ^{\frac{d}{2}}(d-1) d\,\Gamma (d+2)}=\frac{d\Df^2}{4(d-1)}\frac{c^{\text{free}}_T}{c_T}\,.
\label{eq:gravityNormalization}
\ee
In the absence of higher-derivative contact diagrams, the anomalous dimensions $\gamma_{n,\ell}$ with $\ell\geq 2$ would be given by their value in the sum of graviton exchange Witten diagrams in the three channels. Let us denote this pure gravity value as $\gamma^{\text{pure}}_{n,\ell}$. We have for example
\ba
\gamma^{\text{pure}}_{0,2} &=8\pi G \left[-32 \Df ^3+16 (d-6) \Df ^2+8 (4 d-11) \Df+d^3+d^2 +10 d-24\right]\times\\
&\quad \frac{ \Gamma (\Df +1)^2 \Gamma \left(2 \Df-\frac{d}{2} +2\right)}{2 \pi ^{\frac{d}{2}} (d-1)d(d+2) (2 \Df +3)\Gamma (2 \Df +2) \Gamma \left(\Df-\frac{d}{2} +1\right) \Gamma \left(\Df-\frac{d}{2} +2\right)}\,.
\ea
The bounds on couplings derived in this paper constrain the deviation
\be
\delta\gamma_{n,\ell}=\gamma_{n,\ell}-\gamma^{\text{pure}}_{n,\ell}\,.
\ee
If we assume that all couplings have the expected EFT scaling, then for all $\ell \geq 2$ 
\be
\frac{|\delta\gamma_{n,\ell}|}{8\pi G} \leq O(\Dgap^{-2})\,.
\ee
At the leading order at large $\Dgap$, the difference $\delta\gamma_{n,\ell}$ comes from the four-derivative contact diagram, whose Mellin amplitude is $M(\mS,\mT) = g_2 c_2 (\mS^2+\mT^2+\mT^2)$, see section \ref{ssec:mellin}. This diagram contributes only to the $\ell=0,2$ anomalous dimensions. For example,
\be
\gamma_{0,2}|_{g_2} = -\frac{8 \Gamma (\Df +2)^2 \Gamma \left(2 \Df-\frac{d}{2} +2\right)}{\pi ^{\frac{d}{2}} \Gamma (2 \Df +4) \Gamma \left(\Df-\frac{d}{2} +1\right)^2}\,.
\ee
Now, in the above we have derived a lower bound on $g_2$
\be
\frac{g_2}{8\pi G} \geq  \frac{\alpha(d)}{\Dgap^2}\left[1+O(\Dgap^{-2})\right]\,,
\label{eq:g2BoundAdS}
\ee
where $\alpha(d)<0$ is the horizontal location of the left tip of the exclusion plot in Figure~\ref{fig:allowed g2g3} and its generalization to other $d$ \cite{Caron-Huot:2021rmr}. Since $\gamma_{0,2}|_{g_2}$ is always negative, this translates into an upper bound on $\gamma_{0,2}$
\be
\frac{\delta\gamma_{0,2}}{8\pi G} \leq \frac{-\alpha(d)}{\Dgap^2}
\frac{8 \Gamma (\Df +2)^2 \Gamma \left(2 \Df-\frac{d}{2} +2\right)}{\pi ^{\frac{d}{2}} \Gamma (2 \Df +4) \Gamma \left(\Df-\frac{d}{2} +1\right)^2}
\left[1+O(\Dgap^{-2})\right]\,.
\ee

Recall that we have also found a lower and an upper bound on $g_3$ at fixed $g_2$
\ba
\frac{g_3}{8\pi G} &\geq  \left[\frac{\alpha_{-}(d)}{\Dgap^2}\frac{g_2}{8\pi G} + \frac{\beta_{-}(d)}{\Dgap^4}\right]\left[1+O(\Dgap^{-2})\right]\\
\frac{g_3}{8\pi G} &\leq  \left[\frac{\alpha_{+}(d)}{\Dgap^2}\frac{g_2}{8\pi G} + \frac{\beta_{+}(d)}{\Dgap^4}\right]\left[1+O(\Dgap^{-2})\right]\,.
\label{eq:g3BoundsAdS}
\ea
$g_3$ can be measured from the combination of $\delta\gamma_{0,2}$ and $\delta\gamma_{1,2}$ which vanishes in the $g_2$ diagram
\ba
\delta\gamma_{1,2}-&\frac{(d+4) (\Df +2) (d-4 \Df -4)}{2 (2 \Df +5) (d-2 \Df -2)}\delta\gamma_{0,2} = \\
&g_3 \frac{16 (d+4) \Gamma (\Df +3)^2 \Gamma \left(2 \Df-\frac{d}{2} +4\right)}{\pi ^{\frac{d}{2}} \Gamma (2 \Df +6) \Gamma \left(\Df-\frac{d}{2} +1\right) \Gamma \left(\Df -\frac{d}{2}+2\right)}\left[1+O(\Dgap^{-2})\right]\,.
\ea
\eqref{eq:g3BoundsAdS} translate to an upper and lower bound on this combination in terms of $\delta\gamma_{0,2}$ and $\Dgap$. The contribution of pure gravity, $g_2$ and $g_3$ to the anomalous dimensions $\gamma_{0,2}$ and $\gamma_{1,2}$ needed to make these bounds fully explicit are recorded in Appendix \ref{app:anomalousDims}.

\subsection{Classical gravity and the Classical Regge Growth conjecture}\label{ssec:CRG}

The authors of~\cite{Chowdhury:2019kaq} formulated the Classical Regge Growth (CRG) conjecture,
to the effect that any causal classical S-matrix
cannot grow faster than $s^2$ in the Regge limit.
This was effectively proved (for gravity in AdS) in \cite{Chandorkar:2021viw}
by relating CFT correlators in the Rindler wedge and in scattering kinematics in the classical limit of AdS/CFT.

The present paper establishes a slightly stronger, nonperturbative result:
for any S-matrix that can be obtained from a flat space limit of AdS/CFT,
twice-subtracted dispersion relations hold. This effectively amounts to $\lim_{|s|\to\infty}|M(s,t)/s^2|=0$ along any complex $s$ direction, for $t<0$. The argument is that dispersive CFT sum rules with Regge spin-2 converge when acting on any physical correlator. Via~(\ref{eq:holographicbknu}), such sum rules become twice-subtracted flat-space dispersion relations in the flat-space limit $M\to\oo$.

To be fully precise, we established this result for amplitudes integrated against test functions that have compact support in momentum space $p$ and fast decay in impact parameters
(in the Regge limit). Such test functions suffice to prove all our bounds (in $D>4$)
and these properties explain why $\rAdS$ does not appear in the bounds. This could help address concerns raised in \cite{Mizera:2021fap}.

One interesting implication of the CRG conjecture is that (almost) any higher-derivative correction to Einstein gravity must be parametrically suppressed by the Compton wavelength of new higher-spin states. This was first demonstrated in \cite{Camanho:2014apa} for three-point self-interactions, then generalized in \cite{Chowdhury:2019kaq} to four-point graviton self-interactions (this constraints, for example, terms in the effective Lagrangian with up to four powers of the Riemann tensor as well as derivatives).
Roughly, this shows that in any causal classical theory of gravity, such as the Einstein-Hilbert action minimally coupled to matter of spin $2$ or less, birefringence disappears at high energies \cite{Camanho:2014apa}.
This is an important result: we interpret this as saying that
the causal structure of spacetime is unambiguous since it is the same for all high-energy particles.\footnote{A related quantum version of this statement, involving commutativity of shocks, is discussed in \cite{Kologlu:2019bco}.}

The Regge scaling in \cite{Chowdhury:2019kaq} left open one exceptional term: in spacetime dimensions $D\geq 7$, an interaction proportional to
\be \label{Mgggg}
 \mathcal{M}_{gggg} \supset \frac{1}{7!}C\ \big(\e_1\wedge \e_2\wedge \e_3\wedge \e_4\wedge p_1\wedge p_2\wedge p_3\big)^2\,,
\ee
where $h_i^{\mu\nu}=\e_i^\mu \e_i^\nu$ denotes the polarization tensor of the $i$-th graviton.  This structure (also called second Lovelock Lagrangian \cite{Chowdhury:2019kaq}) vanishes identically if $D<7$. Although it contains six derivatives, it only grows like $\sim s^2$ in the Regge limit. This makes it analogous to the $g_3stu$ scalar self-interaction which we discussed above. We thus believe that our techniques can provide sharp lower and upper bounds on all higher-derivative
couplings considered in \cite{Chowdhury:2019kaq}, including the coefficient $C$, in units of the $G$ and the Compton wavelength of heavy higher-spin particles. 

In fact, in some cases, no work is needed: choosing graviton polarizations as the  constant tensor $h_i^{\perp}=\frac12 {\rm diag}(1,1,-1,-1)$ in four transverse dimensions, one finds that \eqref{Mgggg} reduces to
a scalar amplitude in $D-4$ spacetime dimensions with
\be
 \mathcal{M}_{\perp\perp\perp\perp} \supset \frac{-3C}{8} stu\,.
\ee
This effective scalar amplitude has a graviton pole, and bounds on $g_3$ as in eqs.~\eqref{g2_g3_simple_bounds} or \eqref{eq:ads4bounds 1} thus directly apply to $C$.
Of course, using $(D-4)$-dimensional partial waves in this way is suboptimal (and only possible if $D\geq 8$), but this example illustrates that two-sided bounds on $C$ are provable.  The optimal bound are to be found by manipulating twice-subtracted S-matrix dispersions relations in $D$-dimensions.


\section{Conclusions}\label{sec:Conclusions}

In this work, we applied bootstrap methods to derive constraints on the low-energy dynamics of gravitational theories in anti-de Sitter space. We focused on the problem of bounding higher-derivative couplings in the low-energy effective Lagrangian in theories containing a light scalar weakly coupled to gravity. If the mass of the lightest non-EFT state is large in AdS units, i.e.~$\Delta_{\text{gap}} \gg 1$, it is expected that the bulk theory should be local down to distances of order $R_{\text{AdS}}/\Dgap$. This means that any higher-derivative bulk coupling of mass dimension $\delta<0$ should be suppressed by $\Dgap^{\delta}$.

Over the years, strong evidence in favor of this expectation has accumulated. However, previous arguments have only led to parametric bounds, meaning they give no control over the $O(1)$ coefficient in front of $\Dgap^{\delta}$. Furthermore, in some cases (such as direct arguments based on the Lorentzian inversion formula) the bounds scale with the Regge spin rather than the mass dimension of the bulk coupling. In the present work, we overcame these shortcomings and derived sharp bounds that exhibit the correct mass dimension scaling. These sharp bounds remove the last psychological obstacle in embracing the proof of bulk of locality from a CFT perspective. They also offer the hope that interesting theories may live at the boundary of the allowed region in parameter space; a similar 
serendipitous discovery \cite{ElShowk:2012ht} paved the way for  the success of the conformal bootstrap program~\cite{Poland:2018epd}. 

Our main tools were dispersive sum rules in conformal field theory. Applied to holographic CFTs, dispersive sum rules relate the bulk EFT couplings to contributions of states above the UV cutoff $\Dgap$. The UV contributions satisfy positivity constraints as a consequence of unitarity, leading to bounds on the EFT couplings.

The name ``dispersive'' means that states above $\Dgap$ enter the sum rule only through the double discontinuity of the four-point function. As such, dispersive CFT sum rules are an anti-de Sitter analogue of dispersive sum rules for the flat-space S-matrix, in which only the imaginary part of the amplitude enters at high energies.
A distinct advantage of the CFT case is that while S-matrix dispersion relations are not always rigorously established, CFT sum rules are a rigorous consequence of unitarity and the operator product expansion. Indeed, they can be thought of as particular projections of the standard crossing equation equating the $s$- and $t$-channel OPEs. As discussed in Section \ref{ssec:CRG}, our results support the validity of twice-subtracted sum rules for gravitational S-matrices.

Our central conclusion is simple: twice-subtracted S-matrix sum rules uplift
to rigorous CFT sum rules.  Sum rules that are positive for $m>M$ in flat space, remain so for
$\sqrt{\Delta^2-J^2}>\Delta_{\rm gap}$ in CFT.
This is nontrivial and requires looking at regions of large impact parameter. 
Since S-matrix sum rules have previously been constructed which bound generic higher-derivative corrections
to lengths $1/M$, this establishes sharp bounds in AdS effective theories, implying that bulk physics is local
down to the scale $\rAdS/\Delta_{\rm gap}$. In the process, we made a technical assumption \eqref{technical assumption} that infinite towers of higher-derivative couplings cannot conspire to produce contributions enhanced at large $\Dgap$. While we consider the converse possibility very unlikely, it would be nice to remove this assumption.

The $O(1)$ coefficients in the bounds we obtain are identical to the corresponding bounds in flat space.
Given that the relation between CFT and S-matrix sum rules proceeded through physically transparent saddle points, we expect that the same conclusion will remain valid for spinning correlators. For example, rigorous bounds on stress tensor correlators in a CFT with a large gap can now be obtained
by studying graviton S-matrices in flat space. It will be important to determine whether these bounds are optimal, or if AdS is more constraining than flat space.
This would seem necessary if one were to rule out a low-energy theory of pure gravity. One reason to hope that stronger bounds may be possible in AdS is as follows. The condition $\tau\geq \De_\mathrm{gap}$ effectively yields an upper bound on the AdS impact parameter, $\beta < 2\log(m/M)$. Our functionals seem to be positive even above this threshold, so potentially stronger bounds could be obtained by relaxing positivity in this region.

The technical centerpiece of our arguments is the family of dispersive CFT sum rules $C_{k,\nu}$, defined in \eqref{eq:magicmoment}. Here $k$ denotes the number of subtractions, and $\nu$ is the momentum transfer. Crucially, in the flat space limit, these sum rules reduce to simple S-matrix sum rules $\mathcal{C}_{k,u}$ studied in our previous paper \cite{Caron-Huot:2021rmr} and reviewed in Section \ref{Smat sum rules}. That these sum rules indeed agree in the flat space limit can be gleaned from the action of $C_{k,\nu}$ on states with $\Delta\gg 1$ and fixed spin, see \eqref{eq:bulkpointbknu}. Dispersive sum rules enjoy uniqueness properties that make them easy to translate between spaces. For instance, $C_{k,\nu}$ admits a concise representation in Mellin space \eqref{eq:CfromBMellin}. We used it to numerically test expansions around various limits to high accuracy. The sum rules are also related to the Lorentzian inversion formula at negative spin \eqref{eq:cknusuper} and to superconvergence relations \cite{Kologlu:2019mfz}.

The mechanism by which flat space S-matrices are recovered from CFT is interesting.
The $C_{k,\nu}$ sum rules are integrals over a ``spacelike scattering'' region,
where causality is clear (certain commutators vanish), correlators are bounded, and integrals rigorously converge.
However, the image is blurred. The problem of reconstructing the S-matrix from this region is somewhat analogous
to reconstructing an object's shape from the way it diffracts light.
Simple versions of this problem are solved by the Fourier transform,
and one might say that we borrowed a technique from optics to decode the hologram.

Technically, the $S$-matrix is recovered because a highly-oscillatory Fourier transform is
dominated by a complex saddle point where space and time are effectively exchanged, see Figure \ref{fig:spacelike}.
On the saddle point, it becomes causally possible to focus beams
within a distance $\sim \rAdS/\Delta_{\rm gap}$ of a bulk point.
Since AdS curvature appears inessential for this effect,
we believe this method could potentially yield novel proofs of various properties of flat space S-matrices.

Perhaps the most surprising finding of recent developments is the power of (relativistic) causality.
For example, the power-counting rules that underly effective theories have long been justified
by considering example UV completions;  we now know that, in a causal theory,
such rules are implied by causality.  
One wonders to what extent causality as we understand it
is really an exact feature of our world, and what more it will teach us.


\section*{Acknowledgments}
We thank Yue-Zhou Li, David Meltzer, Julio Parra-Martinez, Eric Perlmutter, and Sasha Zhiboedov for discussions. We thank Sandipan Kundu for sharing a draft of \cite{Kundu:2021qpi} prior to publication. 
The work of S. C.-H. is supported by the National Science and Engineering Council of Canada,
the Canada Research Chair program, the Simons Collaboration on the Nonperturbative Bootstrap, and the Sloan Foundation.
D.M. gratefully acknowledges funding provided by Edward and Kiyomi Baird as well as the grant DE-SC0009988 from the U.S. Department of Energy. The work of L.R. is supported in part by NSF grant \# PHY-1915093. D.S.-D. is supported by Simons Foundation grant 488657 (Simons Collaboration on the Nonperturbative Bootstrap) and a DOE Early Career Award under grant no. DE-SC0019085.

\appendix


\section{The space of allowed test functions in impact parameters}
\label{sec:impactparam}

In \eqref{eq:psidefinition} we defined functionals
$\Psi_{k,\gamma}$ whose action is perfectly localized in $\etaAdS$: proportional to $\delta(\gamma-\etaAdS)$.
An important point is that perfect localization is physically impossible.
Here we show that $\Psi_{k,\gamma}$ must be interpreted as a
distribution dual to a certain space of test functions of $\gamma$.
Specifically, consider the integral against a function $f(\gamma)$,
\begin{align}
\label{eq:integralagainsttest}
\int_1^\oo [d\gamma] \, f(\gamma) \Psi_{k,\gamma} &= \int_0^\oo \frac{d\nu}{2\pi} \frac{\gegenbauermeasure(\nu) \tl{f}(\nu)}{\g_{2\De_\f+k-1}(\nu)^2}\p{(-1)^k a_{\De_\f} \widehat{\Pi}_{k,\nu}+\dots},
\end{align}
where $[d\gamma]=2^{d-2}(\gamma^2-1)^{\frac {d-3}{2}} d\gamma$ and $\tl f(\nu)$ is defined by
\begin{align}
f(\gamma) &= \int_0^\oo \frac{d\nu}{2\pi} \gegenbauermeasure(\nu)\tl f(\nu) \cP_{\frac{2-d}{2}+i\nu}(\gamma)\,,
\end{align}
where $\gegenbauermeasure(\nu)$ is the measure appearing in (\ref{harmonic Pi}).

We expect that the moment $\widehat{\Pi}_{k,\nu}$ is dual to Schwarz functions of $\nu$. This follows from the fact that it can be written as a linear combination of applications of the Lorentzian inversion formula (see appendix~\ref{app:superconvergence}), and the Lorentzian inversion formula computes a tempered distribution of $\nu$.\footnote{At least this is true when $\De_\f$ is sufficiently small that the Euclidean four-point function is normalizable with respect to the pairing discussed in \cite{Simmons-Duffin:2017nub}. For general $\De_\f$, one might need to subtract off a finite number of partial waves from the four-point function to make it normalizable. Fully characterizing the type of distribution in $\nu$ produced by the Lorentzian inversion formula is an open problem.} Consequently $\gamma_{2\De_\f+k-1}(\nu)^{-2}\tl{f}(\nu)$ must be a Schwarz function in order for the right-hand side of (\ref{eq:integralagainsttest}) to converge. Since $\gamma_{2\De_\f+k-1}(\nu)^{-2}\sim e^{\pi \nu}$ grows exponentially at large $\nu$, this means that $\tl{f}(\nu)$ must fall off exponentially as $e^{-\pi\nu}$.

An elegant way to express this resulting constraint on $f(\gamma)$ is to parametrize
$\gamma=\frac{\xi+1/\xi}{2}$. At large $\nu$ and $|\xi|>1$, we have
\begin{align}
\label{eq:limitofgegenbauer}
\cP_{\frac{2-d}{2}+i\nu}(\gamma) &\sim \frac{1}{(\xi-1/\xi)^{\frac{d-2}{2}}}\p{\frac{\xi^{i\nu}}{\gegenbauerconst_{\frac{2-d}{2}+i\nu}} +  \frac{\xi^{-i\nu}}{\gegenbauerconst_{\frac{2-d}{2}-i\nu}}}, \qquad (\nu\gg 1,|\xi|>1)\,,
\end{align}
where $\gegenbauerconst_j$ is defined in (\ref{eq:cjdefinition}). For $|\xi|<1$, we can use symmetry under $\xi\to1/\xi$.
The falloff condition on $\tl f(\nu)$ then implies that $f(\gamma)$ is analytic in the cut plane $\xi\in \mathbb{C}\setminus(-\oo,0]$.
This condition can be interpreted by writing the spacetime cross-ratio \eqref{eq:definitionofreta}
in the same form: $\eta=\frac{\xi+\xi^{-1}}{2}$:
the region $\xi<0$ realize the bulk-point kinematics of \cite{Maldacena:2015iua}.
Therefore, the allowed test functions $f(\gamma)$ must admit an unobstructed analytic continuation between the Rindler and bulk point kinematics.

Although the analyticity condition precludes strict $\delta$-function localization,
this result provides us with a sufficiently vast supply of ``bump functions'' $f(\gamma)$
to produce narrow peaks around any desired value of $\etaAdS$.
We expect sharper functionals to exhibit more violent oscillations when written as combinations of moments $\Pi_{k,\eta}$.


\section{Formulas involving the Gegenbauer function}
\label{app:gegenbauerformulas}

In this appendix, we collect some useful formulas involving the Gegenbauer function $\cP_J(\eta)$, defined by
\begin{align}
\cP_J(\eta)= {}_2F_1\p{-J,J+d-2,\tfrac{d-1}{2},\tfrac{1-\eta}{2}}\,.
\end{align}

\subsection{Fourier-Laplace transform of Gegenbauer functions}
\label{app:Fourier Gegen}

In evaluating moments at large-$\Delta$, one finds Fourier-Laplace transforms of Gegenbauer functions.
Here we prove the identity \eqref{Fourier identity}:
\be
 \int_{x>0} d^dx |x|^{a-d} \cP_{\frac{2-d}{2}+i\nu}\p{\frac{-x\.y}{|x||y|}}e^{x\.p}
 = 2^{a-1} \pi^{\frac{d-2}{2}}\gamma_a(\nu)\times |p|^{-a}\cP_{\frac{2-d}{2}+i\nu}\p{\frac{-p\.y}{|p||y|}}\,, \label{Laplace integral to do}
\ee
where $p$ is any future timelike vector.
The Gegenbauer function is proportional to a harmonic function on AdS$_{d-1}$, with $x/|x|$ and $y/|y|$ interpreted as bulk points.
The general form of this result is determined by symmetries (rotational and scale invariance) as well as as Casimir equation
with respect to $y$; the nontrivial output of the following calculation is the proportionality constant.

One method to do the integral is to use the ``split'' representation for the AdS harmonic function as the integral over a boundary point
$z$, interpreted as an embedding space coordinate for CFT${}_{d-2}$:
\begin{align}
\label{eq:integralgegenbauer}
\cP_J\p{\frac{-x\.\bar{y}}{|x||y|}} &= \int \frac{D^{d-2} z}{\vol S^{d-2}} \p{\frac{-z\.x}{|x|}}^J \p{\frac{-z\.\bar{y}}{|y|}}^{2-d-J},
\end{align}
where $D^{d-2} z$ is the measure defined in \eqref{eq:projectivemeasure}.
Substituting this expression into \eqref{Laplace integral to do}, the fact that $z^2=0$ now makes the integral over $x$
elementary (for example, it can be done by separating $x$ into an energy, the angle with the $z$ axis, and the norm of its perpendicular components).
These integrals produce gamma factors:
\be
\int_{x>0} d^d x |x|^{a-d} \p{\frac{-x\.z}{|x|}}^{\Delta+1-d} e^{x\.p} = 2^{a-1}\pi^{\frac{d-2}{2}}\gamma_{a}(\Delta)\times |p|^{-a}\p{\frac{-z\.p}{|p|}}^{\Delta+1-d},
\ee
where
\be
\gamma_a(\nu) \equiv \Gamma\Big(\tfrac{1+a-\frac d 2 - i\nu}{2}\Big) \Gamma\Big(\tfrac{1+a-\frac d 2 + i \nu}{2}\Big)\,. \label{gamma app}
\ee
The proof of \eqref{Laplace integral to do} is completed by using \eqref{eq:integralgegenbauer} in reverse.
The infinite set of poles in the result $\gamma_a(\nu)$ can be understood from the lightcone limit $x^2\to 0$ of the integral,
where $\eta\to\infty$ and $\cP(\eta)\sim \eta^{\frac{2-d}{2}\pm i\nu}$, causing divergences at complex $\nu$.

\subsection{Integral representation and positivity of derivatives}

We now discuss the positivity of derivatives of $\cP_{\frac{2-d}{2}+i\nu}(\cosh\beta)$ with respect to $\nu$,
which is used in Section \ref{sec:derivativesofforward}.
We begin from the split representation \eqref{eq:integralgegenbauer} of $\cP_{\frac{2-d}{2}+i\nu}(\frac{-x\.y}{|x||y|})$ and make the substitution $x=(\cosh\b,\sinh \b, 0,\dots,0)$ and $y=(1,0,\dots,0)$:
\begin{align}
\cP_{\frac{2-d}{2}+i\nu}(\cosh \b) &= \int \frac{D^{d-2} z}{\vol S^{d-2}} (-z\.x)^{\frac{2-d}{2}+i\nu} (-z\.y)^{\frac{2-d}{2}-i\nu} \nn\\
&= \frac{\vol S^{d-3}}{\vol S^{d-2}} \int_0^\pi d\theta \sin^{d-3}\theta (\cosh \b - \sinh \b \cos \theta)^{\frac{2-d}{2}+i \nu}.
\end{align}
Changing variables from $\theta$ to
\begin{align}
u &\equiv \frac{\log(\cosh \b - \sinh \b \cos \theta)}{\b}\,,
\end{align}
we find\footnote{The representation (\ref{eq:newuintegrand}) shows that $\cP_{\frac{2-d}{2}+i\nu}(\cosh\b)$ simplifies when $d=4$, where it becomes proportional to a Bessel function: 
\begin{align}
\cP_{i\nu-1}(\cosh \b) &= \frac{\b}{\sinh \b} \tl J(\nu \b) = \frac{\sin \b \nu}{\nu \sinh \b} \qquad (d=4),
\end{align}
where $\tl J(x)$ is defined in (\ref{eq:impactparamlimitofgeg}). In this case, we have
\begin{align}
\cQ_n(\beta) &= \frac{\sin (\frac{\pi  n}{2}) \Gamma (n)}{\beta^n \sinh\beta} \qquad (d=4).
\end{align}
}
\begin{align}
\label{eq:newuintegrand}
\cP_{\frac{2-d}{2}+i\nu}(\cosh \b) &= 
\frac{\vol S^{d-3}}{\vol S^{d-2}} \p{\frac{\b}{\sinh \b}}^{\frac{d-2}{2}}
\int_{-1}^1 du \p{\frac{2\cosh \b - 2 \cosh u\b}{\b \sinh \b}}^{\frac{d-4}{2}} e^{i u \b \nu}\,.
\end{align}

The formula (\ref{eq:newuintegrand}) yields an expression for derivatives around the forward limit $\nu=0$:
\begin{align}
\label{eq:derivativeforwardlimitgegenbauer}
\left.(-\ptl_\nu^2)^k \cP_{\frac{2-d}{2}+i\nu}(\cosh \b)\right|_{\nu=0} 
&= \frac{\vol S^{d-3}}{\vol S^{d-2}} \p{\frac{\b}{\sinh \b}}^{\frac{d-2}{2}} 2\beta^{2k} R_{2k}(\beta)\,,
\end{align}
where
\begin{align}
R_n(\beta) &\equiv \int_0^1 du\, u^{n} \p{\frac{2\cosh \b - 2 \cosh u\b}{\b \sinh \b}}^{\frac{d-4}{2}}.
\end{align}
The function $R_n(\beta)$ is everywhere positive and interpolates smoothly between the limits
\begin{align}
\lim_{\beta\to 0} R_n(\beta) = \frac{\Gamma (\frac{d-2}{2}) \Gamma (\frac{n+1}{2})}{2 \Gamma (\frac{d+n-1}{2})}\,,
\qquad
\lim_{\beta\to\oo} R_n(\beta) = \frac{1}{n+1}\p{\frac{\beta }{2}}^{\frac{4-d}{2}}.
\end{align}

\subsection{The $\nu$-moments $\cQ_n(\beta)$}
\label{app:qfunction}

From (\ref{eq:newuintegrand}), we can derive a useful integral representation for the function defined in \eqref{eq:defofqn}:
\begin{align}
\label{eq:defofqagain}
\cQ_n(\beta) &= \int_0^\oo d\nu\,\nu^n \cP_{\frac{2-d}{2}+i\nu}(\cosh\beta)\,.
\end{align}
In fact, the integral (\ref{eq:defofqagain}) is naively divergent for $n\notin(-1,0)$, and our formula will serve to define $\cQ_n(\beta)$ outside this range. To obtain it, we integrate (\ref{eq:newuintegrand}) over $\nu\in(0,\oo)$, treating the cases $u>0$ and $u<0$ separately by deforming the $\nu$ contour into the positive and negative imaginary directions, respectively. We then combine these cases to give a single integral over $u\in [0,1]$:
\begin{align}
\cQ_n(\beta) &\equiv \int_0^\oo d\nu\, \nu^n \cP_{\frac{2-d}{2}+i\nu}(\cosh \b) 
\nn\\
&=  
\frac{\vol S^{d-3}}{\vol S^{d-2}} 
\frac{-2\sin(\frac{\pi n}{2})\G(n+1)}{\b^{n+1}}
\p{\frac{\b}{\sinh \b}}^{\frac{d-2}{2}}
\int_0^1 \frac{du}{u^{n+1}} 
\p{
\frac{2\cosh \b - 2 \cosh u\b}{\b \sinh \b}}^{\frac{d-4}{2}
}.
\label{eq:singleintegraoveru}
\end{align}
Although the integral (\ref{eq:singleintegraoveru}) may appear singular at $u=0$, the correct prescription is to define it by analytic regularization. This is clear from the representation (\ref{eq:newuintegrand}), where we can deform $u$ into the upper half-plane to avoid the singularity.

We can now expand the integrand (\ref{eq:singleintegraoveru}) in small or large $\beta$ to determine the asymptotics (\ref{eq:expansionofQ}). We can obtain a quickly convergent approximation to $\cQ_n(\beta)$ by instead expanding the integrand in $u$ and integrating term-by-term:
\begin{align}
\p{\frac{\sinh \b}{\b}}^{\frac{d-2}{2}}\cQ_n(\beta)
&=
\frac{
2^{\frac{d-2}{2}} \Gamma (\frac{d-1}{2}) \Gamma (n) \sin (\frac{\pi  n}{2}) 
}{
\sqrt{\pi } \Gamma (\frac{d-2}{2})
}
\frac{\tanh^{\frac{d-4}{2}} \frac{\beta }{2}}{\beta ^{n+\frac{d-2}{2}}}
\p{
1+\frac{n  (d-4) \beta^2}{8 (2-n) \sinh^2 (\frac{\beta }{2})}
+ \dots
}.
\end{align}
Using the first few terms in this expansion, it is straightforward to plot $\cQ_n(\beta)$, as we do in Figure~\ref{fig:qn}.


\section{Regge moments of heavy blocks}
\label{app:heavy}

Here we present various formulas pertaining to the evaluation of block moments \eqref{eq:inversetransforms}
\begin{align}
 \widehat{\Pi}_{k,\nu}[\cG] &= \int_1^\oo [d\eta] \cP_{\frac{2-d}{2}+i\nu}(\eta)
 \int_0^{r_{\rm max}(\eta)} dr\,r^{k-2} \dDisc_s\,\cG(r,\eta)\,,
 \label{Pi moment appendix}
\end{align}
and corresponding formulas for physical functionals $C_{k,\nu}$.
Here $r=\sqrt{\rho\bar\rho}$ and $\eta=\frac{\rho+\bar\rho}{2\sqrt{\rho\bar\rho}}$ are radial and angular variables in the $u$-channel Regge limit (\ref{eq:definitionofreta}), and $\cP$ is a harmonic function \eqref{def legP}.

\subsection{Integral representation in symmetrical frame}

\label{app:subleading}

In this section, we gauge-fix conformally-invariant integrals like (\ref{conformal measure}) in the conformal frame (\ref{eq:otherframe}), which we reproduce here:
\begin{align}
\label{eq:otherframeagain}
(x_1,x_2,x_3,x_4;x_5) &= \p{-y,\frac{-x}{-x^2},y,\frac{x}{-x^2}; e}, \qquad x,y>0\,.
\end{align}
Recall that the cross-ratios (\ref{eq:definitionofreta}) in this frame are given by
\be
r=|x||y|\,,\qquad  \eta = -\frac{x\.y}{|x||y|}\,.
\ee

To evaluate~\eqref{conformal measure} in this frame,
we need to compute the Fadeev-Popov determinant corresponding to
the gauge-fixing factor $\delta^d(x_1+x_3)\delta^d(x_2^{-1}+x_4^{-1})\delta^d(x_5-e)$
needed to reduce the symmetry to SO($d-1$) rotations acting on $x,y$.
The relevant $\dim \SO(d,2)-\dim \SO(d-1) = 3d$ conformal generators are $d$ translations and $d$ boosts,
together with one rescaling and the $d-1$ rotations acting nontrivially on $e=(1,\vec{0})$.
The latter just give a trivial Jacobian, and the rest gives a $2d\times 2d$ determinant
\be\begin{aligned}
 2^d \det\begin{pmatrix}
 \delta_\mu^\nu & \delta_\mu^\nu x^2-2x_\mu x^\nu \\
 \delta_\mu^\nu y^2-2y_\mu y^\nu & \delta_\mu^\nu \end{pmatrix}
  &= 2^d(1+2x\.y+x^2 y^2)(1-2x\.y+x^2 y^2)(1-x^2y^2)^{d-2}
\\ &= 2^{d+4} r^2\sqrt{u'v'}(1-r^2)^{d-2}\,.
\end{aligned}\ee
Substituting \eqref{Pi as I integral} into \eqref{conformal measure}
then gives a concrete formula for the Regge moment \eqref{Pi moment appendix} of heavy blocks
\be  \label{moment appendix}
\frac{\Pi_{k,\Omega}[G^s_{\De,J}]}{2\sin^2(\pi \frac{\tau-2\De_\f}{2})} = 4^{2\Df}\frac{(-1)^{J}}{b_{\Delta,J}}
\int d^dx d^dy\ \big(|x||y|\big)^{k+d-1} \Omega\p{\frac{x\.y}{|x||y|}}
\mathcal{T}(x,y)\,.
\ee
To complete the formula, we must record the product of three-point functions in the frame \eqref{eq:otherframe}:
\be
\mathcal{T}(x,y)\equiv 2^{5d-2-4\Df}\frac{|\langle \phi_1\phi_2 \mathcal{O}(x_5,n)\rangle|
 |\langle \mathcal{O}^\mathrm{S}(x_5,n)\phi_3\phi_4 \rangle|}{|x|^{4d}(-x_{13}^2)^{\widetilde \De_\f}(-x_{24}^2)^{\widetilde \De_\f}}\,,
\ee
which can be written as: 
\be\begin{aligned}
\mathcal{T}(x,y)&=
\ \frac{1}{(x^2y^2)^{\tl\De_\f}}
\frac{(1 + 2x\.y +x^2y^2)^{d-1-2\Df}}{(e-x)^2(e-y)^2} \left( \frac{(e-x)^2(e-y)^2}{(e+x)^2(e+y)^2}\right)^{\frac{\Delta+J}{2}}
\times \frac{[-n{\cdot}\mathcal{V}_e]^{J}}{[n{\cdot}\mathcal{V}_{-e}]^{J+d-2}}\,,
\\
 \mathcal{V}_e &= \left((1-2e\.x)(1-2e\.y)-x^2y^2\right)e + (e+y)^2x  + (e+x)^2\bar{y}\,,
\end{aligned}\ee
where $\bar{y}=-y-2e(y\.e)$ denotes the space-reflected vector. Expanding at large dimension and spin, with $x,y\sim 1/\Delta$, this reduces to
\be
\lim_{\Delta,J\gg 1} \mathcal{T}(x,y) \approx \frac{e^{2p\.(x+\bar{y})}}{(x^2y^2)^{\tl\De_\f}}\,.
\label{limit symmetrical frame}
\ee
The limit \eqref{limit symmetrical frame} substituted into \eqref{moment appendix}
agrees precisely with that recorded in \eqref{eq:reggemomentspacetimedeltaspace} for
the non-symmetrical frame, after a simple variable change $x\mapsto x/2$, $y\mapsto \bar{y}/2$.
As noted around \eqref{regge subleading}, 
the symmetrical frame is useful to analyze subleading corrections when $x\sim 1/m$ because
only even powers of $1/m$ appear.

\subsection{Expansion of physical functionals as sums of Regge moments}
\label{sec:functionalsexpandreggemoments}

Finally, we consider the expansion of the physical kernel $C_{k,\nu}$ defined
in \eqref{eq:magicmoment} in Regge moments.
We record it here for convenience, after combining it with the formula \eqref{Bkv r eta}
for the $B_{k,v}$ sum rule:
\be
\label{eq:definitionofomegatilde}
 C_{k,\nu}[\cG] =
 \int_1^\infty [d\eta] \int_0^{r_{\rm max}(\eta)} dr\ \widetilde\Omega_{k,\nu}(r,\eta) {\rm dDisc}_s[\cG]\,,
\ee
where the kernel is defined formally by the double integral transform
\begin{align}
 \widetilde\Omega_{k,\nu}(r,\eta) &=
\frac{a_{\Df}}{\g_{2\Df+k-1}(\nu)^2}
\eta\ r^{k-2}(1{-}r^4)
\int_1^\infty d\eta' \frac{(\eta'^2-1)^{\frac{d-3}{2}}}{(\eta^2-1)^{\frac{d-3}{2}}} \cP_{\frac{2-d}{2}+i\nu}(\eta')
\times
\nonumber\\ &\qquad\times
\frac{(-1)^{\frac{k}{2}}(k-1)}{\pi}\int_{\eta'^2}^\infty \frac{dv}{(v-\eta'^2)^{\frac{k+1}{2}}_+}\frac{\theta(\eta^2-v)}{(\eta^2-v)^{\frac{3-k}{2}}}
 \frac{\p{(1+r^2)^2-4v r^2}^{\frac{k-3}{2}}}{\p{(1+r^2)^2-4 \eta^2 r^2}^{k-1}}\,.
\end{align}
Our task is to series-expand this at small $r\sim 1/m$.
A useful first step is to note that the $v$ dependence of the last factor is suppressed as $r\to 0$.
We thus expand out that factor around $v\to\eta^2$:
\be
\frac{\p{(1+r^2)^2-4v r^2}^{\frac{k-3}{2}}}{\p{(1+r^2)^2-4 \eta^2 r^2}^{k-1}}
= \frac{1}{\p{(1+r^2)^2-4 \eta^2 r^2}^{\frac{k+1}{2}}}\sum_{n=0}^\infty
\frac{(\tfrac{3-k}{2})_n}{n!} (\eta^2-v)^n\left(\frac{-4r^2}{(1+r^2)^2-4\eta^2r^2}\right)^n\,.
\ee
The $v$ integral can then be computed term-by-term.
The distributional nature of the integrand makes this somewhat subtle; we find:
\be
\frac{(-1)^{\frac{k}{2}}(k-1)}{\pi} \int_{\eta'^2}^\infty \frac{dv}{(v-\eta'^2)^{\frac{k+1}{2}}_+}\frac{\theta(\eta^2-v)}{(\eta^2-v)^{\frac{3-k}{2}-n}}
= 2\left\{\begin{array}{l@{\quad}l}
\delta(\eta'^2-\eta^2), & n=0, \\
\frac{(\tfrac{k-1}{2})_n}{(n-1)!}(\eta^2-\eta'^2)^{n-1}\theta(\eta-\eta'), & n\geq 1.
\end{array}\right.
\ee
The generic case could be derived simply by evaluating the integral as an analytic function of $k$ and $n$,
in terms of Euler's beta function, and continuing to the desired value.  For $n=0$, that integral naively yields zero, however the singular power-counting of the integral $\sim \frac{1}{\eta'-\eta}$ makes the distributional term possible;
by power-counting, distributional terms cannot appear for $n>0$.
A good way to confirm the above expression is to integrate against test functions which are powers of $\eta$,
for which the left-hand-side evaluates to simple gamma functions.

Combining the preceding two formulas, we express the kernel as a series:
\begin{align} \label{Omega knu as infinite sum}
 \widetilde\Omega_{k,\nu}(r,\eta) &=
\frac{a_{\Df}}{\g_{2\Df+k-1}(\nu)^2} \frac{r^{k-2}(1{-}r^4)}{\p{(1+r^2)^2-4 \eta^2 r^2}^{\frac{k+1}{2}}}\times
\nonumber\\
& \quad
\times\left[\cP_{\frac{2-d}{2}+i\nu}(\eta) 
+\sum\limits_{n=1}^\infty \frac{(\tfrac{k-1}{2})_n}{(n-1)!}\frac{(\tfrac{3-k}{2})_n}{n!}
\left(\frac{-4r^2}{(1+r^2)^2-4\eta^2r^2}\right)^n I_n(\eta)\right]
\end{align}
where
\be
 I_n(\eta)\equiv\eta \int_1^\eta d\eta' \frac{(\eta'^2-1)^{\frac{d-3}{2}}}{(\eta^2-1)^{\frac{d-3}{2}}}
(\eta^2-\eta'^2)^{n-1} \cP_{\frac{2-d}{2}+i\nu}(\eta')\,.
\ee
Remarkably, the integral $I_n(\eta)$ can be written in closed form as a finite sum of $\cP$ with shifted indices!
This may be seen by repeatedly applying the familiar shift identity for multiplication by $\eta$,
together with a similar identity for integration:
\be\begin{aligned} \label{shift identities}
 \eta\,\cP_j(\eta) &= \frac{(j+d-2)\cP_{j+1}(\eta)+j \cP_{j-1}(\eta)}{2j+d-2}\,,
 \\
 \int_1^\eta d\eta' \frac{(\eta'^2-1)^{\frac{d-3}{2}}}{(\eta^2-1)^{\frac{d-3}{2}}} \cP_j(\eta')
 &=\frac{\cP_{j+1}(\eta)-\cP_{j-1}(\eta)}{2j+d-2} \,,
\end{aligned}\ee
both of which can be proven straightforwardly via series expansion around $\eta=1$.

Eqs.~\eqref{Omega knu as infinite sum} and \eqref{shift identities}
allow to express the action of $C_{k,\nu}$ to any desired order in $1/m$
as a finite sum of Regge moments.  For example, 
\begin{align} \frac{\g_{2\Df+k-1}(\nu)^2}{a_{\Df}}C_{k,\nu}&=
\widehat\Pi_{k,\nu} +
\frac{(d{-}2)(2(k{-}3)(k{-}1)+(k{+}1)(d{-}4))}{4(\nu^2+1)} \widehat\Pi_{k+2,\nu}
\nn\\&
\quad-\frac{(d{+}2i\nu)(2(k{-}3)(k{-}1)+(k{+}1)(d{-}2{+}2i\nu))}{8\nu(\nu-i)} \widehat\Pi_{k+2,\nu-2i}
\nn\\&
\quad-\frac{(d{-}2i\nu)(2(k{-}3)(k{-}1)+(k{+}1)(d{-}2{-}2i\nu))}{8\nu(\nu+i)} \widehat\Pi_{k+2,\nu+2i} 
+ O(\Pi_{k+4})\,.
\label{eq:CExpansionInPiHat}
\end{align}
While $\nu$-poles appear in individual terms, they are spurious and cancel between terms:
this is clear since all denominators are generated by eqs.~\eqref{shift identities}.
We also note that subleading terms get shifted away from the principal series.
The enhanced growth at large $\eta$ could cause potential divergences,
however, the imaginary shift in $\nu$ never exceeds the shift in $k$,
and this ensures (see \eqref{gamma app}) that all double-poles continue to cancel against $\gamma^2$:
$C_{k,\nu}$ is an entire function of $\nu$.

A nontrivial check is that the Regge moment expansion of the kernel $\widetilde{\Omega}_{k,\nu}$ coincides with that of
conformal blocks entering the Lorentzian inversion formula at special values of spin corresponding to superconvergence relations. An example is given in (\ref{eq:largenuktwokernel}) below.


\section{The $C_{k,\nu}$ functionals as superconvergence relations}
\label{app:superconvergence}

In this appendix, we establish a relationship between the $C_{k,\nu}$ functionals and ``superconvergence'' sum rules \cite{Kologlu:2019bco,Kologlu:2019mfz}, which express the vanishing of certain null-integrated commutators. This relationship leads to explicit formulas for the kernels $\tl \Omega_{k,\nu}(r,\eta)$ defining $C_{k,\nu}$ in terms of conformal blocks. The relation to superconvergence also suggests a natural way to extend the $C_{k,\nu}$ sum rules for spinning correlators.

We briefly review superconvergence relations. In \cite{Caron-Huot:2020adz}
we showed that the sum rule $B_{2,v}[\cG]=0$ is equivalent to the subtracted superconvergence sum rule 
\begin{align}
\label{eq:vanishingcommutator}
\int_{-\oo}^\oo dx_1^+ \int_{-\oo}^\oo dx_3^+ \frac{v'-u'}{u'v'} \<\Omega|\f(x_4) \left[\f(x_1^+,x_1^-=0,\vec x_1),\f(x_3^+,x^-_3=0,\vec x_3)\right] \f(x_2)|\Omega\> &= 0\,.
\end{align}
Here, we use lightcone coordinates $x=(x^+,x^-,\vec x)$. The cross-ratios $u',v'$ are built from the positions $x_1,x_2,x_3,x_4$ in the usual way (with the $i\e$ prescription appropriate for the given operator ordering). The left-hand side of (\ref{eq:vanishingcommutator}) vanishes because the operators $\f(x_1)$ and $\f(x_3)$ in the commutator are everywhere spacelike-separated and the integral is convergent \cite{Kologlu:2019bco}. The parameter $v$ that labels different sum rules is a function of $x_2,x_4$ and the transverse positions $\vec x_1,\vec x_3$.

By choosing different kinematics for the operators $\f(x_1)\cdots\f(x_4)$, we obtain other equivalent statements of the same set of sum rules.
For our purposes, it is useful to adopt the kinematics of \cite{Kologlu:2019mfz} in which (\ref{eq:vanishingcommutator}) becomes related to the Lorentzian inversion formula \cite{Caron-Huot:2017vep}. To write it down, let us introduce the $u$-channel Lorentzian inversion integral
\begin{align}
\Phi_{k,\nu}[\cG] &\equiv \frac{1}{4^{k} q_{\frac{2-d}{2}+i\nu}} \int_0^{-\oo} \int_0^{-\oo} dw  d\bar w |w-\bar w|^{d-2} G^u_{k+d-1,\frac{2-d}{2}+i\nu}(w,\bar w) \dDisc_s[\cG]\,,
\end{align}
where $q_J$ is defined in (\ref{eq:cjdefinition}). 
We have defined $\Phi_{k,\nu}$ so that it has a Regge moment expansion of the form\footnote{Recall that the leading behavior of a $u$-channel conformal block as $r\to 0$ is $G_{\De,J}^u \sim (4r)^\De q_J \cP_J(\eta)+\dots$. To obtain (\ref{eq:reggemomentofphi}) we must additionally take into account the Jacobian for translating from $w,\bar w$ to $r,\eta$, and a tricky factor of 2 accounting for symmetry under $w\leftrightarrow \bar w$.}
\begin{align}
\label{eq:reggemomentofphi}
\Phi_{k,\nu}[\cG] &\sim \widehat\Pi_{k,\nu} + O(\Pi_{k+2})\,.
\end{align}
For reference, the statement of the Lorentzian inversion formula in terms of $\Phi_{k,\nu}$ is
\begin{align}
C\p{\frac d 2 + i\nu ,J} &= 2^{2J-1}\kappa_{\frac d 2 + i\nu +J} q_{\frac{2-d}{2}+i\nu} \Phi_{J,\nu}[\cG]\,,
\qquad\kappa_\beta \equiv \frac{\G(\b/2)^4}{2\pi^2 \G(\beta-1)\G(\beta)}\,,
\end{align}
where we have analytically continued from even spin $J$, so that the $s$ and $t$ channel dDisc's contribute equally, giving a factor of 2.

A result of \cite{Kologlu:2019mfz,Caron-Huot:2020adz} 
is that (\ref{eq:vanishingcommutator}) is equivalent to vanishing of the Lorentzian inversion integral at the special spin $k=-1$, applied to the modified correlator $\frac{v'-u'}{u'v'}\cG$:
\begin{align}
\label{eq:superconvergencespinminusone}
\Phi_{-1,\nu}\left[\frac{v'-u'}{u'v'} \cG \right] &= 0\,,\qquad \forall\,\nu\in[0,\oo)\,.
\end{align}
We can obtain an even simpler relationship between $C_{2,\nu}$ and Lorentzian inversion using the following observation. Using standard identities for conformal blocks \cite{Dolan:2011dv}, we find that
\begin{align}
\label{eq:magicblockidentity}
(v'-u') \tl G_{-1,\nu}(w,\bar w) &= 
\frac{\frac{2-d}{2}+i\nu}{i\nu} \tl G_{-2,\nu+i}(w,\bar w) - \frac{\frac{2-d}{2}-i\nu}{i\nu} \tl G_{-2,\nu-i}(w,\bar w)\,,
\end{align}
where
\begin{align}
\tl G_{k,\nu}(w,\bar w) &\equiv \frac{G^u_{k+d-1,\frac{2-d}{2}+i\nu}(w,\bar w)}{q_{\frac{2-d}{2}+i\nu}}\,.
\end{align}
The right-hand side of (\ref{eq:magicblockidentity}) is an invertible $\nu$-dependent finite difference operator applied to $\tl G_{-2,\nu}(w,\bar w)$.\footnote{To see that it is invertible, we can study the leading Regge moment of (\ref{eq:magicblockidentity}), which is the elementary Gegenbauer identity
\begin{align}
\eta \cP_{\frac{2-d}{2}+i\nu}(\eta) &= \frac{\frac{2-d}{2}+i\nu}{2i\nu} \cP_{\frac{2-d}{2}+i\nu-1}(\eta) - \frac{\frac{2-d}{2}-i\nu}{2i\nu}\cP_{\frac{2-d}{2}+i\nu+1}(\eta)\,.
\end{align}
This shows that the finite difference operator on the right-hand side is equivalent to multiplication by $\eta$ in $\eta$-space, which is invertible because $\eta\in[1,\oo)$. To obtain its inverse, we transform from $\nu$-space to $\eta$-space, divide by $\eta$, and transform back to $\nu$ space.
} Thus, the subtracted superconvergence relation (\ref{eq:superconvergencespinminusone})
is equivalent to
\begin{align}
\label{eq:bettersuperconvergence}
\Phi_{-2,\nu}\left[\frac{\cG}{u'v'}\right] &= 0\,,\qquad \forall\,\nu\in[0,\oo)\,.
\end{align}
Superconvergence sum rules involving the Lorentzian inversion formula at negative even spins $k=-2,-4,\dots$ were not discussed in \cite{Kologlu:2019mfz}, but they exist as well. We explain how they arise in section~\ref{sec:evenspinsuperconvergence}.

\subsection{The $C_{k,\nu}$ functionals as special cases of Lorentzian inversion}

We can now relate the superconvergence relation
$\Phi_{-2,\nu}$ to $C_{2,\nu}$ using uniqueness properties of dispersive sum rules.
A dispersive functional can be characterized by the list of double-twist
Regge trajectories on which it does not have double zeros (its ``support'') together with its expansion into Regge moments.
For example, ref.~\cite{Caron-Huot:2020adz} showed that any spin-2 convergent sum rule
with double zeros on all but the leading trajectory $\Delta-J=2\Df$, must be a linear combination of $B_{2,v}$ functionals.
The correct combination can then be found simply by matching the leading Regge moment.
This predicts the following exact relation:
\begin{align}
\label{eq:c2intermsofsuperconvergence}
\frac{\g_{2\De_\f+2-1}(\nu)^2}{a_{\De_\f}} C_{2,\nu}[\cG] &= \Phi_{-2,\nu}\left[\frac{\cG}{2^8 u'v'}\right].
\end{align}
In particular, taking into account the Jacobian relating $w,\bar w$ to $r,\eta$, we obtain an explicit expression for the kernel $\tl \Omega_{2,\nu}$ appearing in $C_{2,\nu}$ (\ref{eq:definitionofomegatilde}):
\begin{align}
\label{eq:largenuktwokernel}
\frac{\g_{2\De_\f+2-1}(\nu)^2}{a_{\De_\f}} \tl \Omega_{2,\nu}(r,\eta) &= \frac{(4r)^{3-d} \left(1-r^2\right)^{d-2}}{(1+r^2)^2-4 r^2 \eta ^2}  \frac{G^u_{d-3,\frac{2-d}{2}+i \nu}(w,\bar w)}{q_{\frac{2-d}{2}+i \nu}}\,.
\end{align}
We have verified that this relation holds to high order!

How about higher-subtracted sum rules?  In the main text, we constructed $C_{k,\nu}$ according to the following specifications: it should be a dispersive functional with support on the first $k/2$ double-trace families, and it should be a pure $\widehat \Pi_{k,\nu}$ moment in the Regge limit. 
The result (\ref{eq:c2intermsofsuperconvergence}) reveals another way to construct such functionals using $\Phi_{-2,\nu}$, $\Phi_{-4,\nu}$, etc.. We expect these constructions to be related. Indeed, by matching Regge moments, we find for example
\begin{align}
\label{eq:cfoursuper}
&\frac{\g_{2\De_\f+4-1}(\nu)^2}{a_{\De_\f}}C_{4,\nu}[\cG]= \nn\\
&\quad\Phi _{-4,\nu }[\cG^{(4)}]
-\frac{2(d-2) (\frac{d-8}{2}-i \nu) (\frac{d-8}{2}+i \nu)}{(d-8) (\nu+i)(\nu-i)} \Phi _{-2,\nu }[\cG^{(4)}] \nn\\
&\quad
+\frac{(\frac{d-8}{2}+i \nu) (\frac{d-4}{2}+i \nu) (\frac{d}{2}+i \nu )}{\nu  (\nu -i) (\frac{d-6}{2}+i \nu)} \Phi _{-2,\nu -2 i}[\cG^{(4)}]
+\frac{(\frac{d-8}{2}-i \nu) (\frac{d-4}{2}-i \nu) (\frac{d}{2}-i \nu ) }{\nu  (\nu +i) (\frac{d-6}{2}-i \nu)} \Phi _{-2,\nu +2 i}[\cG^{(4)}]\,,
\end{align}
where we have defined the spin-$2k$-subtracted correlator
\begin{align}
\label{eq:thesubtractedcorrelator}
\cG^{(k)}(u',v') &\equiv \frac{\cG(u',v')}{(2^8 u' v')^{k/2}}\,.
\end{align}
The poles in $\nu$ on the right-hand side of (\ref{eq:cfoursuper}) are spurious --- they cancel among the different terms.  The agreement between the left- and right-hand sides of (\ref{eq:cfoursuper}) can be argued as follows. Firstly,
$\Phi_{-4,\nu}[\cG^{(4)}]$, $\Phi_{-2,\nu}[\cG^{(4)}]$ span the space of dispersive sum rules with spin-4 or higher Regge decay that are nonvanishing on only the first two double-twist trajectories. Any linear combination of these can thus be fixed by its first two Regge moments.  Further Regge moments can then be used as tests of \eqref{eq:cfoursuper}.
In general, we find
\begin{align}
\label{eq:cknusuper}
\frac{\g_{2\De_\f+k-1}(\nu)^2}{a_{\De_\f}}C_{k,\nu}[\cG] &= \Phi _{-k,\nu }[\cG^{(k)}] + \dots\,,
\end{align}
where ``$\dots$'' are a finite sum of higher-spin $\Phi$'s with shifted arguments.

The behavior of functionals at large-$\nu$ is important for our discussion of the bulk point saddle in sections~\ref{sec:heavymomentbulkpoint} and \ref{eq:lightpositionspace}. Any dispersive functional with the same leading Regge moment and the same large-$\nu$ limit would give similar results. 
We note that at large-$\nu$, the kernel for $C_{k,\nu}$ satisfies
\begin{align}
\label{eq:largenukernelrelation}
\g_{2\De_\f+k-1}(\nu)^2 C_{k,\nu}[\cG] &= \g_{2\De_\f+l-1}(\nu)^2 C_{l,\nu}[\cG^{(\frac{k-l}{4})}] \x (1+O(1/\nu^2))\,.
\end{align}
This suggests a possible alternative set of functionals $\tl C_{k,\nu}$ that could play the same role as $C_{k,\nu}$ in this paper:
\begin{align}
\frac{\g_{2\De_\f+k-1}(\nu)^2}{a_{\De_\f}} \tl C_{k,\nu}[\cG]
&= 
\begin{cases}
\Phi_{-2,\nu}[\cG^{(\frac{k-2}{4})}]  & \textrm{if}\ k\equiv 2 \mod 4\\
\frac{\g_{2\De_\f+4-1}(\nu)^2}{a_{\De_\f}} C_{4,\nu}[\cG^{(\frac{k-4}{4})}] & \textrm{if}\ k\equiv 0 \mod 4\,,
\end{cases}
\end{align}
where $C_{4,\nu}$ is given by (\ref{eq:cfoursuper}) (or any similar dispersive functional with the same large-$\nu$ limit). Analogous sum rules could be useful for spinning correlators.

The relation (\ref{eq:largenuktwokernel}) together with (\ref{eq:largenukernelrelation}) also gives a way to compute the large-$\nu$ limit of the kernel for $C_{k,\nu}$. The large-$\nu$ limit of (\ref{eq:largenuktwokernel}) is controlled by the large-spin limit of a conformal block, which was computed in \cite{Kravchuk:2018htv}. Using that result, we find
\begin{align}
\label{eq:largenukernelexpression}
\lim_{\nu\gg 1} \frac{\g_{2\De_\f+k-1}(\nu)^2}{a_{\De_\f}} \tl \Omega_{k,\nu}(r,\eta) &= \frac{(1-r^4)r^{k-2}}{((1+r^2)^2-4 r^2 \eta^2)^{\frac{k+1}{2}}}\cP_{\frac{2-d}{2}+i\nu}(\eta)\,.
\end{align}
Note that $\cP_{\frac{2-d}{2}+i\nu}(\eta)$ is given by (\ref{eq:limitofgegenbauer}) at large-$\nu$, but we have chosen to keep it explicit here.
We could alternatively arrive at this result by noting that at large-$\nu$, the $v$ integral in (\ref{Bkv r eta}) localizes to $v=\eta^2$.

\subsection{Superconvergence at spins $J=-2,-4,\dots$}
\label{sec:evenspinsuperconvergence}

In this section, we given a direct argument for the existence of superconvergence sum rules 
\begin{align}
\Phi_{J,\nu}[\cG]=0\,,\qquad J=-2,-4,\dots\,,
\end{align} 
which are valid whenever the four-point function $\cG$ decays faster than the given spin $J$ in the Regge limit. Such sum rules usually cannot be applied to physical correlators, since physical correlators typically do not possess spin $-2$ or faster Regge decay. However, they can be applied to subtracted correlators like $\cG^{(k)}$ defined in (\ref{eq:thesubtractedcorrelator}).

We follow the notation and conventions of \cite{Kravchuk:2018htv}. We start from the bilocal integral used in the construction of (even signature) light-ray operators
\begin{align}
\label{eq:bilocallightray}
\mathbb{O}_{\De,J}(x_0,z) &= 
\int d^d x_1 d^d x_2 K_{\De,J}(x_1,x_2;x_0,z) \f(x_1)\f(x_2) + (1\leftrightarrow 2)\,.
\end{align}
Here, the kernel $K$ is given by 
\begin{align}
\label{eq:lightraykernel}
K_{\De,J}(x_1,x_2;x_0,z) &= r_{\De,J} \frac{(-2z\.x_{20}x_{10}^2+2z\.x_{10} x_{20}^2)^{1-\De}}{
(x_{12}^2)^{\frac{2d -2\De_\f-(1-J)+(1-\De)}{2}}
(-x_{10}^2)^{\frac{(1-J)+(1-\De)}{2}}
(-x_{20}^2)^{\frac{(1-J)+(1-\De)}{2}}
}\,,
\end{align}
where
\begin{align}
\label{eq:prefactorweneedtocontinue}
r_{\De,J} &= -i\frac{}{} \frac{\G(J+\tfrac d 2)\G(d+J-\De)\G(\De-1)}{\pi^{d-1} \G(J+1)\G(\De-\tfrac d 2)\G(\tfrac{d-\De+J}{2})^2}\,.
\end{align}
See \cite{Kravchuk:2018htv} for a description of the region of integration for (\ref{eq:bilocallightray}).
The key property of $\mathbb{O}_{\De,J}$ is that its (time-ordered) correlation function with a pair of other operators is proportional to $C(\De,J)$:
\begin{align}
\<\f(x_4)\mathbb{O}_{\De,J}^\pm(x_0,z)\f(x_3)\> &= C(\De,J) \x (\textrm{standard factors})\,.
\end{align}

We see immediately from (\ref{eq:prefactorweneedtocontinue}) that the prefactor $r_{\De,J}$ vanishes when $J=0,-1,\dots$. However, this does not imply that $\mathbb{O}_{\De,J}$ itself vanishes, since the zero in $r_{\De,J}$ can be cancelled by a pole in the kinematic factors in (\ref{eq:lightraykernel}), leaving a nontrivial distribution. Explicitly taking the limit $J\to -m$ with $m=2,4,\dots$ using the techniques of \cite{Chang:2020qpj}, we find
\begin{align}
\label{eq:theointegral}
\mathbb{O}_{\De,-m}(x_0,z) &= i \frac{\Gamma (m) \Gamma (\frac{d}{2}-m) 
 \Gamma (d-m-\Delta )
 }{
 \pi ^{d-1} 
 \Gamma (\Delta -\frac{d}{2}) 
 }
 \frac{
 \Gamma (\frac{\Delta-m}{2})^2 
 }{
 \Gamma (\frac{d-\Delta -m}{2})^2
  }\nn\\
  &\quad \x \int d^dx_1 d^d x_2 \frac{(-2z\.x_{20})^{\frac{m-\De}{2}}(-2z\.x_{10})^{\frac{m-\De}{2}}}{
(x_{12}^2)^{\frac{2d-2\De_\f - m-\De}{2}}
}
 \nn\\
 & \quad\quad \x\sum_{k+l=m-1}
(-1)^l\frac{(\frac{\Delta -m}{2})_k (\frac{\Delta -m}{2})_l}{k! l!}
  \frac{\delta^{(k)}(-x_{10}^2) \delta^{(l)}(-x_{20}^2)}{(-2z\.x_{20})^{k}(-2z\.x_{10})^{l}}[\f_1(x_1),\f_2(x_2)]\,.
\end{align}
We see that the integrand contains a distribution with support where $x_{10}^2$ and $x_{20}^2$ simultaneously vanish, so that $\f(x_1)$ and $\f(x_2)$ are constrained to be spacelike-separated. Furthermore, the integrand is proportional to the commutator $[\f(x_1),\f(x_2)]$. It follows that (\ref{eq:theointegral}) vanishes inside a correlation function, as long as the integral converges. The expression (\ref{eq:theointegral}) is valid as long as $d/2 -m$ does not vanish. It would be interesting to understand what replaces the kernel (\ref{eq:theointegral}) when $d/2-m$ vanishes.


\section{Computing $\omega|_\mathrm{light}$ from the Regge moments of $\omega$}
\label{app:omegalight}

In this appendix, we explain how the light contribution to a dispersive sum rule
\begin{align}
\omega|_\mathrm{light} &= \sum_{\De,J\ \mathrm{light}} p_{\De,J} \omega[G_{\De,J}^s] 
\end{align}
can be computed from a contour integral around the $u$-channel Regge limit. Suppose that $\omega$ is a physical functional whose action on blocks is defined by a contour integral of the form
\begin{align}
\w[G_{\De,J}^s] &= \int_{-\oo}^0 \int_{-\oo}^0 \frac{dw d\bar w}{(2\pi i)^2} \mathrm{dDisc}_s \left[\cH(w,\bar w) G_{\De,J}^s(w,\bar w)\right].
\end{align}
The kernel $\cH(w,\bar w)$ should be such that $\mathrm{dDisc}_s$ can be commuted through $\cH(w,\bar w)$ up to distributional terms with support away from the $u$-channel Regge limit. This is indeed the case for the $B_{k,v}$ functionals, see (\ref{Bkv dDisc}).

We begin by approximating the light dimensions and OPE coefficients by those of a tree-level bulk effective field theory whose four-point function is $\cG^\mathrm{EFT}$. This gives
\begin{align}
\left.\w\right|_{\mathrm{light}} &= \int_{-\oo}^0 \int_{-\oo}^0 \frac{dw d\bar w}{(2\pi i)^2} \mathrm{dDisc}_s \cK(w,\bar w),\quad \textrm{where}\quad
\cK(w,\bar w) \equiv  \cH(w,\bar w) \cG^\mathrm{EFT}(w,\bar w)\,.
\label{eq:thingwewanttoevaluate}
\end{align}
Note that $\cG^\mathrm{EFT}$ may grow with spin $>1$ in the Regge limit $w,\bar w\to \oo$. However, we assume that  $\mathrm{dDisc}_s \cG^\mathrm{EFT}(w,\bar w)$ decays sufficiently quickly along the negative $w,\bar w$ axes that (\ref{eq:thingwewanttoevaluate}) is well-defined. This is indeed true when $\cG^\mathrm{EFT}$ is any finite sum of Witten diagrams.

Our goal is to reexpress (\ref{eq:thingwewanttoevaluate}) as a contour integral along an ``arc at infinity'' near the $u$-channel Regge limit. Our strategy will be to deform the $w,\bar w$ contours towards the right cut $w,\bar w\in[1,\oo)$. The integral of $\mathrm{dDisc}_t$ along the right cut is related to $\w|_\mathrm{light}$ by crossing anti-symmetry of $\omega$.

We begin by writing
\begin{align}
\left.\w\right|_{\mathrm{light}} &= \lim_{R\to \oo} I_\mathrm{left}(R)\,,
\end{align}
where
\begin{align}
\label{eq:ileft}
I_\mathrm{left}(R) &= -\frac 1 2 \int_{-R}^{\frac 1 2} \frac{dw}{2\pi i} \int_{-R}^{\frac 1 2} \frac{d\bar w}{2\pi i}\nn\\
&\quad \Big[\cK(w+i\e,\bar w+i\e)+\cK(w-i\e,\bar w-i\e)-\cK(w+i\e,\bar w-i\e)-\cK(w-i\e,\bar w + i\e)\Big]\,.
\end{align}
Here, we used the fact that $\cK(w,\bar w)$ has no branch cut in a neighborhood of $w,\bar w\in [0,1]$ so that we can extend the integration region to include the segment $w,\bar w\in[0,1/2]$.

Now consider the four different terms in the integrand of (\ref{eq:ileft}). The last two terms are in Euclidean kinematics, where the imaginary parts of $w$ and $\bar w$ have opposite signs. The four-point function in this region is controlled by the $u$-channel OPE. The integrand is suppressed at large $w$ or large $\bar w$, so we can safely rotate the $w$ and $\bar w$ contours to lie along $[\frac 1 2,R+1]$, up to a contribution near infinity that is suppressed by a power of $R$,
\begin{align}
& \int_{-R}^{\frac 1 2} \frac{dw}{2\pi i} \int_{-R}^{\frac 1 2} \frac{d\bar w}{2\pi i} \p{-\cK(w+i\e,\bar w-i\e)-\cK(w-i\e,\bar w + i\e)}\nn\\
&= \int_{R+1}^{\frac 1 2} \frac{dw}{2\pi i} \int_{R+1}^{\frac 1 2} \frac{d\bar w}{2\pi i} \p{-\cK(w+i\e,\bar w-i\e)-\cK(w-i\e,\bar w + i\e)} + O(1/R^\#)\,.
\end{align}

Now consider the first two terms in (\ref{eq:ileft}), where the imaginary parts of $w$ and $\bar w$ have the same sign. These terms are in Regge kinematics. Let us change variables to radial and angular coordinates centered around $w=\bar w = \frac 1 2$:
\begin{align}
w-\frac 1 2 = yt\,,\quad \bar w - \frac 1 2 = y/t\,.
\end{align}
The integration region is $t\in[0,\oo]$ and $y\in[0,-R_t]$, where $R_t\equiv-(R+\frac 1 2)\min(t,1/t)$.  Consider the term where $w,\bar w$ are slightly below the left cut. This is equivalent to choosing $y$ slightly below the left cut. We now rotate the $y$ contour to lie below the right cut, keeping $t$ fixed. In doing so, we pick up a contribution from an arc near infinity with radius $R_t$, which we denote $\arcbelowright_{R_t}$:
\begin{align}
& \int_{-R}^{\frac 1 2} \frac{dw}{2\pi i}
\int_{-R}^{\frac 1 2} \frac{d\bar w}{2\pi i} \cK(w-i\e,\bar w-i\e) \nn\\
&=\frac{1}{(2\pi i)^2} \int_0^\oo \frac{dt}{t} \int_{-R_t-i\e}^0 (-2y dy) \cK(w,\bar w) \nn\\
&= \frac{1}{(2\pi i)^2}\int_0^\oo \frac{dt}{t} \p{\int_{R_t-i\e}^0 +\int_{\arcbelowright_{R_t}}} (-2y dy)\,\cK(w,\bar w) \nn\\
&= \int_{R+1}^{\frac 1 2} \frac{dw}{2\pi i} \int_{R+1}^{\frac 1 2} \frac{d\bar w}{2\pi i} \cK(w-i\e,\bar w-i\e) + \frac{1}{(2\pi i)^2}\int_0^\oo \frac{dt}{t}\int_{\arcbelowright_{R_t}} (-2y dy)\,\cK(w,\bar w)\,.
\end{align}
Similarly, we have
\begin{align}
& \int_{-R}^{\frac 1 2} \frac{dw}{2\pi i}
\int_{-R}^{\frac 1 2} \frac{d\bar w}{2\pi i} \cK(w+i\e,\bar w+i\e) \nn\\
&= \int_{R+1}^{\frac 1 2} \frac{dw}{2\pi i} \int_{R+1}^{\frac 1 2} \frac{d\bar w}{2\pi i} \cK(w+i\e,\bar w+i\e) + \frac{1}{(2\pi i)^2}\int_0^\oo \frac{dt}{t}\int_{\arcaboveright_{R_t}} (-2y dy)\,\cK(w,\bar w)\,.
\end{align}
Putting everything together, we find
\begin{align}
I_\mathrm{left}(R) - I_\mathrm{right}(R) &= -\frac 1 2 \frac{1}{(2\pi i)^2}\int_0^\oo \frac{dt}{t}\p{\int_{\arcbelowright_{R_t}}(-2y dy)\,\cK(w,\bar w)+\int_{\arcaboveright_{R_t}}(-2y dy)\,\cK(w,\bar w)} + O(1/R^{\#})\,, 
\end{align}
where $I_\mathrm{right}$ is an analogous integral to $I_\mathrm{left}$ with both $w$ and $\bar w$ wrapping the right cut.

We are interested in applications of this formula to Witten contact and exchange diagrams, which have the property that $\dDisc_{s}$ vanishes close to the $u$-channel Regge limit. We assume that the kernel $\cH$ does not modify this property. Thus, we have
\begin{align}
\cK(w+i\e,\bar w+i\e)+\cK(w-i\e,\bar w-i\e)-\cK(w+i\e,\bar w-i\e)-\cK(w-i\e,\bar w + i\e) &\to 0\nn\\
& (|w|,|\bar w| \gg 1)\,.
\end{align}
Also in the large $|w|,|\bar w|$ regime, the last two terms go to zero, since they are associated to the OPE regime in the $u$ channel. Thus, we actually have
\begin{align}
\cK(w+i\e,\bar w+i\e)+\cK(w-i\e,\bar w-i\e) \to 0 \qquad (|w|,|\bar w| \gg 1)\,.
\end{align}
Consequently, for large $R$, the integral over the upper arc $\arcaboveright_{R_t}$ becomes minus the continuation of the integral over the lower arc $\arcbelowright_{R_t}$. In the limit $R\to \oo$, we can replace the sum of integrals with a single circular contour
\begin{align}
\lim_{R\to\oo}[I_\mathrm{left}(R) - I_\mathrm{right}(R)]
&= 
\lim_{R\to \oo} \frac 1 2\frac{1}{(2\pi i)^2}\int_0^\oo \frac{dt}{t} \int_{\circlearrowleft_{R_t}}(-2y dy)\,\cK_+(w,\bar w) \nn\\
&= \frac 1 2 \frac{1}{(2\pi i)^2} \int_0^\oo \frac{dt}{t} \oint (-2y dy)\, \cK_+(w,\bar w)\,.
\end{align}
where the final contour encircles $y=\oo$. Here, $\cK_+$ indicates that $\cK(w,\bar w)$ should be evaluated for $w,\bar w$ in the upper half-plane, and analytically continued from there.

When $\omega$ is $s$-$t$ antisymmetric, $I_\mathrm{left}=-I_\mathrm{right}$, so we obtain
\begin{align}
\left.\w\right|_\mathrm{light} &= \frac 1 4 \frac{1}{(2\pi i)^2} \int_0^\oo \frac{dt}{t} \oint (-2y dy)\, \cK_+(w,\bar w)\,.
\end{align}
Finally changing variables to $r,\eta$, we find
\begin{align}
\left.\w\right|_\mathrm{light} &= -\frac 1 4 \frac{1}{(2\pi i)^2} \int_1^\oo d\eta \oint dr \frac{\ptl(w,\bar w)}{\ptl(r,\eta)} \cK_+(w,\bar w)\,,
\end{align}
where $\frac{\ptl(w,\bar w)}{\ptl(r,\eta)}=\frac{1+\left(2-4 \eta ^2\right) r^2+r^4}{8  r^3\sqrt{\eta ^2-1}}$ and now the contour encircles $r=0$ counterclockwise. To summarize, our result is that $\omega|_\mathrm{light}$ can be computed by replacing
\begin{align}
\int dr (\cdots) \dDisc\,\cG &\to -\frac 1 4 \oint dr (\cdots) \cG_+\,.
\end{align}
Formula (\ref{eq:eftformula}) follows.


\section{Heavy moments at large $\De,J,\nu$}

In sections~\ref{sec:reggemomentsheavy2} and \ref{sec:heavymomentbulkpoint}, we computed the action of $C_{k,\nu}$ on heavy blocks $G_{\De,J}^s$ in two limits: the bulk point limit of large $\De$ with small $J/\De$ (with arbitrary $\nu\leq \De$), and the Regge limit of large $\De,J$ with fixed $\nu$. In this appendix, we compute $C_{k,\nu}[G_{\De,J}^s]$ in a third limit of simultaneously large $\De,J,\nu$. 

As in section~\ref{sec:heavymomentbulkpoint}, we begin with the exact expression (\ref{moment maintext}) for heavy moments $\widehat\Pi_{k,\nu}[G_{\De,J}^s]$ and look for saddle points. We first plug in the split representation of the Gegenbauer function
\begin{align}
\frac{|\cV_e|^{J}}{|\cV_{-e}|^{J+d-2}} \cP_{J} \p{\frac{\cV_e\.\cV_{-e}}{|\cV_e||\cV_{-e}|}} &= \int \frac{D^{d-2} z}{\vol S^{d-2}} \p{-\cV_e\.z}^{J}\p{\cV_{-e}\.z}^{2-d-J}\,.
\end{align}
This gives a simultaneous integral over $x,y,z$, where $z=(1,\vec n)$ with $\vec n\in S^{d-2}$. 

At large $\De,J,\nu$, we find saddle points where $x,y,z$ all lie in the two-dimensional $\pm$ plane, with vanishing transverse positions $\vec x=\vec y=\vec z=0$. There are two saddles that dominate the integral for $\nu < \De-J$. The first is
\begin{align}
\label{eq:largeJsaddle}
(x^+,x^-,y^+,y^-,)&=\p{
-i\frac{ \taushift-\sqrt{\taushift ^2-\nu ^2}}{\nu },
i\frac{ \bar{\taushift}-\sqrt{\bar{\taushift}^2-\nu ^2}}{\nu },
i\frac{ \bar{\taushift}-\sqrt{\bar{\taushift}^2-\nu ^2}}{\nu },
-i\frac{ \taushift -\sqrt{\taushift ^2-\nu ^2}}{\nu },
},\nn\\
(z^+,z^-) &= (2,0)\,,
\end{align}
where $\taushift,\bar\taushift$ are given by (\ref{eq:taushiftdefinition}). Note that when $J$ is small, the saddle point (\ref{eq:largeJsaddle}) becomes one of the bulk point saddles (\ref{eq:bulkpointsaddle}). 
The second dominant saddle is obtained by swapping the $\pm$ directions and replacing $\nu\to -\nu$. The effect of this second saddle is to add a complex conjugate term to the final formula.

Evaluating the saddle point integral and plugging in the extra factors in the large-$\nu$ limit of kernel for $C_{k,\nu}$ (\ref{eq:largenukernelexpression}), we find
\begin{align}
\lim_{\De,J,\nu\to\oo} C_{k,\nu}[\De,J]
&=
(-1)^J 
(S_\nu(\taushift,\bar\taushift)+\mathrm{h.c.})\,,
\label{eq:largenuJdeltaformula}
\end{align}
where
\begin{align}
\label{eq:formulafors}
S_\nu(\taushift,\bar\taushift) &\equiv
\frac{2^{d-3} \Gamma(\tfrac{d-1}{2})}{\pi^{1/2}}
\p{\frac{i \taushift^2 \bar{\taushift}^2}{\nu(\bar{\taushift}^2-\taushift^2)}}^{\frac{d-2}{2}}
\frac{\taushift  \bar{\taushift}+\sqrt{\taushift ^2-\nu ^2} \sqrt{\bar{\taushift}^2-\nu ^2}}{
(\taushift\bar\taushift)^{\frac{k+1}{2}}
(\taushift ^2-\nu ^2)^{\frac{k+3}{4}}
(\bar{\taushift}^2-\nu ^2)^{\frac{k+3}{4}}
}
\nn\\
&\quad \x
\frac{\left(\frac{\sqrt{\taushift ^2-\nu ^2}+i \nu }{\taushift }\right)^{\taushift } \left(\frac{\sqrt{\bar{\taushift}^2-\nu ^2}-i \nu }{\bar{\taushift}}\right)^{\bar{\taushift}}  \left(\frac{\sqrt{\taushift ^2-\nu ^2}+\taushift }{\sqrt{\bar{\taushift}^2-\nu ^2}+\bar{\taushift}}\right)^{i \nu } 
}{
 \left(\frac{\bar{\taushift} \sqrt{\taushift ^2-\nu ^2}+\taushift  \sqrt{\bar{\taushift}^2-\nu ^2}}{2} \right)^{\frac{d}{2}-2}}\,.
\end{align}
One can check using (\ref{eq:limitofgegenbauer}) that the large $(\De,J)$ limit of (\ref{eq:largenuJdeltaformula}) agrees with the large $\nu$ limit of the Regge limit formula (\ref{eq:bknuinreggelimit}), and that the large $(\De,\nu)$ limit of (\ref{eq:largenuJdeltaformula}) agrees with the large-$J$ limit of the bulk point formula (\ref{eq:bulkpointbknu}).

\subsection{Oscillatory contribution to $\omega_\mathrm{AdS}$}
\label{sec:oscregge}

In section \ref{sec:posunimprovedRegge}, we checked that the oscillatory contribution to $\omega_\mathrm{AdS}^\mathrm{unimproved}$ coming from the $\nu=M$ endpoint of the integral (\ref{eq:cdeltajreggething}) is not large when $m\gg M$.
In this appendix, we check that this conclusion is unchanged for $m\sim M$ and $J \gg 1$. In other words, oscillatory terms are not enhanced outside the regime $m\gg M$, even outside the flat-space region $\beta\ll 1$.

To compute the oscillatory contribution to $\omega_\mathrm{AdS}^\mathrm{unimproved}$, we expand the integral 
\begin{align}
\omega_\mathrm{AdS}^\mathrm{unimproved}[\De,J] &= \int_0^M \frac{d\nu}{M}f\p{\frac\nu M} C_{2,\nu}[\De,J]
\end{align}
near $\nu=M$. We are interested in the case $m,J\gg 1$, but not necessarily $m\gg M$. Since $\nu$ is large, we must use (\ref{eq:largenuJdeltaformula}). Writing $\nu=M(1-x)$, we have
\begin{align}
S_\nu(\taushift,\bar \taushift) \approx S_M(\taushift,\bar \taushift)e^{-iM \f x}\,,
\end{align}
where
\begin{align}
\f &\equiv \log \frac{\sqrt{\taushift ^2-M^2}+\taushift }{\sqrt{{\bar\taushift}^2-M^2}+\bar \taushift}\,.
\end{align}
Using $f(p)\sim f_1(1-p)^l$, the integral near $\nu=M$ can be approximated by
\begin{align}
\left.\omega_\mathrm{AdS}^\mathrm{unimproved}[\De,J]\right|_{\nu=M} &\approx \int_0^\oo dx\,f_1\, x^l S_M(\taushift,\bar\taushift) e^{iM\f x} + \mathrm{h.c.} = \frac{i^{l+1}f_1\Gamma(l+1)}{(M\f)^{l+1}}  S_M(\taushift,\bar\taushift)+\mathrm{h.c.} \nn\\
&\quad (1/M\ll \beta)\,.
\label{eq:reggeosc}
\end{align}
In the limit of large $M$ with $m^2=\taushift\bar\taushift \geq M^2$, this contribution scales like $m^{-4}M^{-\frac d 2 - l}$, which is the same as the oscillatory term (\ref{eq:oscillatoryconttoc2}) in the bulk point limit. In particular, (\ref{eq:reggeosc}) is subdominant to (\ref{eq:cintermsofq}) assuming $l>n+1-\frac d 2$.


\section{Some anomalous dimensions}
\label{app:anomalousDims}
Here we record the anomalous dimensions coming from various terms in the EFT. Let us start with the four-derivative contact diagram, defined by the Mellin amplitude\footnote{Our convention for the Mellin representation is in \eqref{eq:mellinRep2} and the normalization $c_n$ is given in \eqref{eq:cNormalization}.} $M(\mS,\mT) = g_2 c_2 (\mS^2+\mT^2+\mU^2)$. Its $\ell=2$ anomalous dimensions take the form
\ba
&\gamma_{n,2}|_{g_2} = \\
&-\frac{\Gamma\!\left(\frac{d}{2}+n+2\right) \Gamma(n+\Df +2) \Gamma\!\left(\Df-\frac{d}{2}+n +\frac{1}{2}\right) \Gamma\!\left(2 \Df -\frac{d}{2}+n+2\right)}{2^{d} \pi ^{\frac{d}{2}}\Gamma \left(\frac{d+4}{2}\right) \Gamma(n+1) \Gamma\!\left(\Df+n +\frac{5}{2}\right) \Gamma\!\left(\Df-\frac{d}{2}+n +1\right) \Gamma(2 \Df-d+n +1)}\,.
\ea
In the case of the six-derivative contact $M(\mS,\mT) = g_3 c_3\,\mS\mT \mU$, we find
\ba
&\gamma_{n,2}|_{g_3} = \left[\Df  (-d+4 \Df +4)+n (-d+4 \Df +4)+2 n^2\right]\times\\
&\frac{\Gamma \left(\frac{d}{2}+n+2\right) \Gamma (n+\Df +2) \Gamma \left(-\frac{d}{2}+n+\Df +\frac{1}{2}\right) \Gamma \left(-\frac{d}{2}+n+2 \Df +2\right)}{2^{d} \pi ^{\frac{d}{2}}\Gamma \left(\frac{d}{2}+2\right) \Gamma (n+1) \Gamma \left(n+\Df +\frac{5}{2}\right) \Gamma \left(-\frac{d}{2}+n+\Df +1\right) \Gamma (-d+n+2 \Df +1)}\,.
\ea
Finally, let us record the anomalous dimensions in pure gravity, i.e.\ in the sum of the graviton exchange Witten diagram in the three channels, normalized as in \eqref{eq:gravityNormalization}. For $\ell = 2$, we find
\ba
\gamma^{\text{pure}}_{0,2} &=8\pi G \left[-32 \Df ^3+16 (d-6) \Df ^2+8 (4 d-11) \Df+d^3+d^2 +10 d-24\right]\times\\
&\frac{ \Gamma (\Df +1)^2 \Gamma \left(-\frac{d}{2}+2 \Df +2\right)}{2 \pi ^{d/2} (d-1)d(d+2) (2 \Df +3)\Gamma (2 \Df +2) \Gamma \left(-\frac{d}{2}+\Df +1\right) \Gamma \left(-\frac{d}{2}+\Df +2\right)}
\ea
and
\ba
\gamma^{\text{pure}}_{1,2} &=8 \pi  G\left[d^5 (\Df +1)^2+
d^4 \left(-23 \Df ^2-110 \Df -111\right)+\right.\\
&\qquad\qquad\left.2 d^3 \left(8 \Df ^4+160 \Df ^3+721 \Df ^2+1174 \Df +641\right)-\right.\\
&\qquad\qquad\left.8 d^2 (\Df +1) (2 \Df +3) (2 \Df +5) \left(\Df ^2+15 \Df +27\right)+\right.\\
&\qquad\qquad\left.16 d \left(8 \Df ^5+84 \Df ^4+314 \Df ^3+521 \Df ^2+377 \Df +88\right)-\right.\\
&\qquad\qquad\left.128 (\Df +1)^2 (\Df +2) (2 \Df +3) (2 \Df +5)\right]\times\\
&\!\!\!\!\!\!\!\!\!\!\!\!\frac{\Gamma (\Df +1)^2 \Gamma \left(-\frac{d}{2}+2 \Df +3\right)}{2 (d-1) d (d+2) (d+4) \pi ^{d/2} (2 \Df +5)\Gamma (2 \Df +4) \Gamma \left(-\frac{d}{2}+\Df +1\right) \Gamma \left(-\frac{d}{2}+\Df +3\right)}\,.
\ea

\bibliographystyle{./aux/ytphys}
\bibliography{./aux/refs}

\end{document}